# Projective Limits of State Spaces
# III. Toy-Models

**Suzanne Lanéry[1,2] and Thomas Thiemann[1]**

[1] Institute for Quantum Gravity, Friedrich-Alexander University Erlangen-Nürnberg, Germany

[2] Mathematics and Theoretical Physics Laboratory, François-Rabelais University of Tours, France

**Abstract**

In this series of papers, we investigate the projective framework initiated by Jerzy Kijowski [7] and Andrzej Okołów [12, 13], which describes the states of a quantum theory as projective families of density matrices. A strategy to implement the dynamics in this formalism was presented in our first paper [10], which we now test in two simple toy-models. The first one is a very basic linear model, meant as an illustration of the general procedure, and we will only discuss it at the classical level. In the second one, we reformulate the Schrödinger equation, treated as a classical field theory, within this projective framework, and proceed to its (non-relativistic) second quantization. We are then able to reproduce the physical content of the usual Fock quantization.

## Contents

# 1 Introduction

In [10, section 3], we introduced a strategy to deal with dynamical constraints in a projective limit of symplectic manifolds. After having convinced ourselves that a regularization of these constraints will in general be necessary, since we cannot expect them to be adapted to the projective system, we adopted the perspective that a dynamical state can be identified with the family of successive approximations approaching an exact solution of the dynamics. On the one hand, this allows us to put the dynamical state space into a projective form. On the other hand, it also provides a suitable ground for a notion of convergence, that will make it possible to define meaningful physical observables on this state space.

However, applying this procedure demands that one sets up a regularization scheme fulfilling a number of restrictive properties (summarized in [10, prop. 3.23]), which raises the question of its practicability. Hence, we now want to discuss two simple examples, meant as 'proofs of concept' that such schemes can indeed be designed.

Note that the framework in [10, section 3] was purely classical. We have not yet undertaken to formulate a general procedure regarding the resolution of dynamical constraints in projective systems of quantum state spaces [7, 13, 11]. Nevertheless, our second example will explore how analogous ideas can be implemented at the quantum level, and will give us the opportunity to delineate an appropriate course and to underline possible difficulties.

# 2 Linear constraints on a Kähler vector space

This first example is arguably mostly artificial and does not pretend to have great physical relevance. Our motivation here is to illustrate the concepts introduced in [10, sections 2 and 3] in the simplest possible setup. We consider an infinite dimensional Hilbert space $\mathcal{H}$ (which is nothing but a linear Kähler manifold) and form its rendering by a projective structure of finite dimensional Hilbert spaces (to prevent any confusion: the Hilbert spaces in discussion here are the phase spaces of classical systems, there will be nothing quantum in the present section). This rendering is built from an Hilbert basis of $\mathcal{H}$ by considering all the vector subspaces of $\mathcal{H}$ spanned by a finite number of basis vectors and linking them by orthogonal projections (a more satisfactory rendering for $\mathcal{H}$, namely one that does not require the choice of a preferred basis, will be presented in section 3; however we do not want to use it here, since the constraints we will be looking at could be directly formulated as an elementary reduction over a cofinal part of its label set, and it would therefore not be appropriate as an example for the regularization procedure).

**Proposition 2.1** Let $\mathcal{H}, \langle \cdot\,, \cdot \rangle$ be a complex Hilbert space and define:

1. $\forall v \in \mathcal{H},\ J\,v := i\,v$;
2. $\forall v, w \in \mathcal{H},\ \Omega(v,\ w) := 2\,\mathrm{Im}\left(\langle v,\ w\rangle\right)$.

Then, $\mathcal{H}$, $\Omega$, $J$ is a Kähler manifold.

**Proof** The real scalar product $\mathrm{Re}\,\langle\,\cdot\,,\,\cdot\,\rangle$ equips $\mathcal{H}$ (seen as a real vector space) with a structure of real Hilbert space, therefore, any bounded real-valued real-linear form on $\mathcal{H}$ can be written as $\mathrm{Re}\,\langle v,\,\cdot\,\rangle = 2\,\mathrm{Im}\,\left\langle -\frac{i}{2}\,v,\,\cdot\,\right\rangle = \Omega\left(-\frac{i}{2}\,v,\,\cdot\,\right)$ for some $v \in \mathcal{H}$. Hence, $\Omega$ is a strong symplectic structure.

Next, $J$ is by construction a complex structure on $\mathcal{H}$. We have $\forall v, w \in \mathcal{H}$, $\Omega(i\,v,\,i\,w) = \Omega(v,\,w)$, and $v \mapsto \Omega(v,\,i\,v) = 2\,\mathrm{Re}\,\langle v,\,v\rangle$ is positive definite.

The integrability conditions for $\Omega$ and $J$ are trivially satisfied since we actually have a Kähler *vector space*. □

**Proposition 2.2** Let $\mathcal{H}$ be a separable, infinite dimensional Hilbert space (equipped with the strong symplectic structure $\Omega$ defined in prop. 2.1) and let $(e_i)_{i\in\mathbb{N}}$ be an Hilbert basis of $\mathcal{H}$. We define:

1. $\mathcal{L} := \{I \subset \mathbb{N} \mid 0 < \#I < \infty\}$ equipped with the preorder defined by $\subset$;
2. $\forall I \in \mathcal{L}$, $\mathcal{H}_I := \mathrm{Vect}\,\{e_i \mid i \in I\}$ equipped with the induced symplectic structure $\Omega_I$ (which is also the natural symplectic structure on $\mathcal{H}_I$ as a finite dimensional Hilbert space);
3. $\forall I \subset I' \in \mathcal{L}$, $\pi_{I'\to I} := \Pi_I|_{\mathcal{H}_{I'}\to\mathcal{H}_I}$ where $\Pi_I$ is the orthogonal projection on $\mathcal{H}_I$;
4. $\mathcal{H}_{\mathbb{N}} := \mathcal{H}$ and $\forall I \in \mathcal{L}$, $\pi_{\mathbb{N}\to I} := \Pi_I|_{\mathcal{H}\to\mathcal{H}_I}$.

Then, this defines a rendering [10, def. 2.6] of the symplectic manifold $\mathcal{H}$ by the projective system of phase spaces $(\mathcal{L},\,\mathcal{H},\,\pi)^{\downarrow}$. We define $\sigma_{\downarrow} : \mathcal{H} \to \mathcal{S}^{\downarrow}_{(\mathcal{L},\mathcal{H},\pi)}$ as in [10, def. 2.6].

Additionally, defining the dense vector subspace of $\mathcal{H}$, $\mathcal{D} := \mathrm{Vect}\,\{e_i \mid i \in \mathbb{N}\}$ (without completion, ie. the space of finite linear combinations of the $e_i$), we have a bijective antilinear map $\zeta : \mathcal{D}^* \to \mathcal{S}^{\downarrow}_{(\mathcal{L},\mathcal{H},\pi)}$ such that $\zeta^{-1} \circ \sigma_{\downarrow} : \mathcal{H} \to \mathcal{D}^*$ is the canonical identification of $\mathcal{H}$ with $\mathcal{D}' \subset \mathcal{D}^*$ (where $\mathcal{D}^*$ is the algebraical dual of $\mathcal{D}$ and $\mathcal{D}'$ the topological one).

**Proof** $\mathcal{L}$ is a directed set, since $\forall I, I' \in \mathcal{L}$, $I \cup I' \in \mathcal{L}$ and $I, I' \subset I \cup I'$.

Let $I, I' \in \mathcal{L} \sqcup \{\mathbb{N}\}$ with $I \subset I'$. $\pi_{I'\to I}$ is surjective by construction. Next, since $\mathcal{H}_I$ is closed, we have, for any bounded real-valued real-linear form $\upsilon$ on $\mathcal{H}_I$, a vector $\underline{\upsilon} \in \mathcal{H}_I$ such that:

$$\forall v \in \mathcal{H}_I,\ \upsilon(v) = \Omega_I(\underline{\upsilon},\,v) = \mathrm{Re}\,\langle 2i\,\underline{\upsilon},\,v\rangle_I\,.$$

Hence, since $\Pi_I$ is the $\mathbb{C}$-orthogonal projection on the complex vector subspace $\mathcal{H}_I$, it is also the $\mathbb{R}$-orthogonal projection on the real vector subspace $\mathcal{H}_I$, and we have:

$$\forall v \in \mathcal{H}_{I'},\ \upsilon \circ \pi_{I'\to I}(v) = \mathrm{Re}\,\langle 2i\,\underline{\upsilon},\,\Pi_I\,v\rangle_I = \mathrm{Re}\,\langle 2i\,\underline{\upsilon},\,v\rangle_{I'} = \Omega_{I'}(\underline{\upsilon},\,v)\,,$$

and therefore $\pi_{I'\to I}\left(\underline{\upsilon \circ \pi_{I'\to I}}\right) = \pi_{I'\to I}\left(\underline{\upsilon}\right) = \underline{\upsilon}$.

Clearly for $I \in \mathcal{L}$, we have $\pi_{I\to I} = \mathrm{id}_{\mathcal{H}_I}$ and for $I, I', I'' \in \mathcal{L} \sqcup \{\mathbb{N}\}$ with $I \subset I' \subset I''$, $\pi_{I'\to I} \circ \pi_{I''\to I'} = \pi_{I''\to I}$.

Lastly, we define:

$$\begin{array}{rccl} \zeta & : & \mathcal{D}^* & \to & \mathcal{S}^{\downarrow}_{(\mathcal{L},\mathcal{H},\pi)} \\ & & \upsilon & \to & \left(\overline{\upsilon|_{\mathcal{H}_I}}\right)_{I\in\mathcal{L}} \end{array}.$$

where for all $l \in \mathcal{L}$, $\overline{(\,\cdot\,)} : \mathcal{H}_l^* \to \mathcal{H}_l$ is the canonical identification provided by the complex Hilbert space structure on $\mathcal{H}_l$ ($\mathcal{H}_l$ is finite dimensional, hence $\mathcal{H}_l^* = \mathcal{H}_l'$).

The map $\zeta$ is well-defined, since $\forall l \subset l' \in \mathcal{L}$, $\forall v \in \mathcal{H}_l$, $\left\langle \pi_{l' \to l}\left(\overline{\upsilon|_{\mathcal{H}_{l'}}}\right), v \right\rangle_l = \left\langle \overline{\upsilon|_{\mathcal{H}_{l'}}}, v \right\rangle_{l'} = \upsilon(v) = \left\langle \overline{\upsilon|_{\mathcal{H}_l}}, v \right\rangle_l$, hence $\pi_{l' \to l}\left(\overline{\upsilon|_{\mathcal{H}_{l'}}}\right) = \overline{\upsilon|_{\mathcal{H}_l}}$.

On the other hand, we define $\widetilde{\zeta} : \mathcal{S}^{\downarrow}_{(\mathcal{L},\mathcal{H},\pi)} \to \mathcal{D}^*$, by:

$$\forall \left(v_l\right)_{l\in\mathcal{L}}, \forall w \in \mathcal{D}, \widetilde{\zeta}\left(\left(v_l\right)_{l\in\mathcal{L}}\right)(w) = \langle v_l, w\rangle_l \text{ for any } l \in \mathcal{L} \text{ such that } w \in \mathcal{H}_l.$$

The map $\widetilde{\zeta}$ is well-defined since $\mathcal{D} = \bigcup_{l\in\mathcal{L}} \mathcal{H}_l$ and if $l, l' \in \mathcal{L}$ are such that $w \in \mathcal{H}_l \cap \mathcal{H}_{l'}$, then there exists $l'' \in \mathcal{L}$ such that $l, l' \subset l''$ and:

$$\langle v_l, w\rangle_l = \langle \pi_{l''\to l}(v_{l''}), w\rangle_l = \langle v_{l''}, w\rangle_{l''} = \langle v_{l'}, w\rangle_{l'}.$$

Now, we have $\widetilde{\zeta} \circ \zeta = \mathrm{id}_{\mathcal{D}^*}$, $\zeta \circ \widetilde{\zeta} = \mathrm{id}_{\mathcal{S}^{\downarrow}_{(\mathcal{L},\mathcal{H},\pi)}}$ and $\forall v \in \mathcal{H}$, $\forall l \in \mathcal{L}$, $\forall w \in \mathcal{H}_l \subset \mathcal{D}$, $\widetilde{\zeta} \circ \sigma_{\downarrow}(v)(w) = \langle \pi_{\mathbb{N}\to l}(v), w\rangle_l = \langle v, w\rangle_{\mathcal{H}}$. □

We now present the constraint surface of interest, as a real vector subspace of $\mathcal{H}$ admitting a description of a specific form (alternatively, we could characterize it by of a family of linear holomorphic second class constraints and a family of linear first class constraints). Additionally, we anticipate on the regularization of the constraints by providing a rendering (similar to the one we adopted for $\mathcal{H}$) for the corresponding reduced phase space.

**Proposition 2.3** We consider the same objects as in prop. 2.2. Let $\left(f_j\right)_{j\in\mathbb{N}}$ and $\left(g_k\right)_{k\in\mathbb{N}}$ be two, mutually orthogonal, orthonormal families in $\mathcal{H}$. We define:

1. $\mathcal{J} := \overline{\mathrm{Vect}_{\mathbb{C}}\left\{f_j \mid j \in \mathbb{N}\right\}}$ (equipped with the induced symplectic structure $\Omega_{\mathcal{J}}$) and $\mathcal{K}_{\mathbb{R}} := \overline{\mathrm{Vect}_{\mathbb{R}}\left\{g_k \mid k \in \mathbb{N}\right\}}$;
2. $\delta : \mathcal{J} \oplus \mathcal{K}_{\mathbb{R}} \to \mathcal{J}$ by $\delta := \Pi_{\mathcal{J}}|_{\mathcal{J}\oplus\mathcal{K}_{\mathbb{R}}\to\mathcal{J}}$ where $\Pi_{\mathcal{J}}$ is the orthogonal projection on $\mathcal{J}$.

Then $(\mathcal{J}, \mathcal{J} \oplus \mathcal{K}_{\mathbb{R}}, \delta)$ is a phase space reduction of $\mathcal{H}$ [10, def. A.1].

Additionally, we define:

3. $\forall J \in \mathcal{L}$, $\mathcal{J}_J := \mathrm{Vect}_{\mathbb{C}}\left\{f_j \mid j \in J\right\}$ equipped with the induced symplectic structure $\Omega'_J$;
4. $\forall K \in \mathcal{L}$, $\mathcal{K}_K := \mathrm{Vect}_{\mathbb{C}}\left\{g_k \mid k \in K\right\}$ & $\mathcal{K}_{K,\mathbb{R}} := \mathrm{Vect}_{\mathbb{R}}\left\{g_k \mid k \in K\right\}$;
5. $\forall J \subset J' \in \mathcal{L}$, $\pi'_{J'\to J} := \Pi'_J|_{\mathcal{J}_{J'}\to\mathcal{J}_J}$ where $\Pi'_J$ is the orthogonal projection on $\mathcal{J}_J$;
6. $\mathcal{J}_{\mathbb{N}} := \mathcal{J}$ and $\forall J \in \mathcal{L}$, $\pi'_{\mathbb{N}\to J} := \Pi'_J|_{\mathcal{J}\to\mathcal{J}_J}$.

As in prop. 2.2, this provides a rendering of $\mathcal{J}$ by $\left(\mathcal{L}, \mathcal{J}, \pi'\right)^{\downarrow}$ and we define $\sigma'_{\downarrow} : \mathcal{J} \to \mathcal{S}^{\downarrow}_{(\mathcal{L},\mathcal{J},\pi')}$ as well as the bijective antilinear map $\zeta' : \mathcal{F}^* \to \mathcal{S}^{\downarrow}_{(\mathcal{L},\mathcal{J},\pi')}$ where $\mathcal{F} := \mathrm{Vect}\left\{f_j \mid j \in \mathbb{N}\right\}$.

**Proof** $\delta$ is a surjective linear map and for $v \in \mathcal{J}$, we have $\delta^{-1}\langle v\rangle = v + \mathcal{K}_{\mathbb{R}}$, hence $\delta^{-1}\langle v\rangle$ is connected. For $v, w \in \mathcal{J} \oplus \mathcal{K}_{\mathbb{R}}$, we write $v = v' + v''$ and $w = w' + w''$ with $v', w' \in \mathcal{J}$ and $v'', w'' \in \mathcal{K}_{\mathbb{R}}$. Then, we have:

$\Omega(v, w) = 2\,\mathrm{Im}\, \langle v,\, w \rangle_{\mathcal{H}} = 2\,\mathrm{Im}\, \langle v',\, w' \rangle_{\mathcal{H}} + 2\,\mathrm{Im}\, \langle v'',\, w'' \rangle_{\mathcal{H}}$ (since $\mathcal{J} \perp \mathcal{K}_{\mathbb{R}}$)

$= 2\,\mathrm{Im}\, \langle v',\, w' \rangle_{\mathcal{J}} = \Omega_{\mathcal{J}}(\delta(v),\, \delta(w))$ (since $\mathcal{K}_{\mathbb{R}}$ is the real vector subspace generated by an orthonormal family).

Hence, $(\mathcal{J},\, \mathcal{J} \oplus \mathcal{K}_{\mathbb{R}},\, \delta)$ is a phase space reduction of $\mathcal{H}$. □

We are ready to turn to the core of the regularization procedure, namely formulating a set of approached implementations of the constraints (indexed by a label set $\mathcal{E}$), endowing $\mathcal{E}$ with an appropriate preorder, and linking together the approximate dynamics by supplying projecting maps between their reduced phase spaces.

Here we choose $\mathcal{E}$ to enumerate a large class of approximate solutions, ordered by comparing how good they are at approximating the exact solution (the precise definition of $\mathcal{E}$ may at first seem to arise from nowhere but will become transparent when we will actually detail the corresponding approximate constraint surfaces). This way of composing $\mathcal{E}$ will make the study the convergence mostly inexpensive: a large part of the work is actually done beforehand when checking that $\mathcal{E}$ with this preorder is really a directed set.

It also has the advantage of partially getting rid of the arbitrariness inherent of working with an approximating scheme. The philosophy is that an explicit, concretely implemented, approximating scheme will correspond to a specific cofinal part of $\mathcal{E}$, but that we have the option of considering all such particular schemes at the same time, by arranging them into a (huge) set $\mathcal{E}$, provided we carefully tailor its preorder to our purpose.

Besides, note that being quite broad in recruiting suitable approximate theories is, up to a certain extent, forced upon us by the fact that we are dealing with an unphysical and not further specified system, since, in a more realistic example, we could probably, from the physics of the system, infer guiding principles to be more selective.

On the other hand, we could fear that such a loose label set $\mathcal{E}$ will leave us with a disproportionately complicated projective structure for the dynamical theory. But, in fact, this dynamical structure (on $\mathcal{E}\mathcal{L}$) gets spontaneously quotiented down to the projective structure we had already introduced above for the dynamical state space. The idea is that we can transparently match two partial dynamical theories as soon as they have a common ancestor out of which they are carved in the same way (recall this mechanism was presented at the end of [10, subsection 2.2], and expressed precisely in [10, props. 2.8 and 2.9]).

**Definition 2.4** We consider the same objects as in prop. 2.3 and we define $\mathcal{E}$ as the set of all sextuples $\left(I,\, I',\, J,\, K,\, \varphi,\, \epsilon\right)$ such that:

1. $I \subset I' \in \mathcal{L}$ & $J, K \in \mathcal{L}$;
2. $\varphi : \mathcal{J}_J \oplus \mathcal{K}_K \to \mathcal{H}_{I'}$ is a linear application and $\varphi|_{\mathcal{J}_J \oplus \mathcal{K}_K \to \mathrm{Im}\varphi}$ is a unitary map;
3. $\epsilon > 0$ and $\forall v \in \mathcal{J}_J,\ \|v - \varphi(v)\| \leqslant \epsilon\, \|v\|$;
4. $\Pi_I \langle \varphi \langle \mathcal{K}_{K,\mathbb{R}} \rangle \rangle = \Pi_I \langle \mathcal{K}_{K,\mathbb{R}} \rangle$.

On $\mathcal{E}$ we define a preorder $\preccurlyeq$ by $\left(I_1, I'_1, J_1, K_1, \varphi_1, \epsilon_1\right) \preccurlyeq \left(I_2, I'_2, J_2, K_2, \varphi_2, \epsilon_2\right)$ iff:

5. $I_1 \subset I_2$, $I'_1 \subset I'_2$, $J_1 \subset J_2$ & $K_1 \subset K_2$;
6. $\epsilon_2 \leqslant \epsilon_1$.

**Proposition 2.5** We consider the same objects as in def. 2.4. Let $I \in \mathcal{L}$ and $\epsilon > 0$. Let $J, K \in \mathcal{L}$ such that:

$$\dim \Pi_I \langle \mathcal{J}_J \oplus \mathcal{K}_K \rangle = \dim (\mathcal{J}_J \oplus \mathcal{K}_K).$$

Then, there exist $I' \in \mathcal{L}$ and a linear application $\varphi : \mathcal{J}_J \oplus \mathcal{K}_K \to \mathcal{H}_{I'}$ such that $\left(I, I', J, K, \varphi, \epsilon\right) \in \mathcal{E}$.

**Lemma 2.6** Let $\mathcal{H}$ be a Hilbert space and let $F$, $G$ be two finite dimensional vector subspaces of $\mathcal{H}$, such that $\dim \Pi_G \langle F \rangle = \dim F$, where $\Pi_G$ denotes the orthogonal projection on $G$.

Then, there exists a unique linear application $\varphi_{F\to G} : F \to G$ satisfying:

1. $\varphi_{F\to G}\big|_{F\to \mathrm{Im}\varphi_{F\to G}}$ is a unitary map;
2. $\displaystyle\int_{S_F} d\mu_{S_F}(e)\, \left\|e - \varphi_{F\to G}(e)\right\|^2$ is minimal, where $S_F$ is the unit sphere of $F$ equipped with the measure induced by the euclidean structure of $F$.

For $v \in F$, $\displaystyle \|v - \varphi_{F\to G}(v)\| \leqslant 2 \dim F\, \|v\| \sup_{\substack{e\in F\\ \|e\|=1}} \|e - \Pi_G(e)\|$

**Proof** *Existence and uniqueness.* Let $f = \dim F$. From $\dim \Pi_G \langle F \rangle = f$, $\Pi_G$ induces a bijective map $F \to \Pi_G \langle F \rangle$, hence $\langle \Pi_G(\,\cdot\,), \Pi_G(\,\cdot\,)\rangle_G$ defines a positive definite sesquilinear map on $F$. Therefore, there exists an orthonormal basis $(e_i)_{i\in\{1,\dots,f\}}$ such that:

$$\forall i, j \in \{1, \dots, f\}\,,\ \left\langle \Pi_G(e_i), \Pi_G(e_j)\right\rangle = \lambda_i\, \delta_{ij} \text{ with } \lambda_i > 0\,.$$

Let $\varphi$ be a linear application $F \to G$ such that $\varphi|_{F\to\mathrm{Im}\varphi}$ is a unitary map. We define $B_{ij} \in \mathbb{C}$ for $i, j \in \{1, \dots, f\}$ and $w_i \in G \cap (\Pi_G \langle F \rangle)^\perp$ for $i \in \{1, \dots, f\}$ by:

$$\forall i \in \{1, \dots, f\}\,,\ \varphi(e_i) = \frac{1}{\sqrt{\lambda_i}} \Pi_G(e_i) + \sum_j B_{ij} \frac{1}{\sqrt{\lambda_j}} \Pi_G(e_j) + w_i\,.$$

From $\left\langle \varphi(e_i), \varphi(e_j)\right\rangle_G = \delta_{ij}$, we have:

$$\forall i, j \in \{1, \dots, f\}\,,\ B^*_{ij} + B_{ji} + \sum_k B^*_{ik} B_{jk} + \left\langle w_i, w_j\right\rangle = 0\,. \tag{2.6.1}$$

With these notations, we have:

$$\int_{S_F} d\mu_{S_F}(e)\, \|e - \varphi(e)\|^2 = \int_{S_F} d\mu_{S_F}(e)\, \|\Pi_G(e) - \varphi(e)\|^2 + \|e - \Pi_G(e)\|^2$$

$$= \sum_{i,j} \left( \int_{S_{\mathbb{C}^f}} d\mu_{S_{\mathbb{C}^f}}(x)\, x^*_i\, x_j \right) \left\langle \Pi_G(e_i) - \varphi(e_i), \Pi_G(e_j) - \varphi(e_j)\right\rangle + \int_{S_F} d\mu_{S_F}(e)\, \|e - \Pi_G(e)\|^2$$

$$= \sum_i \mathrm{Vol}\,(S_{\mathbb{C}^f}) \left[1 + \lambda_i - 2\sqrt{\lambda_i}\,\mathrm{Re}\,(1 + B_{ii})\right] + \int_{S_F} d\mu_{S_F}(e)\ \|e - \Pi_G(e)\|^2$$

$$= \sum_i \mathrm{Vol}\,(S_{\mathbb{C}^f}) \left[1 + \lambda_i - 2\sqrt{\lambda_i} + \sqrt{\lambda_i}\,\sum_k |B_{ik}|^2 + \sqrt{\lambda_i}\ \|w_i\|^2\right] + \int_{S_F} d\mu_{S_F}(e)\ \|e - \Pi_G(e)\|^2$$

(using eq. (2.6.1)).

Hence, this expression is minimal if and only if $\forall i, j \in \{1, \dots, f\}$ , $B_{ij} = 0$ and $\forall i \in \{1, \dots, f\}$ , $w_i = 0$. Therefore, we define $\varphi_{F\to G}$ by:

$$\forall i \in \{1, \dots, f\}\,,\ \varphi_{F\to G}(e_i) = \frac{1}{\sqrt{\lambda_i}} \Pi_G(e_i)\,.$$

*Bound on* $\|v - \varphi_{F\to G}(v)\|$. Let $v = \sum_{j=1}^{f} v_j\, e_j \in F$. We have:

$$\|v - \varphi_{F\to G}(v)\| \leqslant \sum_{j=1}^{f} \left|v_j\right|\ \left\|e_j - \varphi_{F\to G}(e_j)\right\| \leqslant f\ \|v\|\ \sup_j \left\|e_j - \varphi_{F\to G}(e_j)\right\|.$$

Then, for $j \in \{1, \dots, f\}$, $\left\|e_j - \Pi_G(e_j)\right\|^2 + \lambda_j = 1$ implies:

$$\left|1 - \sqrt{\lambda_j}\right| = \left\|e_j - \Pi_G(e_j)\right\| \frac{\left\|e_j - \Pi_G(e_j)\right\|}{1 + \sqrt{\lambda_j}} \leqslant \left\|e_j - \Pi_G(e_j)\right\|,$$

therefore:

$$\left\|e_j - \varphi_{F\to G}(e_j)\right\| \leqslant \left\|e_j - \Pi_G(e_j)\right\| + \left\|\Pi_G(e_j) - \varphi_{F\to G}(e_j)\right\|$$

$$= \left\|e_j - \Pi_G(e_j)\right\| + \left|\sqrt{\lambda_j} - 1\right| \leqslant 2\ \left\|e_j - \Pi_G(e_j)\right\|.$$

Hence, $\|v - \varphi_{F\to G}(v)\| \leqslant 2f\ \|v\|\ \sup_j \left\|e_j - \Pi_G(e_j)\right\| \leqslant 2f\ \|v\|\ \sup_{\substack{e\in F\\ \|e\|=1}} \|e - \Pi_G(e)\|.$ □

**Proof of prop. 2.5** Since $(e_i)_{i\in\mathbb{N}}$ is an orthonormal basis of $\mathcal{H}$ and $\mathcal{J}_J$ has finite dimension, we can find $I'_1 \in \mathcal{L}$ such that:

1. $\sup_{\substack{e\in\mathcal{J}_J\\ \|e\|=1}} \left\|e - \Pi_{I'_1}(e)\right\| \leqslant \dfrac{\epsilon}{2\#J}$;

and $I'_2$ such that:

2. $I'_2 \cap I = \varnothing \quad \& \quad \dim \Pi_{I'_2} \langle \mathcal{J}_J \rangle + \dim\,(\mathcal{J}_J \cap \mathcal{H}_I) = \dim \mathcal{J}_J$;
3. $\dim \mathcal{H}_{I'_2} \geqslant \dim \Pi_{I'_2} \langle \mathcal{J}_J \rangle + \dim \mathcal{K}_K$.

Let $I' := I \cup I'_1 \cup I'_2$ and $I'_3 := I' \setminus I$. We have $\dim \Pi_{I'_3} \langle \mathcal{J}_J \rangle + \dim\,(\mathcal{J}_J \cap \mathcal{H}_I) = \dim \mathcal{J}_J$ and $\mathcal{K}_K \perp \mathcal{J}_J$, hence for all $k \in K$, there exists $g'_k \in \Pi_{I'_3} \langle \mathcal{J}_J \rangle$ such that:

$$\forall j \in J,\ \left\langle \Pi_I(f_j),\, \Pi_I(g_k)\right\rangle_I + \left\langle \Pi_{I'_3}(f_j),\, g'_k\right\rangle_{I'_3} = 0\,.$$

This holds because, for all families of coefficients $\left(\alpha^j\right)_{j\in J}$ such that $\sum_j \alpha^j \left\langle \Pi_{l'_3}(f_j),\, \cdot\right\rangle_{l'_3} = 0$, we have $\sum_j \alpha^{j*} f_j \in \mathcal{H}_l$ (from point 2.5.2 above), and therefore $\sum_j \alpha^j \left\langle \Pi_l(f_j),\, \Pi_l(g_k)\right\rangle_l = \left\langle \sum_j \alpha^{j*} f_j,\, g_k\right\rangle_{\mathcal{H}} = 0$. So, for all $k \in K$, $\left(\left\langle \Pi_l(f_j),\, \Pi_l(g_k)\right\rangle_l\right)_{j\in J}$ is in the image of $\left(\left\langle \Pi_{l'_3}(f_j),\, \cdot\right\rangle_{l'_3}\right)_{j\in J}$.

Next, using $\dim\mathcal{H}_{l'_3} \geqslant \dim \Pi_{l'_3}\langle \mathcal{J}_J\rangle + \dim\mathcal{K}_K$, there exists a family of vectors $g''_k \in \mathcal{H}_{l'_3} \cap \left(\Pi_{l'_3}\langle \mathcal{J}_J\rangle\right)^\perp$ for all $k \in K$ such that:

$$\forall k, l \in K,\, \langle \Pi_l(g_k),\, \Pi_l(g_l)\rangle_l + \langle g'_k,\, g'_l\rangle_{l'_3} + \langle g''_k,\, g''_l\rangle_{l'_3} = 0\,.$$

Now, we define $\varphi : \mathcal{J}_J \oplus \mathcal{K}_K \to \mathcal{H}_{l'}$ by:

$\forall j \in J,\ \varphi(f_j) := \varphi_{\mathcal{J}_J\to\mathcal{H}_{l'}}(f_j)$ (where $\varphi_{\mathcal{J}_J\to\mathcal{H}_{l'}}$ is defined as in lemma 2.6),

$$\text{and } \forall k \in K,\ \varphi(g_k) := \frac{\Pi_l(g_k) + g'_k + g''_k}{\|\Pi_l(g_k) + g'_k + g''_k\|}\,.$$

From the proof of lemma 2.6, $\varphi_{\mathcal{J}_J\to\mathcal{H}_{l'}}\langle \mathcal{J}_J\rangle = \Pi_{l'}\langle \mathcal{J}_J\rangle$, hence, for all $k \in K$, we have, by construction of $g'_k$ and $g''_k$, $\varphi(g_k) \perp \varphi\langle \mathcal{J}_J\rangle$. Also by construction of $g''_k$, we have, for all $k, l \in K$, $\langle \varphi(g_k),\, \varphi(g_l)\rangle = \delta_{kl}$. Therefore $\varphi$ induces an Hilbert space isomorphism $\mathcal{J}_J \oplus \mathcal{K}_K \to \mathrm{Im}\varphi$.

Finally, we can check that defs. 2.4.3 and 2.4.4 are fulfilled. □

**Proposition 2.7** With the notations of def. 2.4, $\mathcal{E}$, $\preccurlyeq$ is a directed set.

**Proof** Let $\left(l_1,\, l'_1,\, J_1,\, K_1,\, \varphi_1,\, \epsilon_1\right) \in \mathcal{E}$ and $\left(l_2,\, l'_2,\, J_2,\, K_2,\, \varphi_2,\, \epsilon_2\right) \in \mathcal{E}$. We define $\widetilde{l} = l_1 \cup l_2$, $J = J_1 \cup J_2$, $K = K_1 \cup K_2$ and $\epsilon = \min(\epsilon_1,\, \epsilon_2) > 0$. Then, since $(e_i)_{i\in\mathbb{N}}$ is an orthonormal basis of $\mathcal{H}$, we can find $l \in \mathcal{L}$ such that $\widetilde{l} \subset l$ and $\dim \Pi_l\langle \mathcal{J}_J \oplus \mathcal{K}_K\rangle = \dim(\mathcal{J}_J \oplus \mathcal{K}_K)$.

From prop. 2.5, there exist $\widetilde{l}' \in \mathcal{L}$ and $\widetilde{\varphi} : \mathcal{J}_J \oplus \mathcal{K}_K \to \mathcal{H}_{\widetilde{l}'}$ such that $\left(l,\, \widetilde{l}',\, J,\, K,\, \widetilde{\varphi},\, \epsilon\right) \in \mathcal{E}$. We define $l' = l'_1 \cup l'_2 \cup \widetilde{l}'$ and $\varphi : \mathcal{J}_J \oplus \mathcal{K}_K \to \mathcal{H}_{l'}$ by:

$$\forall v \in \mathcal{J}_J \oplus \mathcal{K}_K,\ \varphi(v) := \widetilde{\varphi}(v)\,.$$

Then, $\left(l,\, l',\, J,\, K,\, \varphi,\, \epsilon\right) \in \mathcal{E}$ and $\left(l_1,\, l'_1,\, J_1,\, K_1,\, \varphi_1,\, \epsilon_1\right),\, \left(l_2,\, l'_2,\, J_2,\, K_2,\, \varphi_2,\, \epsilon_2\right) \preccurlyeq \left(l,\, l',\, J,\, K,\, \varphi,\, \epsilon\right)$. □

**Proposition 2.8** We consider the same objects as in def. 2.4. Let $\varepsilon = \left(l,\, l',\, J,\, K,\, \varphi,\, \epsilon\right) \in \mathcal{E}$. We define:

1. $\mathcal{L}^\varepsilon := \{l'' \in \mathcal{L} \mid l' \subset l''\}$;
2. $\forall l'' \in \mathcal{L}^\varepsilon \sqcup \{\mathbb{N}\}$, $\mathcal{J}^\varepsilon_{l''} := \mathcal{J}_J$, equipped with the induced symplectic structure $\Omega'_J$;
3. $\forall l'' \in \mathcal{L}^\varepsilon \sqcup \{\mathbb{N}\}$, $\mathcal{H}^\varepsilon_{l''} := \varphi\langle \mathcal{J}_J \oplus \mathcal{K}_{K,\mathbb{R}}\rangle \subset \mathcal{H}_{l'} \subset \mathcal{H}_{l''}$;
4. $\forall l'' \in \mathcal{L}^\varepsilon \sqcup \{\mathbb{N}\}$, $\delta^\varepsilon_{l''} := \left(\Pi'_J\big|_{\mathcal{J}_J\oplus\mathcal{K}_{K,\mathbb{R}}\to\mathcal{J}_J}\right) \circ \left(\varphi\big|_{\mathcal{J}_J\oplus\mathcal{K}_{K,\mathbb{R}}\to\varphi\langle\mathcal{J}_J\oplus\mathcal{K}_{K,\mathbb{R}}\rangle}\right)^{-1}$ where $\Pi'_J$ is the orthogonal projection on $\mathcal{J}_J$;
5. $\forall l''_1, l''_2 \in \mathcal{L}^\varepsilon \sqcup \{\mathbb{N}\}\ /\ l''_1 \subset l''_2$, $\pi^{\varepsilon\to\varepsilon}_{l''_2\to l''_1} := \mathrm{id}_{\mathcal{J}_J}$.

Then, $\mathcal{L}^\varepsilon$ is cofinal in $\mathcal{L}$ and $\left((\mathcal{J}^\varepsilon_{l''})_{l''\in\mathcal{L}^\varepsilon\sqcup\{\mathbb{N}\}}, (\mathcal{H}^\varepsilon_{l''})_{l''\in\mathcal{L}^\varepsilon\sqcup\{\mathbb{N}\}}, \left(\pi^{\varepsilon\to\varepsilon}_{l''_2\to l''_1}\right)_{l''_1\subset l''_2}, (\delta^\varepsilon_{l''})_{l''\in\mathcal{L}^\varepsilon\sqcup\{\mathbb{N}\}}\right)$ is an elementary reduction [10, def. 3.7] of $(\mathcal{L}^\varepsilon \sqcup \{\mathbb{N}\}, \mathcal{H}, \pi)^\downarrow$.

**Proof** For $\widetilde{l} \in \mathcal{L}$, $\widetilde{l} \cup l' \in \mathcal{L}^\varepsilon$, hence $\mathcal{L}^\varepsilon$ is cofinal in $\mathcal{L}$, so in particular it is directed (and so is $\mathcal{L}^\varepsilon \sqcup \{\mathbb{N}\}$ since it has a greatest element).

Then, it is clear from the definitions that $(\mathcal{L}^\varepsilon \sqcup \{\mathbb{N}\}, \mathcal{J}^\varepsilon, \pi^{\varepsilon\to\varepsilon})$ is a projective system of phase spaces.

Replicating the proof of prop. 2.3, we can show that $\left(\mathcal{J}_J, \mathcal{J}_J \oplus \mathcal{K}_{K,\mathbb{R}}, \Pi'_J|_{\mathcal{J}_J\oplus\mathcal{K}_{K,\mathbb{R}}\to\mathcal{J}_J}\right)$ is a phase space reduction of $\mathcal{J}_J \oplus \mathcal{K}_K$. But since $\varphi|_{\mathcal{J}_J\oplus\mathcal{K}_K\to\mathrm{Im}\varphi}$ is unitary, $(\mathcal{J}_J, \mathcal{H}^\varepsilon_{l''}, \delta^\varepsilon_{l''})$ is a phase space reduction of $\mathrm{Im}\varphi$, hence of $\mathcal{H}_{l''}$, for all $l'' \in \mathcal{L}^\varepsilon \sqcup \{\mathbb{N}\}$.

Let $l''_1, l''_2 \in \mathcal{L}^\varepsilon \sqcup \{\mathbb{N}\}$ such that $l''_1 \subset l''_2$. We have $\pi_{l''_2\to l''_1}\left\langle \mathcal{H}^\varepsilon_{l''_2}\right\rangle = \Pi_{l''_1}\left\langle \mathcal{H}^\varepsilon_{l''_2}\right\rangle = \mathcal{H}^\varepsilon_{l''_1}$ since $\mathcal{H}^\varepsilon_{l''_2} = \mathcal{H}^\varepsilon_{l''_1} \subset \mathcal{H}_{l''_1}$. Lastly, for $x_1 \in \mathcal{H}^\varepsilon_{l''_1}$, $y_2 \in \mathcal{J}^\varepsilon_{l''_2} = \mathcal{J}_J$, we have:

$$\left(\exists\, x_2 \in \mathcal{H}^\varepsilon_{l''_2} \,/\, \delta^\varepsilon_{l''_2}(x_2) = y_2 \quad \& \quad \pi_{l''_2\to l''_1}(x_2) = x_1\right) \Leftrightarrow \left(\exists\, x_2 \in \mathcal{H}^\varepsilon_{l''_1} \,/\, \delta^\varepsilon_{l''_1}(x_2) = y_2 \quad \& \quad x_2 = x_1\right)$$

$$\Leftrightarrow \left(\delta^\varepsilon_{l''_1}(x_1) = y_2\right),$$

therefore def. [10, 3.7.3] is fulfilled. □

**Proposition 2.9** We consider the same objects as in prop. 2.8. We define:

1. $\widetilde{\mathcal{E}} := \mathcal{E} \sqcup \{\mathbb{N}\}$ (we extend the preorder by $\forall\varepsilon \in \mathcal{E}$, $\varepsilon \prec \mathbb{N}$), $\forall\varepsilon \in \mathcal{E}$, $\widetilde{\mathcal{L}}^\varepsilon := \mathcal{L}^\varepsilon \sqcup \{\mathbb{N}\}$, and $\widetilde{\mathcal{L}}^{\mathbb{N}} := \{\mathbb{N}\}$
2. $\forall\varepsilon = \left(l, l', J, K, \varphi, \epsilon\right) \in \mathcal{E}$, $\forall l'' \in \widetilde{\mathcal{L}}^\varepsilon$, $\ell(\varepsilon, l'') := J$, and $\ell(\mathbb{N}, \mathbb{N}) := \mathbb{N}$;
3. $\mathcal{J}^{\mathbb{N}}_{\mathbb{N}} := \mathcal{J}_{\mathbb{N}} = \mathcal{J}$, $\mathcal{H}^{\mathbb{N}}_{\mathbb{N}} := \mathcal{J} \oplus \mathcal{K}_{\mathbb{R}}$ and $\delta^{\mathbb{N}}_{\mathbb{N}} := \delta$;
4. $\forall(\varepsilon_1, l''_1), (\varepsilon_2, l''_2) \in \widetilde{\mathcal{E}}\widetilde{\mathcal{L}} \,/\, \varepsilon_1 \preccurlyeq \varepsilon_2 \quad \& \quad l''_1 \subset l''_2$, $\pi^{\varepsilon_2\to\varepsilon_1}_{l''_2\to l''_1} := \pi'_{\ell(\varepsilon_2,l''_2)\to\ell(\varepsilon_1,l''_1)}$ ($\pi'_{J'\to J}$ for $J, J' \in \mathcal{L}\sqcup\{\mathbb{N}\}$ with $J \subset J'$ has been defined in defs. 2.3.5 and 2.3.6).

Then, $\left(\widetilde{\mathcal{E}}, \left(\widetilde{\mathcal{L}}^\varepsilon\right)_{\varepsilon\in\widetilde{\mathcal{E}}}, (\mathcal{J}^\varepsilon_{l''})_{(\varepsilon,l'')\in\widetilde{\mathcal{E}}\widetilde{\mathcal{L}}}, (\mathcal{H}^\varepsilon_{l''})_{(\varepsilon,l'')\in\widetilde{\mathcal{E}}\widetilde{\mathcal{L}}}, \left(\pi^{\varepsilon_2\to\varepsilon_1}_{l''_2\to l''_1}\right)_{(\varepsilon_1,l''_1)\preccurlyeq(\varepsilon_2,l''_2)}, (\delta^\varepsilon_{l''})_{(\varepsilon,l'')\in\widetilde{\mathcal{E}}\widetilde{\mathcal{L}}}\right)$ is a regularized reduction [10, def. 3.16] of $(\mathcal{L} \sqcup \{\mathbb{N}\}, \mathcal{H}, \pi)^\downarrow$.

Additionaly, we have a bijective map $\kappa : \mathcal{S}^\downarrow_{(\widetilde{\mathcal{E}}\widetilde{\mathcal{L}},\mathcal{J},\pi)} \to \mathcal{S}^\downarrow_{(\mathcal{L}\sqcup\{\mathbb{N}\},\mathcal{J},\pi')}$.

**Proof** $\widetilde{\mathcal{E}}$ is a directed set (for it has a greatest element) and $\left(\widetilde{\mathcal{L}}^\varepsilon\right)_{\varepsilon\in\widetilde{\mathcal{E}}}$ is a family of decreasing cofinal parts of $\mathcal{L} \sqcup \{\mathbb{N}\}$.

Next, we have $\forall(\varepsilon, l'') \in \widetilde{\mathcal{E}}\widetilde{\mathcal{L}}$, $\mathcal{J}^\varepsilon_{l''} = \mathcal{J}_{\ell(\varepsilon,l'')}$ hence for $(\varepsilon_1, l''_1) \preccurlyeq (\varepsilon_2, l''_2) \in \widetilde{\mathcal{E}}\widetilde{\mathcal{L}}$, $\pi^{\varepsilon_2\to\varepsilon_1}_{l''_2\to l''_1}$ is well-defined as a surjective map $\mathcal{J}^{\varepsilon_2}_{l''_2} \to \mathcal{J}^{\varepsilon_1}_{l''_1}$ and it is compatible with the symplectic structures.

Moreover, for $\varepsilon_1 = \varepsilon_2 \in \mathcal{E}$, this definition coincides with the map $\pi^{\varepsilon\to\varepsilon}_{l''_2\to l''_1}$ that has been introduced in prop. 2.8. Hence, for all $\varepsilon \in \mathcal{E}$, we have from prop. 2.8 that $(\mathcal{J}^\varepsilon, \mathcal{H}^\varepsilon, \pi^{\varepsilon\to\varepsilon}, \delta^\varepsilon)$ is an elementary reduction of $\left(\widetilde{\mathcal{L}}^{\varepsilon}, \mathcal{H}, \pi\right)^{\downarrow}$.

And $\left(\mathcal{J}^{\mathbb{N}}, \mathcal{H}^{\mathbb{N}}, \pi^{\mathbb{N}\to\mathbb{N}}, \delta^{\mathbb{N}}\right)$ is an elementary reduction of $\left(\widetilde{\mathcal{L}}^{\mathbb{N}}, \mathcal{H}, \pi\right)^{\downarrow}$ since $\mathcal{L}^{\mathbb{N}\prime}$ has only one element and $\left(\mathcal{J}^{\mathbb{N}}_{\mathbb{N}}, \mathcal{H}^{\mathbb{N}}_{\mathbb{N}}, \delta^{\mathbb{N}}_{\mathbb{N}}\right) = (\mathcal{J}, \mathcal{J} \oplus \mathcal{K}_{\mathbb{R}}, \delta)$ is a phase space reduction of $\mathcal{H}_{\mathbb{N}} = \mathcal{H}$ (prop. 2.3).

Lastly, using $\ell : \widetilde{\mathcal{E}\mathcal{L}} \to \mathcal{L} \sqcup \{\mathbb{N}\}$ ($\ell$ satisfies that $\ell \left\langle \widetilde{\mathcal{E}\mathcal{L}} \right\rangle$ is cofinal in $\mathcal{L} \sqcup \{\mathbb{N}\}$ since it contains $\mathbb{N}$, which is a greatest element in $\mathcal{L} \sqcup \{\mathbb{N}\}$), we have by [10, prop. 2.9] that $\left(\widetilde{\mathcal{E}\mathcal{L}}, \mathcal{J}, \pi\right)^{\downarrow}$ is a projective system of phase spaces, thus $\left(\widetilde{\mathcal{E}}, \left(\widetilde{\mathcal{L}}^{\varepsilon}\right)_{\varepsilon\in\widetilde{\mathcal{E}}}, (\mathcal{J}^{\varepsilon}_{l''})_{(\varepsilon,l'')\in\widetilde{\mathcal{E}\mathcal{L}}}, (\mathcal{H}^{\varepsilon}_{l''})_{(\varepsilon,l'')\in\widetilde{\mathcal{E}\mathcal{L}}}, \left(\pi^{\varepsilon_2\to\varepsilon_1}_{l''_2\to l''_1}\right)_{(\varepsilon_1,l''_1)\preccurlyeq(\varepsilon_2,l''_2)}, (\delta^{\varepsilon}_{l''})_{(\varepsilon,l'')\in\widetilde{\mathcal{E}\mathcal{L}}}\right)$ is a regularized reduction of $(\mathcal{L} \sqcup \{\mathbb{N}\}, \mathcal{H}, \pi)^{\downarrow}$. And, in addition, there exists a bijective map $\kappa : \mathcal{S}^{\downarrow}_{(\widetilde{\mathcal{E}\mathcal{L}},\mathcal{J},\pi)} \to \mathcal{S}^{\downarrow}_{(\mathcal{L}\sqcup\{\mathbb{N}\},\mathcal{J},\pi')}$. □

Lastly, we can investigate the convergence and check that we are indeed in the optimal situation discussed at the end of [10, subsection 3.2] (more precisely in [10, prop. 3.23]). As announced above, the key ingredient for the convergence is the auxiliary result from prop. 2.5, that we proved in the process of establishing the directedness of $\mathcal{E}$.

**Theorem 2.10** We consider the same objects as in prop. 2.9. Let $\psi \in \mathcal{J} = \mathcal{J}^{\mathbb{N}}_{\mathbb{N}}$. For $\varepsilon \in \mathcal{E}$, we define:

1. $\psi^{\varepsilon} := (\delta^{\varepsilon}_{\mathbb{N}})^{-1} \left\langle \pi^{\mathbb{N}\to\varepsilon}_{\mathbb{N}\to\mathbb{N}}(\psi)\right\rangle \subset \mathcal{H}^{\varepsilon}_{\mathbb{N}} \subset \mathcal{H}_{\mathbb{N}} = \mathcal{H}$;
2. $\Psi^{\varepsilon} := \widehat{\sigma}_{\downarrow}(\psi^{\varepsilon})$, where $\widehat{\sigma}_{\downarrow} : \mathcal{P}(\mathcal{H}) \to \widehat{\mathcal{S}}^{\downarrow}_{(\mathcal{L},\mathcal{H},\pi)}$ is defined as in [10, prop. 3.23].

Then, the net $\left(\Psi^{\varepsilon}\right)_{\varepsilon\in\mathcal{E}}$ converges in $\widehat{\mathcal{S}}^{\downarrow}_{(\mathcal{L},\mathcal{H},\pi)}$ to $\widehat{\sigma}_{\downarrow}\left(\delta^{-1}\langle\psi\rangle\right)$ [10, def. 3.21].

**Proof** For $\varepsilon = \left(l, l', J, K, \varphi, \epsilon\right) \in \mathcal{E}$, we have, by putting all definitions together:

$$\psi^{\varepsilon} = \varphi\left\langle \Pi'_J(\psi) + \mathcal{K}_{K,\mathbb{R}}\right\rangle, \text{ where } \Pi'_J \text{ is the orthogonal projection on } \mathcal{J}_J,$$

hence, for all $l_o \in \mathcal{L}$:

$$[\Psi^{\varepsilon}]_{l_o} = \pi_{\mathbb{N}\to l_o}\langle\psi^{\varepsilon}\rangle = \Pi_{l_o}\left\langle \varphi\left\langle \Pi'_J(\psi) + \mathcal{K}_{K,\mathbb{R}}\right\rangle\right\rangle \subset \mathcal{H}_{l_o}.$$

Let $l_o \in \mathcal{L}$ and let $U$ be an open set in $\mathcal{H}_{l_o}$ such that $U \cap \Pi_{l_o}\langle\psi + \mathcal{K}_{\mathbb{R}}\rangle \neq \varnothing$. Let $\psi' \in U \cap \Pi_{l_o}\langle\psi + \mathcal{K}_{\mathbb{R}}\rangle$ and let $\epsilon_1 > 0$ such that $\forall\psi'' \in \mathcal{H}_{l_o}$, $\|\psi'' - \psi'\| \leqslant 3\,\epsilon_1\,(\|\psi\| + 1) \Rightarrow \psi'' \in U$.

Next, there exits $\chi \in \mathcal{K}_{\mathbb{R}}$ such that $\psi' = \Pi_{l_o}\langle\psi + \chi\rangle$. We choose $J_1, K_1 \in \mathcal{L}$ and $\chi' \in \mathcal{K}_{K_1,\mathbb{R}}$ such that $\left\|\psi - \Pi'_{J_1}(\psi)\right\| \leqslant \epsilon_1\,(\|\psi\| + 1)$ and $\|\chi - \chi'\| \leqslant \epsilon_1\,(\|\psi\| + 1)$. And we can find $l_1 \in \mathcal{L}$ with $l_1 \supset l_o$ such that $\dim \Pi_{l_1}\langle\mathcal{J}_{J_1} \oplus \mathcal{K}_{K_1}\rangle = \dim(\mathcal{J}_{J_1} + \mathcal{K}_{K_1})$. So, from prop. 2.5, there exist $l'_1 \in \mathcal{L}$ and $\varphi_1 : \mathcal{J}_{J_1} \oplus \mathcal{K}_{K_1} \to \mathcal{H}_{l'_1}$ such that $\varepsilon_1 := \left(l_1,\, l'_1,\, J_1,\, K_1,\, \varphi_1,\, \epsilon_1\right) \in \mathcal{E}$.

Let $\varepsilon_2 = \left(l_2,\, l'_2,\, J_2,\, K_2,\, \varphi_2,\, \epsilon_2\right) \in \mathcal{E}$ with $\varepsilon_2 \succcurlyeq \varepsilon_1$. Then, we have:

$$\left\|\Pi_{l_o}(\psi) - \Pi_{l_o} \circ \varphi_2 \circ \Pi'_{J_2}(\psi)\right\| \leqslant \left\|\psi - \varphi_2 \circ \Pi'_{J_2}(\psi)\right\|$$

$$\leqslant \left\|\psi - \Pi'_{J_2}(\psi)\right\| + \left\|\Pi'_{J_2}(\psi) - \varphi_2 \circ \Pi'_{J_2}(\psi)\right\|$$

$$\leqslant \left\| \psi - \Pi'_{J_1}(\psi) \right\| + \epsilon_2 \left\| \Pi'_{J_2}(\psi) \right\|$$

$$\leqslant 2\epsilon_1 \left( \|\psi\| + 1 \right).$$

Moreover, we have $\chi' \in \mathcal{K}_{K_1,\mathbb{R}} \subset \mathcal{K}_{K_2,\mathbb{R}}$, so there exists $\chi'' \in \mathcal{K}_{K_2,\mathbb{R}}$ such that $\Pi_{I_2} \circ \varphi_2(\chi'') = \Pi_{I_2}(\chi')$ and we have:

$$\left\| \Pi_{I_o}(\chi) - \Pi_{I_o} \circ \varphi_2\left(\chi''\right) \right\| \leqslant \left\| \Pi_{I_2}(\chi) - \Pi_{I_2} \circ \varphi_2\left(\chi''\right) \right\| \text{ (since } I_o \subset I_1 \subset I_2)$$

$$\leqslant \left\| \Pi_{I_2}(\chi) - \Pi_{I_2}(\chi') \right\| \leqslant \epsilon_1 \left( \|\psi\| + 1 \right).$$

Therefore, $\psi'' := \Pi_{I_o} \circ \varphi_2 \left( \Pi'_{J_2}(\psi) + \chi'' \right) \in U$, but since $\psi'' \in [\Psi^{\varepsilon_2}]_{I_o}$, we have $\forall \varepsilon_2 \succcurlyeq \varepsilon_1$, $[\Psi^{\varepsilon_2}]_{I_o} \cap U \neq \varnothing$.

Let $K$ be a compact set in $\mathcal{H}_{I_o}$ such that $K \cap \Pi_{I_o} \langle \psi + \mathcal{K}_{\mathbb{R}} \rangle = \varnothing$. Hence, there exists $\epsilon_1 > 0$ such that:

$$\forall \psi'' \in \mathcal{H}_{I_o},\ \left( \exists \psi' \in \Pi_{I_o} \langle \psi + \mathcal{K}_{\mathbb{R}} \rangle \,/\, \|\psi' - \psi''\| \leqslant 2\epsilon_1 \left( \|\psi\| + 1 \right) \right) \Rightarrow \left( \psi'' \notin K \right).$$

As above, we can, using prop. 2.5, construct $\varepsilon_1 := \left( I_1, I'_1, J_1, K_1, \varphi_1, \epsilon_1 \right) \in \mathcal{E}$ with $I_o \subset I_1$ and $\left\| \psi - \Pi'_{J_1}(\psi) \right\| \leqslant \epsilon_1 \left( \|\psi\| + 1 \right)$. Let $\varepsilon_2 = \left( I_2, I'_2, J_2, K_2, \varphi_2, \epsilon_2 \right) \in \mathcal{E}$ with $\varepsilon_2 \succcurlyeq \varepsilon_1$ and let $\psi'' \in [\Psi^{\varepsilon_2}]_{I_o}$. Then, there exists $\chi'' \in \mathcal{K}_{K_2,\mathbb{R}}$ such that $\psi'' = \Pi_{I_o} \circ \varphi_2 \left( \Pi'_{J_2}(\psi) + \chi'' \right)$ and there exists $\chi' \in \mathcal{K}_{K_2,\mathbb{R}} \subset \mathcal{K}_{\mathbb{R}}$ such that $\Pi_{I_2} \circ \varphi_2(\chi'') = \Pi_{I_2}(\chi')$. Moreover, we have again:

$$\left\| \Pi_{I_o}(\psi) - \Pi_{I_o} \circ \varphi_2 \circ \Pi'_{J_2}(\psi) \right\| \leqslant 2\epsilon_1 \left( \|\psi\| + 1 \right).$$

We define $\psi' = \Pi_{I_o} \left( \psi + \chi' \right) = \Pi_{I_o}(\psi) + \Pi_{I_o} \circ \varphi_2 \left( \chi'' \right)$ (since $I_o \subset I_1 \subset I_2$) and we have $\|\psi'' - \psi'\| \leqslant 2\epsilon_1 \left( \|\psi\| + 1 \right)$, hence $\psi'' \notin K$, and therefore $\forall \varepsilon_2 \succcurlyeq \varepsilon_1$, $[\Psi^{\varepsilon_2}]_{I_o} \cap K = \varnothing$.

So, the net $\left( [\Psi^{\varepsilon}]_{I_o} \right)_{\varepsilon \in \mathcal{E}}$ converges in $\mathcal{P}(\mathcal{H}_{I_o})$ to $\Pi_{I_o} \langle \psi + \mathcal{K}_{\mathbb{R}} \rangle = \pi_{\mathbb{N} \to I_o} \left\langle \delta^{-1} \langle \psi \rangle \right\rangle$, thus, the net $(\Psi^{\varepsilon})_{\varepsilon \in \mathcal{E}}$ converges in $\widehat{\mathcal{S}}^{\downarrow}_{(\mathcal{L},\mathcal{H},\pi)}$ to $\widehat{\sigma}_{\downarrow} \left( \delta^{-1} \langle \psi \rangle \right)$. □

# 3 Second quantization of the Schrödinger equation

In this section we want to apply the formalism outlined in [10, sections 2 and 3] and [11, section 2] to the second quantization of the Schrödinger equation. In other words, we will consider the one-particle quantum mechanics defined on an Hilbert space $\mathcal{H}$ as a classical field theory (looking at the wave function as a classical field, whose evolution is described by a linear partial differential equation, namely the Schrödinger equation), and we will discuss how this field theory can be quantized. The standard way of doing this leads to the bosonic Fock space build on $\mathcal{H}$ [5, section I.3.4]. Here we want to compare this trusted path with the strategy inspired by [7, 13]: first, look for a rendering of the classical field theory by a collection of finite dimensional partial theories, then come up with a regularizing procedure to implement the dynamics, and, last but not least, take advantage of this classical insight to build a quantization of the theory, thus obtaining a projective system of quantum state spaces. In particular, we want to use this example to illustrate how the classical regularization of the dynamics lays the stage for a corresponding procedure at the quantum level.

## 3.1 Classical theory

In [10, section 3], we only considered dynamics specified by constraints, whereas here we have a theory originally formulated with a 'true' Hamiltonian. However, this is quickly fixed, since there exists a routine trick (discussed in [15, section 1.8] and similar to the more general procedure presented in [9]), that can be physically interpreted as introducing an artificial time parametrization, and allows to transform any theory on $\mathcal{H}$ with an non-vanishing Hamiltonian into a theory on $\mathcal{H} \times \mathbb{R}^2$ with an Hamiltonian constraint (the $\mathbb{R}^2$ part holds the time coordinate and its conjugate momentum, aka. the energy variable).

Note that there is a technical subtlety arising when we try to write the theory on an infinite dimensional symplectic manifold in the naive setup of [10, def. A.1], and we are forced to require the one-particle quantum Hamiltonian to be a bounded operator (we cannot simply restrict the constraint surface so that it is included in an appropriate dense subspace, for it would then cost the reduced phase space its strong symplectic structure, by spoiling the needed non-degeneracy property). However, we will be able to lift this restriction without great efforts when switching to the projective state space formalism.

**Proposition 3.1** Let $\mathcal{H}$ be a separable, infinite dimensional Hilbert space and $H$ be a bounded self-adjoint operator on $\mathcal{H}$. We equip $\mathcal{M}^{\text{KIN}} := \mathcal{H} \times \mathbb{R}^2$ with the strong symplectic structure:

1. $\forall (\varphi_1, u_1, v_1), (\varphi_2, u_2, v_2) \in \mathcal{M}^{\text{KIN}}, \;\; \Omega_{\text{KIN}}(\varphi_1, u_1, v_1; \varphi_2, u_2, v_2) := 2\, \text{Im}\left(\langle \varphi_1, \varphi_2 \rangle\right) + (u_2\, v_1 - u_1\, v_2)$.

We define:

2. $\mathcal{M}^{\text{SHELL}} := \{(\psi, t, E) \in \mathcal{M}^{\text{KIN}} \mid E = \langle \psi, H\psi \rangle\}$;
3. $\mathcal{M}^{\text{DYN}} := \mathcal{H}$ with symplectic structure $\Omega_{\text{DYN}} := 2\, \text{Im}\, \langle \cdot\,, \cdot \rangle$;
4. $\begin{array}{cccc} \delta & : & \mathcal{M}^{\text{SHELL}} & \rightarrow & \mathcal{M}^{\text{DYN}} \\ & & (\psi, t, E) & \mapsto & \exp(i\, t\, H)\, \psi \end{array}$.

Then, $(\mathcal{M}^{\text{DYN}}, \mathcal{M}^{\text{SHELL}}, \delta)$ is a phase space reduction of $\mathcal{M}^{\text{KIN}}$ [10, def. A.1].

**Proof** From prop. 2.1, $\Omega_{\text{KIN}}$, resp. $\Omega_{\text{DYN}}$, defines a strong symplectic structure on $\mathcal{M}^{\text{KIN}}$, resp. $\mathcal{M}^{\text{DYN}}$.

The map $\delta$ is surjective and, for $\psi_o \in \mathcal{M}^{\text{DYN}}$, $\delta^{-1}\langle \psi_o \rangle = \{(\exp(-i\, t\, H)\, \psi_o, t, E_o) \mid t \in \mathbb{R}\}$, where $E_o := \langle \psi_o, H\, \psi_o \rangle$. So $\delta^{-1}\langle \psi_o \rangle$ is in particular connected.

Let $(\psi, t, E) \in \mathcal{M}^{\text{SHELL}}$. We have:

$$T_{(\psi,t,E)}(\mathcal{M}^{\text{SHELL}}) = \{(\varphi, u, 2\, \text{Re}\, \langle \varphi, H\psi \rangle) \mid \varphi \in \mathcal{H}, u \in \mathbb{R}\},$$

and:

$$T_{(\psi,t,E)}\delta \;:\; \begin{array}{rcl} T_{(\psi,t,E)}\left(\mathcal{M}^{\text{SHELL}}\right) & \to & T_{e^{iHt}\psi}\left(\mathcal{M}^{\text{DYN}}\right) \\ (\varphi,\, u,\, 2\,\text{Re}\,\langle\varphi,\, H\psi\rangle) & \mapsto & e^{iHt}\,\varphi + iu\, e^{iHt}\, H\psi \end{array} .$$

Hence, $T_{(\psi,t,E)}\delta$ is surjective and, for $(\varphi_1,\, u_1,\, 2\,\text{Re}\,\langle\varphi_1,\, H\psi\rangle)\,,\, (\varphi_2,\, u_2,\, 2\,\text{Re}\,\langle\varphi_2,\, H\psi\rangle) \in T_{(\psi,t,E)}\left(\mathcal{M}^{\text{SHELL}}\right)$, we have:

$$\Omega_{\text{KIN}}\left(\varphi_1,\, u_1,\, 2\,\text{Re}\,\langle\varphi_1,\, H\psi\rangle;\;\; \varphi_2,\, u_2,\, 2\,\text{Re}\,\langle\varphi_2,\, H\psi\rangle\right) =$$

$$= 2\,\text{Im}\,\langle\varphi_1,\, \varphi_2\rangle + 2\,u_2\,\text{Im}\,\langle\varphi_1,\, iH\psi\rangle - 2\,u_1\,\text{Im}\,\langle\varphi_2,\, iH\psi\rangle + 2\,u_1u_2\,\text{Im}\,\langle i\,H\psi,\, i\,H\psi\rangle$$

$$= 2\,\text{Im}\,\left\langle T_{(\psi,t,E)}\delta\left(\varphi_1,\, u_1,\, 2\,\text{Re}\,\langle\varphi_1,\, H\psi\rangle\right),\, T_{(\psi,t,E)}\delta\left(\varphi_2,\, u_2,\, 2\,\text{Re}\,\langle\varphi_2,\, H\psi\rangle\right)\right\rangle,$$

therefore $\Omega_{\text{KIN}}\big|_{T_{(\psi,t,E)}\left(\mathcal{M}^{\text{SHELL}}\right)} = \left[\delta^*\,\Omega_{\text{DYN}}\right]_{(\psi,t,E)}$. □

On $\mathcal{H}$ viewed as a phase space, we can define some remarkable observables (this defines the algebra that we will latter endeavor to quantize): of interest are for us the scalar product with a vector $e \in \mathcal{H}$ (that will give rise in the quantum theory to the corresponding creation and annihilation operators) and the expectation value of an operator on $\mathcal{H}$. Additionally the Heisenberg (ie. time-dependent) operators of the first-quantized theory can be seen in a natural way as dynamical observables associated (in the sense of [10, def. A.2]) to particular kinematical observables (up to a technical artefact: we restrict the support of the considered observables to spheres in $\mathcal{H}$ because we had defined the map $(\,\cdot\,)^{\text{DYN}}$ translating a kinematical observable into its dynamical version only for bounded observables; note that, alternatively, we could just weaken this requirement, for it would be enough to only demand the kinematical observables to be bounded on orbits of the dynamics).

**Proposition 3.2** We consider the same objects as in prop. 3.1. Let $e \in \mathcal{H}$. On $\mathcal{H}$ we can define the observables:

$$\text{a}_e \;:\; \begin{array}{rcl} \mathcal{H} & \to & \mathbb{C} \\ \psi & \mapsto & \langle e,\, \psi\rangle \end{array} \quad \text{and} \quad \text{a}^*_e \;:\; \begin{array}{rcl} \mathcal{H} & \to & \mathbb{C} \\ \psi & \mapsto & \langle \psi,\, e\rangle \end{array} .$$

We have, for all $e,\, f \in \mathcal{H}$:

$$\{\text{a}_e,\, \text{a}_f\}_{\mathcal{H}} = 0\,,\; \{\text{a}^*_e,\, \text{a}^*_f\}_{\mathcal{H}} = 0\,,\; \text{and}\; \{\text{a}_e,\, \text{a}^*_f\}_{\mathcal{H}} = i\,\langle e,\, f\rangle\,.$$

Let $A$ be a bounded self-adjoint operator on $\mathcal{H}$. We define on $\mathcal{H}$ the observable $\langle A\rangle$ by:

$$\forall\psi \in \mathcal{H},\; \langle A\rangle(\psi) := \langle\psi,\, A\,\psi\rangle\,.$$

We have, for all $A$, $B$ bounded self-adjoint operators on $\mathcal{H}$ and $e \in \mathcal{H}$:

$$\left\{\langle A\rangle\,,\, \langle B\rangle\right\}_{\mathcal{H}} = i\,\left\langle [A,\, B]_{\mathcal{H}}\right\rangle,\; \left\{\text{a}_e,\, \langle A\rangle\right\}_{\mathcal{H}} = i\,\text{a}_{Ae}\,,\; \text{and}\; \left\{\text{a}^*_e,\, \langle A\rangle\right\}_{\mathcal{H}} = -i\,\text{a}^*_{Ae}\,.$$

Lastly, for $A$ a bounded self-adjoint operator on $\mathcal{H}$, $N > 0$ and $t_o \in \mathbb{R}$, we can define on $\mathcal{M}^{\text{KIN}}$ the observable:

$$\langle A,\, N,\, t_o\rangle \;:\; \begin{array}{rcl} \mathcal{M}^{\text{KIN}} & \to & \mathbb{R} \\ (\psi,\, t,\, E) & \mapsto & \begin{cases} \langle A\rangle(\psi) & \text{if}\quad \|\psi\| = N \quad \& \quad t = t_o \\ 0 & \text{else} \end{cases} \end{array} ,$$

and we have:

$$\forall \psi_o \in \mathcal{M}^{\text{DYN}},\ \langle A,\, N,\, t_o\rangle^{\text{DYN}}(\psi_o) := \sup_{(\psi,t,E)\in\delta^{-1}\langle\psi_o\rangle} \langle A,\, N,\, t_o\rangle(\psi,\, t,\, E)$$

$$= \begin{cases} \left\langle e^{it_oH} A\, e^{-it_oH}\right\rangle(\psi_o) & \text{if} \quad \|\psi_o\| = N \\ 0 & \text{else} \end{cases} . \qquad (3.2.1)$$

**Proof** In order to compute the Poisson brackets between observables of the type $\mathrm{a}_e$ and $\mathrm{a}_e^*$, we have to be careful not to mix up the complex structure on $\mathcal{H}$ with the complex structure coming from $\mathrm{a}_e$ and $\mathrm{a}_e^*$ being $\mathbb{C}$-valued. Therefore, we will write $J\varphi$ for the scalar multiplication of $\varphi$ by $i$ (in $\mathcal{H}$ seen as a $\mathbb{C}$-vector space) and $\imath\,\varphi$ for the vector $(i \otimes_{\mathbb{R}} \varphi)$ in $\mathbb{C}\otimes_{\mathbb{R}}\mathcal{H} \approx T_{\mathbb{C}}(\mathcal{H})$ (for $\mathcal{H}$ seen as a real manifold). Extending $\operatorname{Im}\langle\,\cdot\,,\,\cdot\,\rangle$ and $\operatorname{Re}\langle\,\cdot\,,\,\cdot\,\rangle$ by $\mathbb{C}$-bilinearity on $T_{\mathbb{C}}(\mathcal{H})$ (because we want $\{\,\cdot\,,\,\cdot\,\}_{\mathcal{H}}$ to be $\mathbb{C}$-bilinear), we then have:

$$\operatorname{Im}\langle\varphi',\, J\varphi\rangle = -\operatorname{Im}\langle J\varphi',\, \varphi\rangle = \operatorname{Re}\langle\varphi',\, \varphi\rangle$$

$$\& \quad \operatorname{Im}\langle\varphi',\, \imath\,\varphi\rangle = \operatorname{Im}\langle\imath\,\varphi',\, \varphi\rangle = i\operatorname{Im}\langle\varphi',\, \varphi\rangle .$$

With this we can compute the Hamiltonian vector fields at $\psi\in\mathcal{H}$ of $\mathrm{a}_e$ and $\mathrm{a}_e^*$, for $e\in\mathcal{H}$:

$$[d\mathrm{a}_e]_\psi(\varphi) = \langle e,\, \varphi\rangle = 2\operatorname{Im}\left\langle -\frac{J\,e}{2} + \frac{\imath\,e}{2},\, \varphi\right\rangle$$

$$\& \quad [d\mathrm{a}_e^*]_\psi(\varphi) = \langle\varphi,\, e\rangle = 2\operatorname{Im}\left\langle -\frac{J\,e}{2} - \frac{\imath\,e}{2},\, \varphi\right\rangle .$$

Hence, for $e,\, f\in\mathcal{H}$:

$$\{\mathrm{a}_e,\, \mathrm{a}_f\}_{\mathcal{H},\psi} = 2\operatorname{Im}\langle X_{\mathrm{a}_f,\psi},\, X_{\mathrm{a}_e,\psi}\rangle = 2\operatorname{Im}\left\langle -\frac{J\,f}{2} + \frac{\imath\,f}{2},\, -\frac{J\,e}{2} + \frac{\imath\,e}{2}\right\rangle = 0\,,$$

$$\{\mathrm{a}_e^*,\, \mathrm{a}_f^*\}_{\mathcal{H},\psi} = 2\operatorname{Im}\left\langle X_{\mathrm{a}_f^*,\psi},\, X_{\mathrm{a}_e^*,\psi}\right\rangle = 2\operatorname{Im}\left\langle -\frac{J\,f}{2} - \frac{\imath\,f}{2},\, -\frac{J\,e}{2} - \frac{\imath\,e}{2}\right\rangle = 0\,,$$

$$\text{and} \quad \{\mathrm{a}_e,\, \mathrm{a}_f^*\}_{\mathcal{H},\psi} = 2\operatorname{Im}\left\langle X_{\mathrm{a}_f^*,\psi},\, X_{\mathrm{a}_e,\psi}\right\rangle = 2\operatorname{Im}\left\langle -\frac{J\,f}{2} - \frac{\imath\,f}{2},\, -\frac{J\,e}{2} + \frac{\imath\,e}{2}\right\rangle$$

$$= i\left(\operatorname{Re}\langle f,\, e\rangle - i\operatorname{Im}\langle f,\, e\rangle\right) = i\,\langle e,\, f\rangle .$$

Similarly, we have for any $A$ bounded self-adjoint operator on $\mathcal{H}$ and at every $\psi\in\mathcal{H}$:

$$[d\langle A\rangle]_\psi = 2\operatorname{Re}\langle A\,\psi,\, \varphi\rangle = 2\operatorname{Im}\langle -J\,A\,\psi,\, \varphi\rangle ,$$

hence, for $A$, $B$ bounded self-adjoint operators on $\mathcal{H}$, $e\in\mathcal{H}$, and $\psi\in\mathcal{H}$:

$$\{\langle A\rangle,\, \langle B\rangle\}_{\mathcal{H},\psi} = 2\operatorname{Im}\left\langle X_{\langle B\rangle,\psi},\, X_{\langle A\rangle,\psi}\right\rangle = 2\operatorname{Im}\langle -J\,B\,\psi,\, -J\,A\,\psi\rangle$$

$$= -i\left(\langle B\,\psi,\, A\,\psi\rangle - \langle A\,\psi,\, B\,\psi\rangle\right) = i\left\langle [A,\, B]_{\mathcal{H}}\right\rangle(\psi) ,$$

$$\{\mathrm{a}_e,\, \langle A\rangle\}_{\mathcal{H},\psi} = 2\operatorname{Im}\left\langle X_{\langle A\rangle,\psi},\, X_{\mathrm{a}_e,\psi}\right\rangle = 2\operatorname{Im}\left\langle -J\,A\,\psi,\, -\frac{J\,e}{2} + \frac{\imath\,e}{2}\right\rangle$$

$$= i\,(-i \operatorname{Im}\langle\psi, A\,e\rangle + \operatorname{Re}\langle\psi, A\,e\rangle) = i\,\mathrm{a}_{Ae}(\psi)\,,$$

$$\text{and } \left\{\mathrm{a}^*_e, \langle A\rangle\right\}_{\mathcal{H},\psi} = 2\operatorname{Im}\left\langle X_{\langle A\rangle,\psi}, X_{\mathrm{a}^*_e,\psi}\right\rangle = 2\operatorname{Im}\left\langle -J\,A\,\psi, -\frac{J\,e}{2} - \frac{\mathrm{i}\,e}{2}\right\rangle$$

$$= -i\,(i \operatorname{Im}\langle\psi, A\,e\rangle + \operatorname{Re}\langle\psi, A\,e\rangle) = -i\,\mathrm{a}^*_{Ae}(\psi)\,.$$

Lastly, eq. (3.2.1) comes from:

$$\forall\psi_o \in \mathcal{M}^{\text{DYN}},\ \delta^{-1}\langle\psi_o\rangle = \{(\exp(-i\,t\,H)\,\psi_o,\, t,\, \langle H\rangle(\psi_o)) \mid t \in \mathbb{R}\}\,.$$

□

The projective system we will use here differs significantly from the one we were using in the previous section (prop. 2.2), for we do not rely any more on the choice of a particular basis to define a family of vector subspaces: instead, we simply take as label set the set of *all* finite dimensional vector subspaces of $\mathcal{H}$ (this structure is of course more satisfactory from a physical point of view; as mentioned at the beginning of section 2, we could not use it in the previous example, for our aim was to illustrate the regularizing strategy, while this larger label set contains a cofinal family on which the linear constraints we were considering form an elementary reduction).

Note that the space of states of this projective system can be naturally identified with the algebraic dual on $\mathcal{H}$, in such a way that the injection of $\mathcal{H}$ into the projective state space (in the sense of a rendering, as introduced in [10, def. 2.6]) corresponds to the identification with its topological dual.

**Proposition 3.3** Let $\mathcal{H}$ be a separable, infinite dimensional Hilbert space. We define $\mathcal{L}$ as the set of all finite dimensional vector subspaces of $\mathcal{H}$ and we equip it with the preorder $\subset$. We define:

1. $\forall \mathcal{I} \in \mathcal{L} \sqcup \{\mathcal{H}\}$, $\mathcal{M}^{\text{KIN}}_{\mathcal{I}} := \mathcal{I} \times \mathbb{R}^2$, equipped with the symplectic structure $\Omega_{\text{KIN},\mathcal{I}}$ induced from $\mathcal{M}^{\text{KIN}}$, $\Omega_{\text{KIN}}$;
2. $\forall \mathcal{I}, \mathcal{I}' \in \mathcal{L} \sqcup \{\mathcal{H}\}$, with $\mathcal{I} \subset \mathcal{I}'$, $\pi^{\text{KIN}}_{\mathcal{I}'\to\mathcal{I}} := \Pi_{\mathcal{I}}|_{\mathcal{I}'\to\mathcal{I}} \times \mathrm{id}_{\mathbb{R}^2}$ where $\Pi_{\mathcal{I}}$ is the orthogonal projection on $\mathcal{I}$

Then, this defines a rendering [10, def. 2.6] of $\mathcal{M}^{\text{KIN}}$ by the projective system of phase spaces $(\mathcal{L}, \mathcal{M}^{\text{KIN}}, \pi^{\text{KIN}})^{\downarrow}$. We define $\sigma^{\text{KIN}}_{\downarrow} : \mathcal{M}^{\text{KIN}} \to \mathcal{S}^{\downarrow}_{(\mathcal{L},\mathcal{M}^{\text{KIN}},\pi^{\text{KIN}})}$ as in [10, def. 2.6].

Additionally, we have a bijective map $\zeta^{\text{KIN}} : \mathcal{H}^* \times \mathbb{R}^2 \to \mathcal{S}^{\downarrow}_{(\mathcal{L},\mathcal{M}^{\text{KIN}},\pi^{\text{KIN}})}$ such that $\zeta^{\text{KIN},-1} \circ \sigma^{\text{KIN}}_{\downarrow} : \mathcal{M}^{\text{KIN}} \to \mathcal{H}^* \times \mathbb{R}^2$ corresponds to the canonical identification of $\mathcal{H}$ with $\mathcal{H}' \subset \mathcal{H}^*$.

**Proof** The proof works in the same way as the proof of prop. 2.2. □

We are ready to go on to the formulation of an approximating scheme for the dynamics. The approximation here will take place in two different directions. First, we introduce a deformation of the constraint surface, controlled by a small parameter $\epsilon > 0$, to replace the non-compact orbits of the exact dynamics (going in time from $-\infty$ to $+\infty$) by compact orbits (running only through a finite time interval): the rough idea is that instead of having a 'free particle' in the energy-time variable, we put an harmonic oscillator, thus preventing the time variable to grow for ever. This will be more comfortable when switching to the quantum theory: having compact orbits is closely related to having well-normalized states solving the quantum constraints (heuristically, quantum solutions of the constraints have much in common with classical statistical states, supported by the constraint surface and constant on the gauge orbits, and these will only exist as properly normalized probability measures if the orbits are compact).

The other aspect of the approximation is what will allow us to build, for the approximated dynamics, a corresponding elementary reduction on a cofinal part of the projective system introduced previously. For this, we truncate the exact Hamiltonian $H$ of the first-quantized theory as $\Pi_{\mathcal{J}} H \Pi_{\mathcal{J}}$ where $\mathcal{J}$ is a finite vector subspace of $\mathcal{H}$, such that $H$ is bounded on $\mathcal{J}$ (from now on, we can indeed relax the requirement we had above, and we allow $H$ to be an unbounded, densely defined, operator on $\mathcal{H}$). In other words, we project the Hamiltonian flow on the symplectic submanifold $\mathcal{J} \times \mathbb{R}^2$ of $\mathcal{H} \times \mathbb{R}^2$. Moreover, we include in the approximated dynamics additional second class constraints, forcing the wave function $\psi$ to belong to $\mathcal{J}$ (by definition of the truncated Hamiltonian these additional constraints are preserved by the truncated evolution): the point is that it does not make sense to keep the degrees of freedom orthogonal to the subspace $\mathcal{J}$ since with the truncated Hamiltonian we would not evolve them at all and they would soon lie very far away from their correct values (note that the degrees of freedom along $\mathcal{J}$ are not evolved exactly either, but at least they are evolved approximately; the error comes from neglecting the backreaction of the degrees of freedoms along $\mathcal{J}^\perp$, due to the cross-terms of the exact Hamiltonian $H$ between $\mathcal{J}$ and $\mathcal{J}^\perp$).

The side effect of these additional second class constraints is to make the approximated reduced phase space finite dimensional (aka. $\mathcal{M}^{\text{DYN},\mathcal{E}}_\infty$, using the notations of [10, prop. 3.24]): this is not needed for the construction (in general only the 'partial' reduced phase space $\mathcal{M}^{\text{DYN},\mathcal{E}}_\eta$, arising from the constraint surface projected on $\mathcal{M}^{\text{KIN}}_\eta$ for some $\eta \in \mathcal{L}^{\mathcal{E}}$, is expected to be finite dimensional), but it will simplify the structure of the dynamical projective system.

**Definition 3.4** We consider the same objects as in prop. 3.3. Let $H$ be a densely defined (possibly unbounded) self-adjoint operator on $\mathcal{H}$. We define $\mathcal{E}$ as the set of all pairs $(\mathcal{J}, \epsilon)$ such that:

1. $\mathcal{J} \in \mathcal{L}$ and $\forall \psi \in \mathcal{J},\ \|H\psi\| < \infty$;
2. $\epsilon > 0$.

On $\mathcal{E}$ we will use the preorder:

3. $(\mathcal{J}, \epsilon) \preccurlyeq (\mathcal{J}', \epsilon') \quad \Leftrightarrow \quad \big(\mathcal{J} \subset \mathcal{J}' \quad \& \quad \epsilon \geqslant \epsilon'\big)$.

**Proposition 3.5** $\mathcal{E}, \preccurlyeq$ is a directed preordered set.

**Proof** Let $(\mathcal{J}_1, \epsilon_1), (\mathcal{J}_2, \epsilon_2) \in \mathcal{E}$. Then, we have $\mathcal{J}_1 + \mathcal{J}_2 \in \mathcal{L}$, $\forall \psi \in \mathcal{J}_1 + \mathcal{J}_2$, $\|H\psi\| < \infty$ and $\min(\epsilon_1, \epsilon_2) > 0$. Hence, $(\mathcal{J}_1 + \mathcal{J}_2, \min(\epsilon_1, \epsilon_2)) \in \mathcal{E}$ and $(\mathcal{J}_1, \epsilon_1), (\mathcal{J}_2, \epsilon_2) \preccurlyeq (\mathcal{J}_1 + \mathcal{J}_2, \min(\epsilon_1, \epsilon_2))$. □

**Proposition 3.6** We consider the same objects as in def. 3.4. Let $\varepsilon = (\mathcal{J}, \epsilon) \in \mathcal{E}$. We define:

1. $\mathcal{L}^{\varepsilon} := \{\mathcal{I} \in \mathcal{L} \mid \mathcal{J} \subset \mathcal{I}\}$;
2. $\forall \mathcal{I} \in \mathcal{L}^{\varepsilon} \sqcup \{\mathcal{H}\}$, $\mathcal{M}^{\text{DYN},\varepsilon}_{\mathcal{I}} := \mathcal{J}$ equipped with the symplectic structure $\Omega_{\text{DYN},\mathcal{J}}$ induced from $\mathcal{M}^{\text{DYN}}, \Omega_{\text{DYN}}$;
3. $\forall \mathcal{I} \in \mathcal{L}^{\varepsilon} \sqcup \{\mathcal{H}\}$, $\mathcal{M}^{\text{SHELL},\varepsilon}_{\mathcal{I}} := \Big\{(\psi, t, E) \in \mathcal{M}^{\text{KIN}}_{\mathcal{J}} \subset \mathcal{M}^{\text{KIN}}_{\mathcal{I}} \,\Big|\, (E - \langle \psi, H\psi\rangle)^2 + \epsilon^4 t^2 = \epsilon^2\Big\}$;
4. $\forall \mathcal{I} \in \mathcal{L}^{\varepsilon} \sqcup \{\mathcal{H}\}$, $\forall (\psi, t, E) \in \mathcal{M}^{\text{SHELL},\varepsilon}_{\mathcal{I}}$, $\delta^{\varepsilon}_{\mathcal{I}}(\psi, t, E) := \exp(i\, t\, \Pi_{\mathcal{J}} H)\, \psi \in \mathcal{J}$;

5. $\forall \mathcal{I}, \mathcal{I}' \in \mathcal{L}^\varepsilon \sqcup \{\mathcal{H}\}$, with $\mathcal{I} \subset \mathcal{I}'$, $\pi^{\text{DYN},\varepsilon\to\varepsilon}_{\mathcal{I}'\to\mathcal{I}} := \text{id}_{\mathcal{J}}$.

Then, $\mathcal{L}^\varepsilon$ is cofinal in $\mathcal{L}$ and $\left(\left(\mathcal{M}^{\text{DYN},\varepsilon}_{\mathcal{I}}\right)_{\mathcal{I}\in\mathcal{L}^\varepsilon\sqcup\{\mathcal{H}\}}, \left(\mathcal{M}^{\text{SHELL},\varepsilon}_{\mathcal{I}}\right)_{\mathcal{I}\in\mathcal{L}^\varepsilon\sqcup\{\mathcal{H}\}}, \left(\pi^{\text{DYN},\varepsilon\to\varepsilon}_{\mathcal{I}'\to\mathcal{I}}\right)_{\mathcal{I}\subset\mathcal{I}'}, (\delta^\varepsilon_{\mathcal{I}})_{\mathcal{I}\in\mathcal{L}^\varepsilon\sqcup\{\mathcal{H}\}}\right)$ is an elementary reduction [10, def. 3.7] of $(\mathcal{L}^\varepsilon \sqcup \{\mathcal{H}\}, \mathcal{M}^{\text{KIN}}, \pi^{\text{KIN}})^\downarrow$.

**Proof** For $\mathcal{I} \in \mathcal{L}$, $\mathcal{I} + \mathcal{J} \in \mathcal{L}^\varepsilon$, hence $\mathcal{L}^\varepsilon$ is cofinal in $\mathcal{L}$, so in particular it is directed (and so is $\mathcal{L}^\varepsilon \sqcup \{\mathcal{H}\}$ since it has a greatest element).

Then, it is clear from the definitions that $(\mathcal{L}^\varepsilon \sqcup \{\mathcal{H}\}, \mathcal{M}^{\text{DYN},\varepsilon}, \pi^{\text{DYN},\varepsilon\to\varepsilon})$ is a projective system of phase spaces.

Now, we define:

6. $\mathcal{M}^\varepsilon := \left\{(\psi, t, E) \in \mathcal{M}^{\text{KIN}}_{\mathcal{J}} \subset \mathcal{M}^{\text{KIN}} \,\middle|\, (E - \langle\psi, H\psi\rangle)^2 + \epsilon^4 t^2 = \epsilon^2\right\}$;

7. $\forall (\psi, t, E) \in \mathcal{M}^\varepsilon$, $\delta^\varepsilon(\psi, t, E) := \exp(i\, t\, \Pi_{\mathcal{J}} H)\, \psi \in \mathcal{J}$;

and we want to show that $(\mathcal{J}, \mathcal{M}^\varepsilon, \delta^\varepsilon)$ is a phase space reduction of $\mathcal{M}^{\text{KIN}}$.

$\Pi_{\mathcal{J}} H|_{\mathcal{J}\to\mathcal{J}}$ is a bounded (by definition of $\mathcal{E}$), self-adjoint operator on $\mathcal{J}$. Therefore, the map $\delta^\varepsilon$ is surjective and, for $\psi_o \in \mathcal{J}$, $\delta^{\varepsilon,-1}\langle\psi_o\rangle = \left\{\left(e^{-\frac{i}{\epsilon}\sin\theta\,\Pi_{\mathcal{J}} H}\,\psi_o, \frac{1}{\epsilon}\sin\theta, E_o + \epsilon\cos\theta\right) \,\middle|\, \theta \in [0, 2\pi[\right\}$, where $E_o := \langle\psi_o, H\psi_o\rangle$. So $\delta^{\varepsilon,-1}\langle\psi_o\rangle$ is in particular connected.

Let $(\psi, t, E) \in \mathcal{M}^\varepsilon$. $T_{(\psi,t,E)}(\mathcal{M}^\varepsilon)$ is given by:

$$\left\{(\varphi, u, v) \in \mathcal{M}^{\text{KIN}}_{\mathcal{J}} \,\middle|\, \varphi \in \mathcal{J} \;\;\&\;\; 2\epsilon\sqrt{1-\epsilon^2 t^2}\,(v - 2\,\text{Re}\,\langle\varphi, H\psi\rangle) + 2\,\epsilon^4\, t\, u = 0\right\}, \tag{3.6.1}$$

and we have:

$$\begin{array}{rcl} T_{(\psi,t,E)}\,\delta^\varepsilon \;:\; T_{(\psi,t,E)}(\mathcal{M}^\varepsilon) & \to & \mathcal{J} \\ (\varphi, u, v) & \mapsto & i\, u e^{i\,t\,\Pi_{\mathcal{J}} H}\,\Pi_{\mathcal{J}} H\,\psi + e^{i\,t\,\Pi_{\mathcal{J}} H}\,\varphi \end{array}.$$

Hence, $T_{(\psi,t,E)}\,\delta^\varepsilon$ is surjective and, for $(\varphi_1, u_1, v_1), (\varphi_2, u_2, v_2) \in T_{(\psi,t,E)}(\mathcal{M}^\varepsilon)$, we have:

$\Omega_{\text{KIN}}(\varphi_1, u_1, v_1; \varphi_2, u_2, v_2) =$

$$= 2\,\text{Im}\,\langle\varphi_1, \varphi_2\rangle + u_2\left(2\,\text{Re}\,\langle\varphi_1, H\psi\rangle - \frac{\epsilon^3\, t}{\sqrt{1-\epsilon^2 t^2}}\,u_1\right) - u_1\left(2\,\text{Re}\,\langle\varphi_2, H\psi\rangle - \frac{\epsilon^3\, t}{\sqrt{1-\epsilon^2 t^2}}\,u_2\right)$$

(using eq. (3.6.1))

$$= 2\,\text{Im}\,\langle T_{(\psi,t,E)}\,\delta^\varepsilon(\varphi_1, u_1, v_1), T_{(\psi,t,E)}\,\delta^\varepsilon(\varphi_2, u_2, v_2)\rangle$$ (like in the proof of prop. 3.1),

therefore $\Omega_{\text{KIN}}|_{T_{(\psi,t,E)}(\mathcal{M}^\varepsilon)} = [\delta^{\varepsilon,*}\,\Omega_{\text{DYN},\mathcal{J}}]_{(\psi,t,E)}$.

Thus, for all $\mathcal{I} \in \mathcal{L}^\varepsilon \sqcup \{\mathcal{H}\}$, $\left(\mathcal{M}^{\text{DYN},\varepsilon}_{\mathcal{I}}, \mathcal{M}^{\text{SHELL},\varepsilon}_{\mathcal{I}}, \delta^\varepsilon_{\mathcal{I}}\right) = (\mathcal{J}, \mathcal{M}^\varepsilon, \delta^\varepsilon)$ is a phase space reduction of $\mathcal{M}^{\text{KIN}}$, hence of $\mathcal{M}^{\text{KIN}}_{\mathcal{I}}$.

Let $\mathcal{I}, \mathcal{I}' \in \mathcal{L}^\varepsilon \sqcup \{\mathcal{H}\}$, with $\mathcal{I} \subset \mathcal{I}'$. We have:

$$\pi^{\text{KIN}}_{\mathcal{I}'\to\mathcal{I}}\langle\mathcal{M}^{\text{SHELL},\varepsilon}_{\mathcal{I}'}\rangle = \left\{(\Pi_{\mathcal{I}}\,\psi, t, E) \,\middle|\, \psi \in \mathcal{J} \;\;\&\;\; (E - \langle\psi, H\psi\rangle)^2 + \epsilon^4 t^2 = \epsilon^2\right\} = \mathcal{M}^{\text{SHELL},\varepsilon}_{\mathcal{I}},$$

since $\mathcal{J} \subset \mathcal{I}$. Lastly, for $(\psi, t, E) \in \mathcal{M}^{\text{SHELL},\varepsilon}_{\mathcal{I}}$, $\psi'_o \in \mathcal{M}^{\text{DYN},\varepsilon}_{\mathcal{I}'} = \mathcal{J}$, we get:

$$\left(\exists\,(\psi', t', E') \in \mathcal{M}^{\text{SHELL},\varepsilon}_{\mathcal{I}'} \,/\, \delta^\varepsilon_{\mathcal{I}'}(\psi', t', E') = \psi'_o \;\;\&\;\; \pi^{\text{KIN}}_{\mathcal{I}'\to\mathcal{I}}(\psi', t', E') = (\psi, t, E)\right) \Leftrightarrow$$

$$\Leftrightarrow \left(\exists\,(\psi', t', E') \in \mathcal{M}_{\mathcal{J}}^{\text{SHELL},\mathcal{E}} \,/\, \delta_{\mathcal{J}}^{\mathcal{E}}(\psi', t', E') = \psi'_o \quad \& \quad (\psi', t', E') = (\psi, t, E)\right)$$

$$\Leftrightarrow \left(\delta_{\mathcal{J}}^{\mathcal{E}}(\psi, t, E) = \psi'_o = \pi_{\mathcal{J}' \to \mathcal{J}}^{\text{DYN},\mathcal{E} \to \mathcal{E}}(\psi'_o)\right),$$

therefore def. [10, 3.7.3] is fulfilled. □

Now, as we did in the previous section (prop. 2.3), we introduce a more concise dynamical projective system, that we will be able to identify with the one on the label set $\mathcal{EL}$ using the facility developed in [10, prop. 2.8]. This dynamical projective system could be thought of as a rendering [10, def. 2.6] of the dense domain $\mathcal{D}$ of the operator $H$, except for the fact that $\mathcal{D}$ is actually *not* a strong symplectic manifold (unless $H$ is bounded, in which case $\mathcal{D} = \mathcal{H}$). For the same reason, the assertion in prop. 3.8 could not be put in the form of [10, prop. 3.24], since we are lacking a phase space reduction at the level of the infinite dimensional manifold $\mathcal{M}^{\text{KIN}} = \mathcal{H} \times \mathbb{R}^2$ when $H$ is unbounded. Instead, we collect in prop. 3.9 a set of properties imitating the framework of [10, prop. 3.24], and we will formulate the convergence on this substitute ground.

It is worth mentioning that here, as in the previous example, we are able to directly give a projective system rendering the space of dynamical states, and more generally being able to find a regularizing scheme in the sense of [10, subsection 3.2] implies that one can construct such a dynamical projective structure. This perhaps requires a few comments. At first it sounds as if implementing and solving the constraints requires to already know completely the structure of the dynamical theory. However, one should keep in mind that solving the dynamics and obtaining the dynamical theory is *not* simply constructing the space of physical states: the more crucial part is to construct the dynamical observables, not simply as a space of functions on the reduced phase space, but as a family of *non functionally independent* elementary observables, each of which should be linked to a physical meaning (aka. an experimental protocol).

This point is transparently illustrated by the toy model we are studying in the present section. The submanifold $t = 0$ is obviously a gauge fixing surface of the theory we are considering, and this is what allows us to obtain immediately a description of the reduced phase space. But, clearly, having realized this property of the dynamics does not mean we have solved the theory: if we want to know how a given system will evolve we need to define dynamical observables associated to kinematical ones with support on other constant time surfaces. Indeed, the dynamical observables associated with time $t = 0$ are the only ones that can be directly defined on the reduced phase space defined through the aforementioned gauge fixing. And, although they provide a parametrization of the dynamical state space, they do not allow us to compute predictions for any arbitrary experiment, since, as underlined many times in the discussion of the handling of constraints [10, section 3], the predictive content of the theory is encoded in the functional relations among an overcomplete set of dynamical observables, arising from functionally independent kinematical observables.

Note that in any theory admitting some obvious gauge fixing (which needs not to be singled out nor preferred in any sense: in the example at hand, selecting $t = 0$ rather than any other time surface is an arbitrary choice), we can use this gauge fixing surface as a starting point to design an approximating scheme: it provides an explicit description of the reduced phase space, and we can use it as a pivot to define projections between the successive approximated dynamical theories (for we can relate approximated orbits depending on their intersection with the gauge fixing surface, as we indeed do in the present example). In particular, this suggests that such approximating schemes could be obtained without many difficulties within the so called 'deparametrization' framework [4].

**Proposition 3.7** Under the same hypotheses as in def. 3.4, we define:

1. $\mathcal{L}_H := \{\mathcal{J} \in \mathcal{L} \mid \forall \psi \in \mathcal{J},\ \|H\psi\| < \infty\}$ with the preorder defined by $\subset$;
2. $\forall \mathcal{J} \in \mathcal{L}_H$, $\mathcal{M}^{\text{DYN}}_{\mathcal{J}} := \mathcal{J}$, equipped with the symplectic structure $\Omega_{\text{DYN},\mathcal{J}}$ induced from $\mathcal{M}^{\text{DYN}}$, $\Omega_{\text{DYN}}$;
3. $\mathcal{D} := \{\psi \in \mathcal{H} \mid \|H\psi\| < \infty\}$ ($\mathcal{D}$ is the dense domain of the self-adjoint possibly unbounded operator $H$) and $\mathcal{M}^{\text{DYN}}_{\mathcal{D}} := \mathcal{D}$;
4. $\forall \mathcal{J}, \mathcal{J}' \in \mathcal{L}_H$, with $\mathcal{J} \subset \mathcal{J}'$, $\pi^{\text{DYN}}_{\mathcal{J}' \to \mathcal{J}} := \Pi_{\mathcal{J}}|_{\mathcal{J}' \to \mathcal{J}}$ where $\Pi_{\mathcal{J}}$ is the orthogonal projection on $\mathcal{J}$;
5. $\forall \mathcal{J} \in \mathcal{L}_H$, $\pi^{\text{DYN}}_{\mathcal{D} \to \mathcal{J}} := \Pi_{\mathcal{J}}|_{\mathcal{D} \to \mathcal{J}}$.

Then, $(\mathcal{L}_H, \mathcal{M}^{\text{DYN}}, \pi^{\text{DYN}})^{\downarrow}$ is a projective system of phase spaces and we can define (in analogy to [10, def. 2.6]) a map $\sigma^{\text{DYN}}_{\downarrow}$ as:

$$\begin{array}{rccl} \sigma^{\text{DYN}}_{\downarrow} : & \mathcal{M}^{\text{DYN}}_{\mathcal{D}} & \to & \mathcal{S}^{\downarrow}_{(\mathcal{L}_H, \mathcal{M}^{\text{DYN}}, \pi^{\text{DYN}})} \\ & \psi & \mapsto & \left(\pi^{\text{DYN}}_{\mathcal{D} \to \mathcal{J}}(\psi)\right)_{\mathcal{J} \in \mathcal{L}_H} \end{array}.$$

Additionally, we have a bijective antilinear map $\zeta^{\text{DYN}} : \mathcal{D}^* \to \mathcal{S}^{\downarrow}_{(\mathcal{L}_H, \mathcal{M}^{\text{DYN}}, \pi^{\text{DYN}})}$ such that $\zeta^{\text{DYN},-1} \circ \sigma^{\text{DYN}}_{\downarrow} : \mathcal{M}^{\text{DYN}}_{\mathcal{D}} \to \mathcal{D}^*$ is the restriction to $\mathcal{D}$ of the canonical identification of $\mathcal{H}$ with $\mathcal{D}' \subset \mathcal{D}^*$.

**Proof** We prove that $\mathcal{L}_H$ is directed like in the proof of prop. 3.5.

For $\mathcal{J} \in \mathcal{L}_H$, we have that $\pi^{\text{DYN}}_{\mathcal{D} \to \mathcal{J}}$ is surjective (since $\mathcal{J} \subset \mathcal{D}$) and, for $\mathcal{J} \subset \mathcal{J}' \in \mathcal{L}_H$, $\pi^{\text{DYN}}_{\mathcal{J}' \to \mathcal{J}} \circ \pi^{\text{DYN}}_{\mathcal{D} \to \mathcal{J}'} = \pi^{\text{DYN}}_{\mathcal{D} \to \mathcal{J}}$ (but speaking of compatibility with symplectic structure does not make sense for $\pi^{\text{DYN}}_{\mathcal{D} \to \mathcal{J}}$ since $\mathcal{D}$ is not a strong symplectic manifold).

The rest of the proof works as for prop. 2.2. □

**Proposition 3.8** We consider the objects introduced in props. 3.6 and 3.7. We define:

1. $\forall \varepsilon \in \mathcal{E}$, $\mathcal{L}'^{\varepsilon} := \mathcal{L}^{\varepsilon} \sqcup \{\mathcal{H}\}$;
2. $\forall \varepsilon = (\mathcal{J}, \epsilon) \in \mathcal{E}$, $\forall \mathcal{I} \in \mathcal{L}'^{\varepsilon}$, $\ell^{\text{DYN}}(\varepsilon, \mathcal{I}) := \mathcal{J}$;
3. $\forall (\varepsilon_1, \mathcal{I}_1), (\varepsilon_2, \mathcal{I}_2) \in \mathcal{E}\mathcal{L}' \,/\, (\varepsilon_1, \mathcal{I}_1) \preccurlyeq (\varepsilon_2, \mathcal{I}_2)$, $\pi^{\text{DYN},\varepsilon_2 \to \varepsilon_1}_{\mathcal{I}_2 \to \mathcal{I}_1} := \pi^{\text{DYN}}_{\ell^{\text{DYN}}(\varepsilon_2,\mathcal{I}_2) \to \ell^{\text{DYN}}(\varepsilon_1,\mathcal{I}_1)}$.

Then, $\left(\mathcal{E},\ (\mathcal{L}'^{\varepsilon})_{\varepsilon \in \mathcal{E}},\ (\mathcal{M}^{\text{DYN},\varepsilon}_{\mathcal{I}})_{(\varepsilon,\mathcal{I}) \in \mathcal{E}\mathcal{L}'},\ (\mathcal{M}^{\text{SHELL},\varepsilon}_{\mathcal{I}})_{(\varepsilon,\mathcal{I}) \in \mathcal{E}\mathcal{L}'},\ (\pi^{\text{DYN},\varepsilon_2 \to \varepsilon_1}_{\mathcal{I}_2 \to \mathcal{I}_1})_{(\varepsilon_1,\mathcal{I}_1) \preccurlyeq (\varepsilon_2,\mathcal{I}_2)},\ (\delta^{\varepsilon}_{\mathcal{I}})_{(\varepsilon,\mathcal{I}) \in \mathcal{E}\mathcal{L}'}\right)$ is a regularized reduction of $(\mathcal{L} \sqcup \{\mathcal{H}\}, \mathcal{M}^{\text{KIN}}, \pi^{\text{KIN}})^{\downarrow}$ and we have a bijective map $\kappa^{\text{DYN}} : \mathcal{S}^{\downarrow}_{(\mathcal{E}\mathcal{L}', \mathcal{M}^{\text{DYN}}, \pi^{\text{DYN}})} \to \mathcal{S}^{\downarrow}_{(\mathcal{L}_H, \mathcal{M}^{\text{DYN}}, \pi^{\text{DYN}})}$.

**Proof** $\mathcal{E}$ is a directed set (prop. 3.5) and $\left(\mathcal{L}'^{\varepsilon}\right)_{\varepsilon \in \mathcal{E}}$ is a family of decreasing cofinal parts of $\mathcal{L} \sqcup \{\mathcal{H}\}$.

Next, we have $\forall (\varepsilon, \mathcal{I}) \in \mathcal{E}\mathcal{L}'$, $\mathcal{M}^{\text{DYN},\varepsilon}_{\mathcal{I}} = \mathcal{M}^{\text{DYN}}_{\ell^{\text{DYN}}(\varepsilon,\mathcal{I})}$ hence for $(\varepsilon_1, \mathcal{I}_1) \preccurlyeq (\varepsilon_2, \mathcal{I}_2) \in \mathcal{E}\mathcal{L}'$, $\pi^{\text{DYN},\varepsilon_2 \to \varepsilon_1}_{\mathcal{I}_2 \to \mathcal{I}_1}$ is well-defined as a surjective map $\mathcal{M}^{\text{DYN},\varepsilon_2}_{\mathcal{I}_2} \to \mathcal{M}^{\text{DYN},\varepsilon_1}_{\mathcal{I}_1}$ and it is compatible with the symplectic structures.

Moreover, for $\varepsilon_1 = \varepsilon_2 \in \mathcal{E}$, this definition coincides with the map $\pi^{\text{DYN},\varepsilon \to \varepsilon}_{\mathcal{I}_2 \to \mathcal{I}_1}$ that has been introduced in prop. 3.6. Hence, for all $\varepsilon \in \mathcal{E}$, we have from prop. 3.6 that $(\mathcal{M}^{\text{DYN},\varepsilon}, \mathcal{M}^{\text{SHELL},\varepsilon}, \pi^{\text{DYN},\varepsilon \to \varepsilon}, \delta^{\varepsilon})$ is an elementary reduction of $\left(\mathcal{L}'^{\varepsilon}, \mathcal{M}^{\text{KIN}}, \pi^{\text{KIN}}\right)^{\downarrow}$.

Lastly, using $\ell^{\text{DYN}} : \mathcal{E}\mathcal{L}' \to \mathcal{L}_H$ (with $\ell^{\text{DYN}}\langle \mathcal{E}\mathcal{L}' \rangle = \mathcal{L}_H$), we have by [10, prop. 2.8] that $\left(\mathcal{E}\mathcal{L}', \mathcal{M}^{\text{DYN}}, \pi^{\text{DYN}}\right)^{\downarrow}$ is a projective system of phase spaces, thus:

$$\left(\mathcal{E},\ \left(\mathcal{L}'^{\varepsilon}\right)_{\varepsilon\in\mathcal{E}},\ \left(\mathcal{M}^{\text{DYN},\varepsilon}_{\mathcal{I}}\right)_{(\varepsilon,\mathcal{I})\in\mathcal{E}\mathcal{L}'},\ \left(\mathcal{M}^{\text{SHELL},\varepsilon}_{\mathcal{I}}\right)_{(\varepsilon,\mathcal{I})\in\mathcal{E}\mathcal{L}'},\ \left(\pi^{\text{DYN},\varepsilon_2\to\varepsilon_1}_{\mathcal{I}_2\to\mathcal{I}_1}\right)_{(\varepsilon_1,\mathcal{I}_1)\preccurlyeq(\varepsilon_2,\mathcal{I}_2)},\ \left(\delta^{\varepsilon}_{\mathcal{I}}\right)_{(\varepsilon,\mathcal{I})\in\mathcal{E}\mathcal{L}'}\right)$$

is a regularized reduction of $(\mathcal{L}\sqcup\{\mathcal{H}\},\ \mathcal{M}^{\text{KIN}},\ \pi^{\text{KIN}})^{\downarrow}$. And, in addition, there exists a bijective map $\kappa^{\text{DYN}} : \mathcal{S}^{\downarrow}_{(\mathcal{E}\mathcal{L}',\mathcal{M}^{\text{DYN}},\pi^{\text{DYN}})} \to \mathcal{S}^{\downarrow}_{(\mathcal{L}_H,\mathcal{M}^{\text{DYN}},\pi^{\text{DYN}})}$. □

**Proposition 3.9** We consider the same objects as in prop. 3.8 and we additionally define:

1. $\mathcal{M}^{\text{DYN},\mathcal{D}}_{\mathcal{H}} := \mathcal{M}^{\text{DYN}}_{\mathcal{D}} = \mathcal{D}$;
2. $\mathcal{M}^{\text{SHELL},\mathcal{D}}_{\mathcal{H}} := \{(\psi,\ t,\ E)\in\mathcal{M}^{\text{KIN}}_{\mathcal{H}} \mid \psi\in\mathcal{D} \quad \& \quad E = \langle\psi,\ H\psi\rangle\}$;
3. $\begin{array}{rccl}\delta^{\mathcal{D}}_{\mathcal{H}} : & \mathcal{M}^{\text{SHELL},\mathcal{D}}_{\mathcal{H}} & \to & \mathcal{M}^{\text{DYN},\mathcal{D}}_{\mathcal{H}} \\ & (\psi,\ t,\ E) & \mapsto & \exp(i\,t\,H)\ \psi\end{array}$;
4. $\forall(\varepsilon,\ \mathcal{I})\in\mathcal{E}\mathcal{L}',\ \pi^{\text{DYN},\mathcal{D}\to\varepsilon}_{\mathcal{H}\to\mathcal{I}} := \pi^{\text{DYN}}_{\mathcal{D}\to\ell^{\text{DYN}}(\varepsilon,\mathcal{I})}$.

Then, we have:

5. $\delta^{\mathcal{D}}_{\mathcal{H}}$ is surjective and, for all $\psi_o\in\mathcal{M}^{\text{DYN},\mathcal{D}}_{\mathcal{H}}$, $\left(\delta^{\mathcal{D}}_{\mathcal{H}}\right)^{-1}\langle\psi_o\rangle$ is connected;
6. for all $(\varepsilon,\ \mathcal{I})\in\mathcal{E}\mathcal{L}'$, $\pi^{\text{DYN},\mathcal{D}\to\varepsilon}_{\mathcal{H}\to\mathcal{I}}$ is surjective and, for all $(\varepsilon_1,\ \mathcal{I}_1)\preccurlyeq(\varepsilon_2,\ \mathcal{I}_2)\in\mathcal{E}\mathcal{L}'$, $\pi^{\text{DYN},\varepsilon_2\to\varepsilon_1}_{\mathcal{I}_2\to\mathcal{I}_1}\circ\pi^{\text{DYN},\mathcal{D}\to\varepsilon_2}_{\mathcal{H}\to\mathcal{I}_2} = \pi^{\text{DYN},\mathcal{D}\to\varepsilon_1}_{\mathcal{H}\to\mathcal{I}_1}$.

**Proof** Since $H$ is self-adjoint, $\exp(-i\,t\,H)$ defines a unitary operator on $\mathcal{H}$, and this operator stabilizes $\mathcal{D}$, for $\forall\psi\in\mathcal{D}$, $\|H\exp(-i\,t\,H)\ \psi\| = \|\exp(-i\,t\,H)\ H\,\psi\| = \|H\,\psi\| < \infty$. Hence, for $\psi_o\in\mathcal{M}^{\text{DYN},\mathcal{D}}_{\mathcal{H}}$:

$$\left(\delta^{\mathcal{D}}_{\mathcal{H}}\right)^{-1}\langle\psi_o\rangle = \{(\psi,\ t,\ E)\in\mathcal{M}^{\text{KIN}}_{\mathcal{H}} \mid t\in\mathbb{R},\ \psi = \exp(-i\,t\,H)\ \psi_o\in\mathcal{D} \quad \& \quad E = \langle\psi_o,\ H\psi_o\rangle < \infty\}.$$

Next, the statements 3.9.6 follows from the proof of prop. 3.7 (for $\forall(\varepsilon,\ \mathcal{I})\in\mathcal{E}\mathcal{L}'$, $\mathcal{M}^{\text{DYN},\varepsilon}_{\mathcal{I}} = \mathcal{M}^{\text{DYN}}_{\ell^{\text{DYN}}(\varepsilon,\mathcal{I})}$ and $\mathcal{M}^{\text{DYN},\mathcal{D}}_{\mathcal{H}} = \mathcal{M}^{\text{DYN}}_{\mathcal{D}}$). □

We close the discussion of the classical part of this toy model by proving that we indeed have suitable convergence at least for the dynamical states corresponding to vectors in $\mathcal{D}$, and, more precisely, that the successive projective families of orbits approximating such a state on the kinematical side correctly converge to the family arising from its associated orbit in the pseudo phase space reduction of $\mathcal{M}^{\text{KIN}}$ (that we introduced in prop. 3.9).

**Theorem 3.10** We consider the same objects as in prop. 3.9. Let $\psi_o\in\mathcal{D}$. For $\varepsilon\in\mathcal{E}$, we define:

1. $\psi^{\varepsilon} := (\delta^{\varepsilon}_{\mathcal{H}})^{-1}\left\langle\pi^{\text{DYN},\mathcal{D}\to\varepsilon}_{\mathcal{H}\to\mathcal{H}}(\psi_o)\right\rangle \subset \mathcal{M}^{\text{SHELL},\varepsilon}_{\mathcal{H}} \subset \mathcal{M}^{\text{KIN}}_{\mathcal{H}} = \mathcal{M}^{\text{KIN}}$;
2. $\Psi^{\varepsilon} := \widehat{\sigma}^{\text{KIN}}_{\downarrow}(\psi^{\varepsilon})$, where $\widehat{\sigma}^{\text{KIN}}_{\downarrow} : \mathcal{P}(\mathcal{M}^{\text{KIN}}) \to \widehat{\mathcal{S}}^{\downarrow}_{(\mathcal{L},\mathcal{M}^{\text{KIN}},\pi^{\text{KIN}})}$ is defined as in [10, prop. 3.23].

Then, the net $(\Psi^{\varepsilon})_{\varepsilon\in\mathcal{E}}$ converges in $\widehat{\mathcal{S}}^{\downarrow}_{(\mathcal{L},\mathcal{M}^{\text{KIN}},\pi^{\text{KIN}})}$ to $\widehat{\sigma}^{\text{KIN}}_{\downarrow}\left(\left(\delta^{\mathcal{D}}_{\mathcal{H}}\right)^{-1}\langle\psi_o\rangle\right)$ [10, def. 3.21].

**Proof** For $\varepsilon = (\mathcal{J},\ \epsilon)\in\mathcal{E}$, we have, from the proof of prop. 3.6:

$$\psi^{\varepsilon} = \left\{ \left( e^{-\frac{i}{\epsilon}\sin\theta\,\Pi_{\mathcal{J}} H}\,\Pi_{\mathcal{J}}(\psi_o),\ \tfrac{1}{\epsilon}\sin\theta,\ E_{\mathcal{J}} + \epsilon\cos\theta \right) \;\middle|\; \theta \in [0, 2\pi[ \right\},$$

where $E_{\mathcal{J}} := \langle \Pi_{\mathcal{J}}(\psi_o),\ H\,\Pi_{\mathcal{J}}(\psi_o) \rangle$. Hence, for all $\mathcal{I} \in \mathcal{L}$:

$$[\Psi^{\varepsilon}]_{\mathcal{I}} = \left\{ \left( \Pi_{\mathcal{I}}\, e^{-\frac{i}{\epsilon}\sin\theta\,\Pi_{\mathcal{J}} H}\,\Pi_{\mathcal{J}}(\psi_o),\ \tfrac{1}{\epsilon}\sin\theta,\ E_{\mathcal{J}} + \epsilon\cos\theta \right) \;\middle|\; \theta \in [0, 2\pi[ \right\}.$$

And, from the proof of prop. 3.9:

$$\left[ \widehat{\sigma}_{\downarrow}^{\text{KIN}} \left( \left( \delta_{\mathcal{H}}^{\mathcal{D}} \right)^{-1} \langle \psi_o \rangle \right) \right]_{\mathcal{I}} = \left\{ \left( \Pi_{\mathcal{I}}\, e^{-i\,t\,H}\,\psi_o,\ t,\ E_{\mathfrak{D}} \right) \;\middle|\; t \in \mathbb{R} \right\} \text{ where } E_{\mathfrak{D}} := \langle \psi_o,\ H\,\psi_o \rangle.$$

Let $\mathcal{I} \in \mathcal{L}$ and let $U$ be an open set in $\mathcal{M}^{\text{KIN}}_{\mathcal{I}}$, such that $U \cap \left[ \widehat{\sigma}_{\downarrow}^{\text{KIN}} \left( \left( \delta_{\mathcal{H}}^{\mathcal{D}} \right)^{-1} \langle \psi_o \rangle \right) \right]_{\mathcal{I}} \neq \varnothing$. Let $t \in \mathbb{R}$ such that:

$$\left( \Pi_{\mathcal{I}}\, e^{-i\,t\,H}\,\psi_o,\ t,\ E_{\mathfrak{D}} \right) \in U,$$

and let $\epsilon_1 > 0$ such that:

$$\begin{aligned} &\forall \psi \in \mathcal{I},\ \forall E \in \mathbb{R}, \\ &\qquad \left( \left\| \psi - \Pi_{\mathcal{I}}\, e^{-it\,H}\,\psi_o \right\| \leqslant \epsilon_1\ (\|\psi_o\| + 1) \quad \& \quad |E - E_{\mathfrak{D}}| \leqslant \epsilon_1 \right) \quad \Rightarrow \quad ((\psi,\ t,\ E) \in U). \end{aligned}$$

Let $\epsilon_2 = \frac{1}{1+|t|}\log\left(1 + \frac{\epsilon_1}{2}\right) > 0$ and $\epsilon_3 = \min\left(\epsilon_1,\ \frac{1}{1+|t|}\right) > 0$.

By spectral resolution, we can define $H^{\epsilon_2} := \epsilon_2 \left\lfloor \frac{1}{\epsilon_2} H \right\rfloor$ (where $\lfloor\,\cdot\,\rfloor$ denotes the floor function). Then, we have $[H,\ H^{\epsilon_2}] = 0$, $\|H - H^{\epsilon_2}\| \leqslant \epsilon_2$ and $H^{\epsilon_2}$ has discrete spectrum included in $\epsilon_2\,\mathbb{Z}$. Hence, there exists $N \in \mathbb{N}$ such that:

$$\left\| \psi_o - \sum_{k=-N}^{+N} \psi_o^{\epsilon_2, k} \right\| \leqslant \frac{\epsilon_1}{2},$$

where, for $k \in \mathbb{Z}$, $\psi_o^{\epsilon_2,k}$ is the projection of $\psi_o$ on the eigenspace of $H^{\epsilon_2}$ with eigenvalue $\epsilon_2\,k$ (defining these eigenspace to be $\{0\}$ if $\epsilon_2\,k$ is not in the spectrum of $H^{\epsilon_2}$).

We define $\mathcal{J}_2 = \text{Vect}\left\{ \psi_o^{\epsilon_2,k} \;\middle|\; k \in \{-N, \dots, N\} \right\}$, $\mathcal{J}_2$ is finite dimensional, and $\forall \psi \in \mathcal{J}_2$, $\|H\psi\| \leqslant \epsilon_2\ \|\psi\| + N\epsilon_2\ \|\psi\| < \infty$, hence $\mathcal{J}_2 \in \mathcal{L}_H$. Moreover, $\mathcal{J}_2$ is stabilized by $H^{\epsilon_2}$ and $\|\psi_o - \Pi_{\mathcal{J}_2}\,\psi_o\| \leqslant \frac{1}{2}\,\epsilon_1$. Next, we define $\mathcal{J}_3 = \mathcal{J}_2 + \text{Vect}\ \{\psi_o\} \in \mathcal{L}_H$ (because $\psi_o \in \mathfrak{D}$).

Now, we consider $\varepsilon = (\mathcal{J},\ \epsilon) \in \mathcal{E}$, with $(\mathcal{J},\ \epsilon) \succcurlyeq (\mathcal{J}_3,\ \epsilon_3)$. We choose $\theta \in [0, 2\pi[$ such that $\sin\theta = \epsilon\,t$ ($|\epsilon t| \leqslant \epsilon_3\,|t| \leqslant 1$) and we define:

$$\psi := \Pi_{\mathcal{I}}\, e^{-\frac{i}{\epsilon}\sin\theta\,\Pi_{\mathcal{J}} H}\,\Pi_{\mathcal{J}}(\psi_o).$$

We have:

$$\begin{aligned} \left\| \psi - \Pi_{\mathcal{I}}\, e^{-it\,H}\,\psi_o \right\| &\leqslant \left\| e^{-it\,\Pi_{\mathcal{J}} H}\,\Pi_{\mathcal{J}}(\psi_o) - e^{-it\,H}\,\psi_o \right\| \\ &\leqslant \left\| e^{-it\,\Pi_{\mathcal{J}}\,(H - H^{\epsilon_2})} - \text{id}_{\mathcal{J}} \right\|\ \left\| e^{-it\,\Pi_{\mathcal{J}}\,H^{\epsilon_2}}\,\Pi_{\mathcal{J}}(\psi_o) \right\| + \left\| e^{-it\,\Pi_{\mathcal{J}}\,H^{\epsilon_2}}\,\Pi_{\mathcal{J}}(\psi_o) - e^{-it\,H^{\epsilon_2}}\,\psi_o \right\| + \\ &\quad + \left\| e^{-it\,(H - H^{\epsilon_2})} - \text{id}_{\mathcal{H}} \right\|\ \left\| e^{-it\,H^{\epsilon_2}}\,(\psi_o) \right\| \\ &\leqslant 2\ \left| e^{|t|\,\|H - H^{\epsilon_2}\|} - 1 \right|\ \|\psi_o\| + \left\| e^{-it\,\Pi_{\mathcal{J}}\,H^{\epsilon_2}}\,\Pi_{\mathcal{J}_2}(\psi_o) - e^{-it\,H^{\epsilon_2}}\,\Pi_{\mathcal{J}_2}(\psi_o) \right\| + \end{aligned}$$

$$+ \left\| e^{-it\,\Pi_{\mathcal{J}} H^{\epsilon_2}} \left(\psi_o - \Pi_{\mathcal{J}_2}(\psi_o)\right) - e^{-it\,H^{\epsilon_2}} \left(\psi_o - \Pi_{\mathcal{J}_2}(\psi_o)\right) \right\|$$

$$\leqslant 2 \left| e^{|t|\,\epsilon_2} - 1 \right| \, \|\psi_o\| + \left\| e^{-it\,H^{\epsilon_2}}\, \Pi_{\mathcal{J}_2}(\psi_o) - e^{-it\,H^{\epsilon_2}}\, \Pi_{\mathcal{J}_2}(\psi_o) \right\| + 2\, \|\psi_o - \Pi_{\mathcal{J}_2}(\psi_o)\|$$

$$\leqslant 2\,\frac{\epsilon_1}{2}\, \|\psi_o\| + 2\,\frac{\epsilon_1}{2} \leqslant \epsilon_1\, (1 + \|\psi_o\|)\,,$$

and:

$$|E_{\mathcal{J}} + \epsilon\, \cos\theta - E_{\mathfrak{D}}| \leqslant |\langle \psi_o, \, (H - \Pi_{\mathcal{J}}\, H\, \Pi_{\mathcal{J}})\, \psi_o \rangle| + \epsilon_1 = \epsilon_1 \text{ (since } \psi_o \in \mathcal{J}_3 \subset \mathcal{J}).$$

Therefore, $\left(\psi, \, \frac{1}{\epsilon}\, \sin\theta, \, E_{\mathcal{J}} + \epsilon\cos\theta\right) \in U$, but since $\left(\psi, \, \frac{1}{\epsilon}\, \sin\theta, \, E_{\mathcal{J}} + \epsilon\cos\theta\right) \in [\psi^{\varepsilon}]_{\mathfrak{I}}$, we have $\forall \varepsilon = (\mathcal{J}, \, \epsilon) \succcurlyeq (\mathcal{J}_3, \, \epsilon_3)\,, \, [\psi^{\varepsilon}]_{\mathfrak{I}} \cap U \neq \varnothing$.

Let $K$ be a compact subset of $\mathcal{M}^{\text{KIN}}_{\mathfrak{I}}$, such that $K \cap \left[ \widehat{\sigma}^{\text{KIN}}_{\downarrow} \left( \left(\delta^{\mathfrak{D}}_{\mathcal{H}}\right)^{-1} \langle \psi_o \rangle \right) \right]_{\mathfrak{I}} = \varnothing$. Hence, there exist $T > 0$ and $\epsilon_1 > 0$ such that:

$$\forall \psi \in \mathfrak{I}, \, \forall t \in \mathbb{R}, \, \forall E \in \mathbb{R},$$

$$\left[ \left( \left\| \psi - \Pi_{\mathfrak{I}}\, e^{-it\,H}\, \psi_o \right\| \leqslant \epsilon_1\, (\|\psi_o\| + 1) \quad \& \quad |E - E_{\mathfrak{D}}| \leqslant \epsilon_1 \right) \quad \text{or} \quad |t| \geqslant T \right] \;\Rightarrow\; ((\psi, \, t, \, E) \notin K)\,.$$

Following the same path as above, we can define:

$$\epsilon_2 = \tfrac{1}{1+T} \log\left(1 + \tfrac{\epsilon_1}{2}\right) > 0,$$

and construct a vector subspace $\mathcal{J}_2 \in \mathcal{L}_H$ such that:

$$\forall \mathcal{J} \in \mathcal{L}_H \,/\, \mathcal{J} \supset \mathcal{J}_2, \, \forall t \in \mathbb{R}, \; \left\| e^{-it\,\Pi_{\mathcal{J}} H}\, \Pi_{\mathcal{J}}(\psi_o) - e^{-it\,H}\, \psi_o \right\| \leqslant 2 \left| e^{|t|\,\epsilon_2} - 1 \right| \, \|\psi_o\| + \epsilon_1\,.$$

Analogously, we define $\epsilon_3 = \epsilon_1$ and $\mathcal{J}_3 = \mathcal{J}_2 + \text{Vect}\, \{\psi_o\} \in \mathcal{L}_H$.

Now, we consider $\varepsilon = (\mathcal{J}, \, \epsilon) \in \mathcal{E}$, with $(\mathcal{J}, \, \epsilon) \succcurlyeq (\mathcal{J}_3, \, \epsilon_3)$, and $\theta \in [0, 2\pi[$. If $\left|\frac{1}{\epsilon} \sin\theta\right| < T$, then we have, with $t = \frac{1}{\epsilon} \sin\theta$:

$$\left\| \Pi_{\mathfrak{I}}\, e^{-\frac{i}{\epsilon}\, \sin\theta\, \Pi_{\mathcal{J}} H}\, \Pi_{\mathcal{J}}(\psi_o) - \Pi_{\mathfrak{I}}\, e^{-it\,H}\, \psi_o \right\| \leqslant \epsilon_2\, (1 + \|\psi_o\|)\,,$$

and:

$$|E_{\mathcal{J}} + \epsilon\, \cos\theta - E_{\mathfrak{D}}| \leqslant \epsilon_1\,.$$

Therefore, $\forall \varepsilon = (\mathcal{J}, \, \epsilon) \succcurlyeq (\mathcal{J}_3, \, \epsilon_3)\,, \, [\psi^{\varepsilon}]_{\mathfrak{I}} \cap K = \varnothing$.

So, for every $\mathfrak{I} \in \mathcal{L}$, the net $\left([\Psi^{\varepsilon}]_{\mathfrak{I}}\right)_{\varepsilon \in \mathcal{E}}$ converges in $\mathcal{P}\left(\mathcal{M}^{\text{KIN}}_{\mathfrak{I}}\right)$ to $\left[ \widehat{\sigma}^{\text{KIN}}_{\downarrow} \left( \left(\delta^{\mathfrak{D}}_{\mathcal{H}}\right)^{-1} \langle \psi_o \rangle \right) \right]_{\mathfrak{I}}$, thus, the net $(\Psi^{\varepsilon})_{\varepsilon \in \mathcal{E}}$ converges in $\widehat{\mathcal{S}}^{\downarrow}_{(\mathcal{L}, \mathcal{M}^{\text{KIN}}, \pi^{\text{KIN}})}$ to $\widehat{\sigma}^{\text{KIN}}_{\downarrow} \left( \left(\delta^{\mathfrak{D}}_{\mathcal{H}}\right)^{-1} \langle \psi_o \rangle \right)$. □

## 3.2 Quantum theory

We now want to implement this construction at the quantum level, with the aim of using this simple toy model to get a first hold on the implementation of constraints in projective systems of quantum state spaces.

To fix the notations, we begin by summarizing the main properties of (bosonic) Fock spaces [5, section I.3.4].

**Definition 3.11** Let $\mathcal{H}$ be a separable Hilbert space. We define the Fock space $\widehat{\mathcal{H}}$ by:

$\widehat{\mathcal{H}} := \overline{\bigoplus_{n\in\mathbb{N}} \mathcal{H}^{\otimes n,\,\mathrm{sym}}}$ where $\mathcal{H}^{\otimes n,\,\mathrm{sym}}$ is the symmetric vector subspace of $\mathcal{H}^{\otimes n}$.

For $(e_i)_{i\in I}$ an orthonormal basis of $\mathcal{H}$ ($I \subset \mathbb{N}$), we define:

$\Lambda_I := \{(n_i)_{i\in I} \mid \forall i \in I,\, n_i \in \mathbb{N} \quad \& \quad \sum_i n_i < \infty\}$,

indexing the orthonormal basis $\left( |(n_i)_{i\in I}\,,\, (e_i)_{i\in I}\rangle \right)_{(n_i)_{i\in I}\in\Lambda_I}$ of $\widehat{\mathcal{H}}$:

$$\forall\, (n_i)_{i\in I} \in \Lambda_I,\ |(n_i)_{i\in I}\,,\, (e_i)_{i\in I}\rangle := \sqrt{\frac{\prod_{i\in I}(n_i!)}{N!}} \sum_{\substack{k_1,\dots,k_N \\ \forall i\in I,\, \#\{l \,\mid\, k_l=i\}=n_i}} |e_{k_1}\rangle \otimes \ldots \otimes |e_{k_N}\rangle\,,$$

where $N = \sum_{i\in I} n_i$.

If $\left(f_j\right)_{j\in I}$ is an other orthonormal basis of $\mathcal{H}$, we have:

$$\forall\, (n_i)_{i\in I}\,,\, \left(m_j\right)_{j\in I} \in \Lambda_I,\ \left\langle (n_i)_{i\in I}\,,\, (e_i)_{i\in I} \,\middle|\, \left(m_j\right)_{j\in I}\,,\, \left(f_j\right)_{j\in I}\right\rangle =$$

$$= \sum_{\substack{l_{i,j}\in\mathbb{N}^{I\times I} \\ \forall i,\, n_i=\sum_j l_{i,j} \\ \forall j,\, m_j=\sum_i l_{i,j}}} \prod_{i\in I}\left(\sqrt{\frac{n_i!}{\prod_j l_{i,j}!}}\right) \prod_{j\in I}\left(\sqrt{\frac{m_j!}{\prod_i l_{i,j}!}}\right) \prod_{i,j} \left\langle e_i\,,\, f_j\right\rangle^{l_{i,j}}. \tag{3.11.1}$$

**Definition 3.12** We consider the same objects as in def. 3.11. Let $e \in \mathcal{H}$, $N \geqslant 1$ and $l \in \{1, \ldots, N\}$. We define the operators $\widehat{\mathrm{a}}_e^{N,l} : \mathcal{H}^{\otimes N} \to \mathcal{H}^{\otimes N-1}$ and $\left(\widehat{\mathrm{a}}_e^{N,l}\right)^+ : \mathcal{H}^{\otimes N-1} \to \mathcal{H}^{\otimes N}$ by:

$$\forall \varphi_1, \ldots, \varphi_N \in \mathcal{H},\ \ \widehat{\mathrm{a}}_e^{N,\,l}\, \varphi_1^{(1)} \otimes \ldots \otimes \varphi_N^{(N)} := \frac{\langle e\,,\, \varphi_l\rangle}{\sqrt{N}}\, \varphi_1^{(1)} \otimes \ldots \otimes \not{\varphi}_l^{(l)} \otimes \ldots \otimes \varphi_N^{(N-1)}\,,$$

$$\text{and}\quad \forall \varphi_1, \ldots, \varphi_{N-1} \in \mathcal{H},\ \ \left(\widehat{\mathrm{a}}_e^{N,\,l}\right)^+ \varphi_1^{(1)} \otimes \ldots \otimes \varphi_{N-1}^{(N-1)} := \frac{1}{\sqrt{N}} \varphi_1^{(1)} \otimes \ldots \otimes e^{(l)} \otimes \ldots \otimes \varphi_{N-1}^{(N)}\,.$$

Then, on $\widehat{\mathcal{H}}$ we can define (unbounded) operators $\widehat{\mathrm{a}}_e$ and $\widehat{\mathrm{a}}_e^+$, such that:

$$\forall \psi \in \mathcal{H}^{\otimes N,\,\mathrm{sym}},\ \ \widehat{\mathrm{a}}_e\, \psi = \sum_{l=1}^{N} \widehat{\mathrm{a}}_e^{N,\,l}\, \psi\ \in \mathcal{H}^{\otimes N-1,\,\mathrm{sym}}\,,$$

$$\text{and}\quad \forall \psi \in \mathcal{H}^{\otimes N-1,\,\mathrm{sym}},\ \ \widehat{\mathrm{a}}_e^+\, \psi = \sum_{l=1}^{N} \left(\widehat{\mathrm{a}}_e^{N,\,l}\right)^+ \psi\ \in \mathcal{H}^{\otimes N,\,\mathrm{sym}}\,.$$

Let $A$ be a bounded self-adjoint operator on $\mathcal{H}$. We can define an (unbounded) operator $\widehat{A}$ on $\widehat{\mathcal{H}}$ such that:

$$\forall \psi \in \mathcal{H}^{\otimes N,\,\mathrm{sym}},\ \ \widehat{A}\,\psi = \sum_{l=1}^{N} \mathrm{id}_{\mathcal{H}}^{(1)} \otimes \ldots \otimes A^{(l)} \otimes \ldots \otimes \mathrm{id}_{\mathcal{H}}^{(N)}\, \psi \ \in H^{\otimes N,\,\mathrm{sym}} .$$

For $(e_i)_{i \in I}$ is an orthonormal basis of $\mathcal{H}$, we have:

$$\widehat{A} = \sum_{i,j \in I} \left\langle e_i ,\, A\, e_j \right\rangle\, \widehat{\mathrm{a}}^{+}_{e_i}\, \widehat{\mathrm{a}}_{e_j} .$$

Lastly, let $e,\, f \in \mathcal{H}$ and let $A,\, B$ be bounded self-adjoint operator on $\mathcal{H}$. The commutators between the operators defined above are given by:

$$[\widehat{\mathrm{a}}_e ,\, \widehat{\mathrm{a}}_f] = 0,\ \left[\widehat{\mathrm{a}}^{+}_e ,\, \widehat{\mathrm{a}}^{+}_f\right] = 0, \text{ and } \left[\widehat{\mathrm{a}}_e ,\, \widehat{\mathrm{a}}^{+}_f\right] = \langle e,\, f\rangle\, \mathrm{id}_{\widehat{\mathcal{H}}} ,$$

$$\left[\widehat{A},\, \widehat{B}\right] = \widehat{[A,\, B]_{\mathcal{H}}} ,\ \left[\widehat{\mathrm{a}}_e ,\, \widehat{A}\right] = \widehat{\mathrm{a}}_{Ae} , \text{ and } \left[\widehat{\mathrm{a}}^{+}_e ,\, \widehat{A}\right] = -\widehat{\mathrm{a}}^{+}_{Ae} .$$

Before going on to the quantization using projective structures, we recall the more conventional quantization of $\mathcal{M}^{\mathrm{DYN}}$, $\Omega_{\mathrm{DYN}}$ (ie. a reduced phase space quantization for the theory we are considering) using Fock spaces techniques. The notable fact is that this direct quantization of the Schrödinger equation (considered as a classical field theory, aka. second quantization) can be identified with the (bosonic) Fock space describing an arbitrary number of independent, indistinguishable quantum particles of the corresponding first quantized theory [2]. This identification is not merely a naive matching of the Hilbert spaces: we can check that the quantized observables correspond in a natural way to the observables built on the Fock space.

**Proposition 3.13** We consider the objects introduced in prop. 3.2 and def. 3.12. We define the Fock quantization of $\mathcal{M}^{\mathrm{DYN}}$ as $\widehat{\mathcal{M}}^{\mathrm{DYN}}_{\mathrm{Fock}} := \widehat{\mathcal{H}}$. For $A$ a bounded self-adjoint operator on $\mathcal{H}$ and $e \in \mathcal{H}$ we define the following quantizations for the observables on $\mathcal{M}^{\mathrm{DYN}}$:

$$\widehat{\langle A \rangle}_{\mathrm{Fock}} := \widehat{A},\ \widehat{(\mathrm{a}_e)}_{\mathrm{Fock}} := \widehat{\mathrm{a}}_e , \text{ and } \widehat{(\mathrm{a}^{*}_e)}_{\mathrm{Fock}} := \widehat{\mathrm{a}}^{+}_e .$$

Then, we have:

$$\forall O,\, O' \in \{\mathrm{a}_e \mid e \in \mathcal{H}\} \cup \{\mathrm{a}^{*}_e \mid e \in \mathcal{H}\} \cup \{\langle A \rangle \mid A \text{ bounded, self-adj on } \mathcal{H}\} ,$$

$$\left[\widehat{O}_{\mathrm{Fock}},\, \widehat{O}'_{\mathrm{Fock}}\right] = -i \left(\widehat{\{O,\, O'\}}_{\mathrm{DYN}}\right)_{\mathrm{Fock}} .$$

**Proof** This can be directly checked by comparing prop. 3.2 with def. 3.12. □

The key tool for constructing a projective system of quantum state spaces reproducing the classical structure from prop. 3.3 is the realization that the Fock space arising from a direct orthogonal sum of two Hilbert space can be naturally identified with the tensor product of the two corresponding Fock spaces. This is in fact a special case of the well-known property of quantization, that translates a Cartesian product of symplectic manifold into a tensor product of Hilbert spaces (for a direct sum is indeed a Cartesian product).

**Proposition 3.14** Let $\mathfrak{I}$ be a separable Hilbert space. Let $\mathfrak{J}$ be a vector subspace of $\mathfrak{I}$ and $\mathfrak{J}^\perp$ the orthogonal complement of $\mathfrak{J}$ in $\mathfrak{I}$. Let $(e_i)_{i\in J}$ be an orthonormal basis of $\mathfrak{J}$ and $(e_i)_{i\in I\setminus J}$ be an orthonormal basis of $\mathfrak{J}^\perp$ (with $I \supset J$). Hence, $(e_i)_{i\in I}$ is an orthonormal basis of $\mathfrak{I} = \mathfrak{J} \oplus \mathfrak{J}^\perp$.

We consider the corresponding Fock spaces $\widehat{\mathfrak{I}}$, $\widehat{\mathfrak{J}}$ & $\widehat{\mathfrak{J}^\perp}$ (def. 3.11) and we define the linear application $\widehat{\varphi}_{\mathfrak{I}\to\mathfrak{J}} : \widehat{\mathfrak{I}} \to \widehat{\mathfrak{J}^\perp} \otimes \widehat{\mathfrak{J}}$ by its action on the orthonormal basis $\left(|(n_i)_{i\in I}, (e_i)_{i\in I}\rangle\right)_{(n_i)_{i\in I}\in\Lambda_I}$ of $\widehat{\mathfrak{I}}$:

$$\widehat{\varphi}_{\mathfrak{I}\to\mathfrak{J}} \, |(n_i)_{i\in I}, (e_i)_{i\in I}\rangle := \left|(n_i)_{i\in I\setminus J}, (e_i)_{i\in I\setminus J}\right\rangle \otimes |(n_i)_{i\in J}, (e_i)_{i\in J}\rangle . \tag{3.14.1}$$

Then, $\widehat{\varphi}_{\mathfrak{I}\to\mathfrak{J}}$ is an Hilbert space isomorphism. Moreover, $\widehat{\varphi}_{\mathfrak{I}\to\mathfrak{J}}$ does not depend on the choice of the bases $(e_i)_{i\in J}$ and $(e_i)_{i\in I\setminus J}$.

**Proof** $\widehat{\varphi}_{\mathfrak{I}\to\mathfrak{J}}$ sends an orthonormal basis to an orthonormal basis, since the map:

$$\begin{array}{rcl} \Lambda_I & \to & \Lambda_{J\setminus I} \times \Lambda_J \\ (n_i)_{i\in I} & \mapsto & (n_i)_{i\in I\setminus J}, (n_i)_{i\in J} \end{array},$$

is bijective.

Then, if $(f_i)_{i\in J}$ is an other orthonormal basis of $\mathfrak{J}$ and $(f_i)_{i\in I\setminus J}$ is an other orthonormal basis of $\mathfrak{J}^\perp$, we have, using eq. (3.11.1) for $\left(m_j\right) j \in I \in \Lambda_I$:

$$\widehat{\varphi}_{\mathfrak{I}\to\mathfrak{J}} \left|\left(m_j\right)_{j\in I}, \left(f_j\right)_{j\in I}\right\rangle =$$

$$= \sum_{\substack{l_{i,j}\in\mathbb{N}^{I\times I} \\ \forall j\in I,\, m_j=\sum_i l_{i,j}}} \prod_{i\in I} \left(\sqrt{\left(\sum_{j\in I} l_{i,j}\right)!}\right) \prod_{j\in I} \left(\sqrt{m_j!}\right) \prod_{i,j\in I} \frac{\left\langle e_i, f_j\right\rangle^{l_{i,j}}}{l_{i,j}!} \, \widehat{\varphi}_{\mathfrak{I}\to\mathfrak{J}} \left|\left(\sum_{j\in I} l_{i,j}\right)_{i\in I}, (e_i)_{i\in I}\right\rangle .$$

Now, for $i, j \in J \times (I \setminus J)$ or $(I \setminus J) \times J$, $\left\langle e_i, f_j\right\rangle = 0$ since $\left\langle \mathfrak{J}, \mathfrak{J}^\perp\right\rangle = 0$. Therefore, the only non-vanishing terms in the sum above are such that $l_{i,j} = \mathbb{1}_{(i,j)\in(I\setminus J)^2} k_{i,j} + \mathbb{1}_{(i,j)\in J^2} p_{i,j}$ with $k_{i,j} \in \mathbb{N}^{(I\setminus J)\times(I\setminus J)}$ and $p_{i,j} \in \mathbb{N}^{J\times J}$. Hence, using eq. (3.14.1):

$$\widehat{\varphi}_{\mathfrak{I}\to\mathfrak{J}} \left|\left(m_j\right)_{j\in I}, \left(f_j\right)_{j\in I}\right\rangle =$$

$$= \sum_{\substack{k_{i,j}\in\mathbb{N}^{(I\setminus J)\times(I\setminus J)} \\ \forall j\in(I\setminus J),\, m_j=\sum_i k_{i,j}}} \sum_{\substack{p_{i,j}\in\mathbb{N}^{J\times J} \\ \forall j\in J,\, m_j=\sum_i p_{i,j}}} \prod_{i\in I\setminus J} \left(\sqrt{\left(\sum_{j\in I\setminus J} k_{i,j}\right)!}\right) \prod_{i\in J} \left(\sqrt{\left(\sum_{j\in J} p_{i,j}\right)!}\right) \prod_{j\in I\setminus J} \left(\sqrt{m_j!}\right) \times$$

$$\times \prod_{j\in J} \left(\sqrt{m_j!}\right) \prod_{i,j\in I\setminus J} \frac{\left\langle e_i, f_j\right\rangle^{k_{i,j}}}{k_{i,j}!} \prod_{i,j\in J} \frac{\left\langle e_i, f_j\right\rangle^{p_{i,j}}}{p_{i,j}!} \left|\left(\sum_{j\in I\setminus J} k_{i,j}\right)_{i\in I\setminus J}, (e_i)_{i\in I\setminus J}\right\rangle \otimes \left|\left(\sum_{j\in J} p_{i,j}\right)_{i\in J}, (e_i)_{i\in J}\right\rangle$$

$$= \left|\left(m_j\right)_{j\in I\setminus J}, \left(f_j\right)_{j\in I\setminus J}\right\rangle \otimes \left|\left(m_j\right)_{j\in J}, \left(f_j\right)_{j\in J}\right\rangle ,$$

where we used again eq. (3.11.1), both in $\widehat{\mathcal{J}^\perp}$ and in $\widehat{\mathcal{J}}$. □

**Proposition 3.15** We consider the objects introduced in props. 3.3 and 3.14. We define:

1. $\forall \mathcal{I} \in \mathcal{L}$, $\widehat{\mathcal{M}}^{\text{KIN}}_{\mathcal{I}} := \widehat{\mathcal{I}} \otimes \mathcal{T}$ where $\mathcal{T} := L_2(\mathbb{R}, d\mu)$ ($\mu$ being the Lebesgue measure on $\mathbb{R}$);
2. $\forall \mathcal{I} \subset \mathcal{I}' \in \mathcal{L}$, $\widehat{\mathcal{M}}^{\text{KIN}}_{\mathcal{I}' \to \mathcal{I}} := \widehat{\mathcal{I}^\perp \cap \mathcal{I}'}$ (with the convention that $\widehat{\mathcal{M}}^{\text{KIN}}_{\mathcal{I}' \to \mathcal{I}} = \mathbb{C}$ if $\mathcal{I}' = \mathcal{I}$);
3. $\forall \mathcal{I} \subset \mathcal{I}' \in \mathcal{L}$, $\widehat{\varphi}^{\text{KIN}}_{\mathcal{I}' \to \mathcal{I}} := \widehat{\varphi}_{\mathcal{I}' \to \mathcal{I}} \otimes \text{id}_{\mathcal{T}} : \widehat{\mathcal{I}'} \otimes \mathcal{T} \to \widehat{\mathcal{I}^\perp \cap \mathcal{I}'} \otimes \left(\widehat{\mathcal{I}} \otimes \mathcal{T}\right)$;
4. $\forall \mathcal{I} \subset \mathcal{I}' \subset \mathcal{I}'' \in \mathcal{L}$, $\widehat{\varphi}^{\text{KIN}}_{\mathcal{I}'' \to \mathcal{I}' \to \mathcal{I}} := \widehat{\varphi}_{\left(\mathcal{I}^\perp \cap \mathcal{I}''\right) \to \left(\mathcal{I}^\perp \cap \mathcal{I}'\right)} : \widehat{\mathcal{I}^\perp \cap \mathcal{I}''} \to \widehat{\mathcal{I}'^\perp \cap \mathcal{I}''} \otimes \widehat{\mathcal{I}^\perp \cap \mathcal{I}'}$ (note that $\left(\mathcal{I}^\perp \cap \mathcal{I}'\right)^\perp \cap \left(\mathcal{I}^\perp \cap \mathcal{I}''\right) = \mathcal{I}'^\perp \cap \mathcal{I}''$ since $\mathcal{I} \subset \mathcal{I}'$).

$\left(\mathcal{L}, \widehat{\mathcal{M}}^{\text{KIN}}, \widehat{\varphi}^{\text{KIN}}\right)^{\otimes}$ is a projective system of quantum state spaces [11, def. 2.1].

**Proof** $\mathcal{L}$ is directed since for $\mathcal{I}, \mathcal{I}' \in \mathcal{L}$, $\mathcal{I} + \mathcal{I}' \in \mathcal{L}$. And, for $\mathcal{I} \subset \mathcal{I}' \subset \mathcal{I}'' \in \mathcal{L}$, $\widehat{\varphi}^{\text{KIN}}_{\mathcal{I}' \to \mathcal{I}}$ and $\widehat{\varphi}^{\text{KIN}}_{\mathcal{I}'' \to \mathcal{I}' \to \mathcal{I}}$ are Hilbert space isomorphisms.

Let $\mathcal{I} \subset \mathcal{I}' \subset \mathcal{I}'' \in \mathcal{L}$. We choose an orthonormal basis $(e_i)_{i \in I}$ of $\mathcal{I}$, an orthonormal basis $(e_i)_{i \in I' \setminus I}$ of $\mathcal{I}' \cap \mathcal{I}^\perp$ (with $I' \supset I$) and an orthonormal basis $(e_i)_{i \in I'' \setminus I'}$ of $\mathcal{I}'' \cap \mathcal{I}'^\perp$ (with $I'' \supset I'$). Since eq. (3.14.1) is valid for any choice of orthonormal bases, we have for $(n_i)_{i \in I''} \in \Lambda_{I''}$:

$$\left(\text{id}_{\widehat{\mathcal{I}'' \cap \mathcal{I}'^\perp}} \otimes \widehat{\varphi}_{\mathcal{I}' \to \mathcal{I}}\right) \circ \widehat{\varphi}_{\mathcal{I}'' \to \mathcal{I}'} \left|(n_i)_{i \in I''}, (e_i)_{i \in I''}\right\rangle =$$

$$= \left|(n_i)_{i \in I'' \setminus I'}, (e_i)_{i \in I'' \setminus I'}\right\rangle \otimes \left|(n_i)_{i \in I' \setminus I}, (e_i)_{i \in I' \setminus I}\right\rangle \otimes \left|(n_i)_{i \in I}, (e_i)_{i \in I}\right\rangle$$

$$= \left(\widehat{\varphi}_{\left(\mathcal{I}'' \cap \mathcal{I}^\perp\right) \to \left(\mathcal{I}' \cap \mathcal{I}^\perp\right)} \otimes \text{id}_{\widehat{\mathcal{I}}}\right) \circ \widehat{\varphi}_{\mathcal{I}'' \to \mathcal{I}} \left|(n_i)_{i \in I''}, (e_i)_{i \in I''}\right\rangle. \tag{3.15.1}$$

Hence, [11, eq. (2.1.1)] is fulfilled:

$$\left(\text{id}_{\widehat{\mathcal{M}}^{\text{KIN}}_{\mathcal{I}'' \to \mathcal{I}'}} \otimes \widehat{\varphi}^{\text{KIN}}_{\mathcal{I}' \to \mathcal{I}}\right) \circ \left(\widehat{\varphi}^{\text{KIN}}_{\mathcal{I}'' \to \mathcal{I}'}\right) = \left(\widehat{\varphi}^{\text{KIN}}_{\mathcal{I}'' \to \mathcal{I}' \to \mathcal{I}} \otimes \text{id}_{\widehat{\mathcal{M}}^{\text{KIN}}_{\mathcal{I}}}\right) \circ \widehat{\varphi}^{\text{KIN}}_{\mathcal{I}'' \to \mathcal{I}}.$$

□

**Proposition 3.16** We consider the objects introduced in props. 3.7 and 3.14. We define:

1. $\forall \mathcal{J} \in \mathcal{L}_H$, $\widehat{\mathcal{M}}^{\text{DYN}}_{\mathcal{J}} := \widehat{\mathcal{J}}$;
2. $\forall \mathcal{J} \subset \mathcal{J}' \in \mathcal{L}_H$, $\widehat{\mathcal{M}}^{\text{DYN}}_{\mathcal{J}' \to \mathcal{J}} := \widehat{\mathcal{J}^\perp \cap \mathcal{J}'}$ & $\widehat{\varphi}^{\text{DYN}}_{\mathcal{J}' \to \mathcal{J}} := \widehat{\varphi}_{\mathcal{J}' \to \mathcal{J}}$;
3. $\forall \mathcal{J} \subset \mathcal{J}' \subset \mathcal{J}'' \in \mathcal{L}_H$, $\widehat{\varphi}^{\text{DYN}}_{\mathcal{J}'' \to \mathcal{J}' \to \mathcal{J}} := \widehat{\varphi}_{\left(\mathcal{J}^\perp \cap \mathcal{J}''\right) \to \left(\mathcal{J}^\perp \cap \mathcal{J}'\right)}$.

$\left(\mathcal{L}_H, \widehat{\mathcal{M}}^{\text{DYN}}, \widehat{\varphi}^{\text{DYN}}\right)^{\otimes}$ is a projective system of quantum state spaces.

Let $A$ be a bounded self-adjoint operator on $\mathcal{H}$ and suppose that $(\text{Ker} A)^\perp \in \mathcal{L}_H$. For $\mathcal{J} \in \mathcal{L}_H$ such that $(\text{Ker} A)^\perp \subset \mathcal{J}$, we define $\widehat{A}_{\mathcal{J}} := \widehat{\left(A|_{\mathcal{J} \to \mathcal{J}}\right)}$ (def. 3.12). For $\mathcal{J}, \mathcal{J}' \in \mathcal{L}_H$ such that $(\text{Ker} A)^\perp \subset \mathcal{J}, \mathcal{J}'$, we have:

$\widehat{A}_{\mathcal{J}} \sim \widehat{A}_{\mathcal{J}'}$ (with the equivalence relation $\sim$ defined in [11, eq. (2.3.2)]),

hence, we can define $\widehat{A}_{\mathcal{L}_H} := \left[\widehat{A}_{\mathcal{J}}\right]_\sim \in \mathcal{O}^{\otimes}_{(\mathcal{L}_H, \widehat{\mathcal{M}}^{\text{DYN}}, \widehat{\varphi}^{\text{DYN}})}$ [11, prop. 2.5].

**Proof** We know from prop. 3.7 that $\mathcal{L}_H$ is a directed set. Then, we can show that $\left(\mathcal{L}_H, \widehat{\mathcal{M}}^{\text{DYN}}, \widehat{\varphi}^{\text{DYN}}\right)^{\otimes}$ is a projective system of quantum state spaces exactly like in the proof of prop. 3.15.

Let $\mathcal{J} \subset \mathcal{J}'' \in \mathcal{L}_H$, $(e_i)_{i\in J}$ be an orthonormal basis of $\mathcal{J}$ and $(e_i)_{i\in J''\setminus J}$ (with $J'' \supset J$) an orthonormal basis of $\mathcal{J}'' \cap \mathcal{J}^\perp$. For $k, l \in J$ and $(n_i)_{i\in J''} \in \Lambda_{J''}$, we have:

$$\widehat{\varphi}^{-1}_{\mathcal{J}''\to\mathcal{J}} \left(\text{id}_{\widehat{\mathcal{J}''\cap\mathcal{J}^\perp}} \otimes \widehat{a}^{\mathcal{J},+}_{e_k}\, \widehat{a}^{\mathcal{J}}_{e_l}\right) \widehat{\varphi}_{\mathcal{J}''\to\mathcal{J}} \left|(n_i)_{i\in J''}, (e_i)_{i\in J''}\right\rangle =$$

$$= \sqrt{n_l}\sqrt{n_k + 1 - \delta_{lk}}\, \left|(n_i - \delta_{li} + \delta_{ki})_{i\in J''}, (e_i)_{i\in J''}\right\rangle$$

$$= \widehat{a}^{\mathcal{J}'',+}_{e_k}\, \widehat{a}^{\mathcal{J}''}_{e_l} \left|(n_i)_{i\in J''}, (e_i)_{i\in J''}\right\rangle .$$

Now, if $(\text{Ker}A)^\perp \subset \mathcal{J}$, we have $\mathcal{J}'' \cap \mathcal{J}^\perp \subset \mathcal{J}^\perp \subset \text{Ker}A$, therefore:

$$\left(\widehat{A|_{\mathcal{J}''\to\mathcal{J}''}}\right) = \sum_{k,l\in J''} \langle e_k, A\, e_l\rangle\, \widehat{a}^{\mathcal{J}'',+}_{e_k}\, \widehat{a}^{\mathcal{J}''}_{e_l}$$

$$= \sum_{k,l\in J} \langle e_k, A\, e_l\rangle\, \widehat{a}^{\mathcal{J}'',+}_{e_k}\, \widehat{a}^{\mathcal{J}''}_{e_l}$$

$$= \widehat{\varphi}^{-1}_{\mathcal{J}''\to\mathcal{J}} \left[\text{id}_{\widehat{\mathcal{J}''\cap\mathcal{J}^\perp}} \otimes \left(\sum_{k,l\in J} \langle e_k, A\, e_l\rangle\, \widehat{a}^{\mathcal{J},+}_{e_k}\, \widehat{a}^{\mathcal{J}}_{e_l}\right)\right] \widehat{\varphi}_{\mathcal{J}''\to\mathcal{J}}$$

$$= \widehat{\varphi}^{-1}_{\mathcal{J}''\to\mathcal{J}} \left[\text{id}_{\widehat{\mathcal{J}''\cap\mathcal{J}^\perp}} \otimes \left(\widehat{A|_{\mathcal{J}\to\mathcal{J}}}\right)\right] \widehat{\varphi}_{\mathcal{J}''\to\mathcal{J}} . \tag{3.16.1}$$

Hence, $\widehat{A}_{\mathcal{J}''} = \widehat{\varphi}^{\text{DYN},-1}_{\mathcal{J}''\to\mathcal{J}} \left[\text{id}_{\widehat{\mathcal{M}}^{\text{DYN}}_{\mathcal{J}''\to\mathcal{J}}} \otimes \widehat{A}_{\mathcal{J}}\right] \widehat{\varphi}^{\text{DYN}}_{\mathcal{J}''\to\mathcal{J}}$ .

Finally, if $\mathcal{J}, \mathcal{J}' \in \mathcal{L}_H$ are such that $(\text{Ker}A)^\perp \subset \mathcal{J}, \mathcal{J}'$, we can find $\mathcal{J}'' \in \mathcal{L}_H$ such that $\mathcal{J}, \mathcal{J}' \subset \mathcal{J}''$ (because $\mathcal{L}_H$ is directed), so $\widehat{A}_{\mathcal{J}} \sim \widehat{A}_{\mathcal{J}'}$ . □

Using the general result derived in [11, theorem 2.9], we are able to embed the space of density matrices on the Fock space into the larger quantum state space constructed by projective techniques, and to precisely characterize the image of this embedding, by giving a condition for a projective state to be representable as a density matrix on $\widehat{\mathcal{H}}$.

**Proposition 3.17** We consider the same objects as in prop. 3.16. There exists an injective map $\widehat{\sigma}_\downarrow : \overline{\mathcal{S}}_{\text{Fock}} \to \overline{\mathcal{S}}^{\otimes}_{(\mathcal{L}_H, \widehat{\mathcal{M}}^{\text{DYN}}, \widehat{\varphi}^{\text{DYN}})}$ (where $\overline{\mathcal{S}}_{\text{Fock}}$ is the space of (self-adjoint) positive semi-definite, traceclass operators over $\widehat{\mathcal{M}}^{\text{DYN}}_{\text{Fock}}$ and $\overline{\mathcal{S}}^{\otimes}_{(\mathcal{L}_H, \widehat{\mathcal{M}}^{\text{DYN}}, \widehat{\varphi}^{\text{DYN}})}$ was defined in [11, def. 2.1]) satisfying, for any bounded self-adjoint operator $A$ on $\mathcal{H}$ with $(\text{Ker}A)^\perp \in \mathcal{L}_H$, and any $\rho \in \overline{\mathcal{S}}_{\text{Fock}}$:

$$\text{Tr}_{\widehat{\mathcal{M}}^{\text{DYN}}_{\text{Fock}}} \left[\rho\ \mathbb{I}_W\left(\widehat{\langle A\rangle}_{\text{Fock}}\right)\right] = \text{Tr}\left[\widehat{\sigma}_\downarrow(\rho)\ \mathbb{I}_W\left(\widehat{A}_{\mathcal{L}_H}\right)\right], \tag{3.17.1}$$

where $W$ is a measurable subset in the spectrum of $\widehat{\langle A \rangle}_{\text{Fock}}$, and $\mathbb{I}_W(\,\cdot\,)$ denotes the corresponding spectral projectors.

Moreover, we have:

$$\widehat{\sigma}_\downarrow \langle \mathcal{S}_{\text{Fock}} \rangle = \left\{ (\rho_{\mathcal{J}})_{\mathcal{J}\in\mathcal{L}_H} \;\middle|\; \sup_{\mathcal{J}\in\mathcal{L}_H} \inf_{\mathcal{J}'\supset\mathcal{J}} \operatorname{Tr}_{\widehat{\mathcal{M}}^{\text{DYN}}_{\mathcal{J}'}} \left( \rho_{\mathcal{J}'}\, \widehat{\Pi}_{\mathcal{J}'|\mathcal{J}} \right) = 1 \right\},$$

where $\mathcal{S}_{\text{Fock}}$ is the space of density matrices over $\widehat{\mathcal{M}}^{\text{DYN}}_{\text{Fock}}$ and:

$$\forall N \in \mathbb{N}, \forall \psi \in \mathcal{J}'^{\otimes N,\text{sym}},\; \widehat{\Pi}_{\mathcal{J}'|\mathcal{J}}\, \psi := (\Pi_{\mathcal{J}})^{\otimes N}\, \psi \in \mathcal{J}'^{\otimes N,\text{sym}},$$

$\Pi_{\mathcal{J}}$ being the orthogonal projection on $\mathcal{J}$.

**Proof** For $\mathcal{J} \subset \mathcal{J}' \in \mathcal{L}_H$, we define $\zeta_{\mathcal{J}'\to\mathcal{J}} \in \widehat{\mathcal{M}}^{\text{DYN}}_{\mathcal{J}'\to\mathcal{J}}$ as the vacuum state of $\widehat{\mathcal{M}}^{\text{DYN}}_{\mathcal{J}'\to\mathcal{J}} = \widehat{\mathcal{J}' \cap \mathcal{J}^\perp}$ (ie. $\zeta_{\mathcal{J}'\to\mathcal{J}} = |(O)_{i\in I}\,,\,(e_i)_{i\in I}\rangle$ for any basis $(e_i)_{i\in I}$ of $\mathcal{J}' \cap \mathcal{J}^\perp$). The family of vectors $(\zeta_{\mathcal{J}'\to\mathcal{J}})_{\mathcal{J}\subset\mathcal{J}'}$ fulfills the hypotheses of [11, theorem 2.9].

Next, for all $\mathcal{J} \in \mathcal{L}_H$, we define an injection $\tau_{\text{Fock}\leftarrow\mathcal{J}}$ from $\widehat{\mathcal{M}}^{\text{DYN}}_{\mathcal{J}} = \widehat{\mathcal{J}}$ into $\widehat{\mathcal{M}}^{\text{DYN}}_{\text{Fock}} = \widehat{\mathcal{H}}$ by:

$$\tau_{\text{Fock}\leftarrow\mathcal{J}} = \widehat{\varphi}^{-1}_{\mathcal{H}\to\mathcal{J}} \left(\zeta_{\text{Fock}\to\mathcal{J}} \otimes (\,\cdot\,)\right),$$

where $\zeta_{\text{Fock}\to\mathcal{J}}$ is the vacuum state of $\widehat{\mathcal{J}^\perp}$. Using eq. (3.15.1), we can show that $\forall \mathcal{J} \subset \mathcal{J}' \in \mathcal{L}_H$, $\tau_{\text{Fock}\leftarrow\mathcal{J}'} \circ \tau_{\mathcal{J}'\leftarrow\mathcal{J}} = \tau_{\text{Fock}\leftarrow\mathcal{J}}$ (where $\tau_{\mathcal{J}'\leftarrow\mathcal{J}}$ is defined from $\zeta_{\mathcal{J}'\to\mathcal{J}}$ as in [11, theorem 2.9]).

Now, we can choose an orthonormal basis $(e_i)_{i\in\mathbb{N}}$ of $\mathcal{H}$ such that $\forall i \in \mathbb{N}$, $\|H\, e_i\| < \infty$ and consider for $N \geqslant 1$, $\mathcal{J}_N := \operatorname{Vect}\{e_i \mid i \leqslant N\} \in \mathcal{L}_H$. Using eq. (3.14.1) with the orthonormal basis $(e_i)_{i\leqslant N}$ of $\mathcal{J}_N$ and the orthonormal basis $(e_i)_{i>N}$ of $\mathcal{J}_N^\perp$, we get:

$$\tau_{\text{Fock}\leftarrow\mathcal{J}_N} \left\langle \widehat{\mathcal{J}_N} \right\rangle = \overline{\operatorname{Vect}\left\{ |(n_i)_{i\in\mathbb{N}}\,,\,(e_i)_{i\in\mathbb{N}}\rangle \;\middle|\; (n_i)_{i\in\mathbb{N}} \in \Lambda^N_{\mathbb{N}} \right\}},$$

where $\Lambda^N_{\mathbb{N}} := \{(n_i)_{i\in\mathbb{N}} \in \Lambda_{\mathbb{N}} \mid \forall i > N,\, n_i = 0\}$. Hence, from $\Lambda_{\mathbb{N}} = \bigcup_{N\geqslant 1} \Lambda^N_{\mathbb{N}}$, we have:

$$\widehat{\mathcal{M}}^{\text{DYN}}_{\text{Fock}} = \overline{\bigcup_{\mathcal{J}\in\mathcal{L}_H} \operatorname{Im}\tau_{\text{Fock}\leftarrow\mathcal{J}}}.$$

Therefore, we can identify $\widehat{\mathcal{M}}^{\text{DYN}}_{\text{Fock}}$ with the inductive limit $\widehat{\mathcal{M}}^{\text{DYN}}_{\zeta}$ introduced in [11, theorem 2.9], so we have an injection $\widehat{\sigma}_\downarrow : \overline{\mathcal{S}}_{\text{Fock}} \to \overline{\mathcal{S}}^{\otimes}_{(\mathcal{L}_H, \widehat{\mathcal{H}}^{\text{DYN}}, \widehat{\varphi}^{\text{DYN}})}$, satisfying:

$$\widehat{\sigma}_\downarrow \langle \mathcal{S}_{\text{Fock}} \rangle = \left\{ (\rho_{\mathcal{J}})_{\mathcal{J}\in\mathcal{L}_H} \;\middle|\; \sup_{\mathcal{J}\in\mathcal{L}_H} \inf_{\mathcal{J}'\supset\mathcal{J}} \operatorname{Tr}_{\widehat{\mathcal{M}}^{\text{DYN}}_{\mathcal{J}'}} \left( \rho_{\mathcal{J}'}\, \widehat{\Pi}_{\mathcal{J}'|\mathcal{J}} \right) = 1 \right\},$$

where:

$$\forall \mathcal{J} \subset \mathcal{J}' \in \mathcal{L}_H,\; \widehat{\Pi}_{\mathcal{J}'|\mathcal{J}} = \widehat{\varphi}^{-1}_{\mathcal{J}'\to\mathcal{J}} \circ \left( |\zeta_{\mathcal{J}'\to\mathcal{J}}\,\rangle\langle\, \zeta_{\mathcal{J}'\to\mathcal{J}}| \otimes \mathrm{id}_{\widehat{\mathcal{J}}} \right) \circ \widehat{\varphi}_{\mathcal{J}'\to\mathcal{J}}.$$

Let $\mathcal{J} \subset \mathcal{J}' \in \mathcal{L}_H$ and let $(e_i)_{i\in J}$, resp. $(e_i)_{i\in J'\setminus J}$ be an orthonormal basis of $\mathcal{J}$, resp. $\mathcal{J}' \cap \mathcal{J}^\perp$. For $(n_i)_{i\in}\, J' \in \Lambda_{J'}$, we have:

$$\widehat{\Pi}_{\mathcal{J}'|\mathcal{J}}\, |(n_i)_{i\in J'}\,,\,(e_i)_{i\in J'}\rangle = \begin{cases} |(n_i)_{i\in J'}\,,\,(e_i)_{i\in J'}\rangle & \text{if } \forall i \in J' \setminus J,\; n_i = 0 \\ 0 & \text{otherwise} \end{cases}$$

$$= \Pi_{\mathcal{J}}^{\otimes \sum_{i\in J'} n_i} \left|(n_i)_{i\in J'}, (e_i)_{i\in J'}\right\rangle ,$$

therefore $\forall N \in \mathbb{N}$, $\forall \psi \in \mathcal{J}'^{\otimes N,\text{sym}}$, $\widehat{\Pi}_{\mathcal{J}'|\mathcal{J}}\, \psi = (\Pi_{\mathcal{J}})^{\otimes N}\, \psi$.

Lastly, let $A$ be a bounded self-adjoint operator on $\mathcal{H}$ such that $\mathcal{J} := (\text{Ker}A)^{\perp} \in \mathcal{L}_H$. Eq. (3.17.1) is then an application of [11, prop. 2.5], using the definition of $\widehat{\sigma}_{\downarrow}$ (given in the proof of [11, theorem 2.9]) together with:

$$\widehat{\langle A\rangle}_{\text{Fock}} := \widehat{A} = \widehat{\varphi}^{-1}_{\mathcal{H}\to\mathcal{J}} \left[\text{id}_{\widehat{\mathcal{J}^{\perp}}} \otimes \widehat{A}_{\mathcal{J}}\right] \widehat{\varphi}_{\mathcal{H}\to\mathcal{J}} ,$$

which can be shown like in the proof of prop. 3.16 (eq. (3.16.1) ). □

We can now implement and solve in the quantum theory the approximated constraints we had on the classical side, and thus define a family of maps (indexed by the regularization parameter $\varepsilon$) from the dynamical projective system of quantum state spaces introduced above into the kinematical one.

**Proposition 3.18** We consider the objects introduced in def. 3.4 and prop. 3.15. Let $\varepsilon = (\mathcal{J}, \epsilon) \in \mathcal{E}$ and let $\mathfrak{I} \in \mathcal{L}^{\varepsilon}$. We define the map:

$$\begin{array}{rcl} \widehat{\delta}^{\varepsilon}_{\mathfrak{I}} : \widehat{\mathcal{J}} & \to & \widehat{\mathfrak{I}} \otimes \mathfrak{T} \\ \psi & \mapsto & \left(\widehat{\varphi}^{-1}_{\mathfrak{I}\to\mathcal{J}} \otimes \text{id}_{\mathfrak{T}}\right) \left[\zeta_{\mathfrak{I}\to\mathcal{J}} \otimes \exp\left(-i\left(\left(\widehat{\Pi_{\mathcal{J}} H \Pi_{\mathcal{J}})|_{\mathcal{J}\to\mathcal{J}}}\right) \otimes \widehat{T}\right) (\psi \otimes \delta_{\epsilon})\right] \end{array},$$

where $\zeta_{\mathfrak{I}\to\mathcal{J}}$ is the vacuum state in $\widehat{\mathfrak{I} \cap \mathcal{J}^{\perp}}$, $\Pi_{\mathcal{J}}$ is the orthogonal projection on $\mathcal{J}$, $\widehat{T}$ is the position operator on $\mathfrak{T} = L_2(\mathcal{R}, d\mu)$ and $\delta_{\epsilon} \in \mathfrak{T}$ is defined by:

$$\forall t \in \mathbb{R},\ \delta_{\epsilon}(t) = \frac{\sqrt{\epsilon}}{\pi^{1/4}} \exp\left(-\frac{\epsilon^2\, t^2}{2}\right).$$

Then, we have:

$$\widehat{\delta}^{\varepsilon}_{\mathfrak{I}} \left\langle \widehat{\mathcal{J}} \right\rangle = \left\{\psi \in \mathfrak{I} \,\middle|\, \left(\widehat{\Pi}_{\mathfrak{I}|\mathcal{J}} \otimes \text{id}_{\mathfrak{T}}\right) \psi = \psi \quad \& \quad \widehat{C}^{\varepsilon}\, \psi = \psi\right\},$$

$$\text{with } \widehat{C}^{\varepsilon} = \frac{1}{\epsilon^2} \left(\text{id}_{\widehat{\mathfrak{I}}} \otimes \widehat{E} - \left(\left(\widehat{\Pi_{\mathcal{J}} H \Pi_{\mathcal{J}})|_{\mathfrak{I}\to\mathcal{J}}}\right) \otimes \text{id}_{\mathfrak{T}}\right)^2 + \epsilon^2\, \text{id}_{\widehat{\mathfrak{I}}} \otimes \widehat{T}^2 ,$$

where $\widehat{\Pi}_{\mathfrak{I}|\mathcal{J}}$ is defined as in prop. 3.17 and $\widehat{E}$ is the operator $i\,\partial_t$ on $\mathfrak{T}$. Moreover, $\widehat{\delta}^{\varepsilon}_{\mathfrak{I}}\Big|_{\widehat{\mathcal{J}}\to\widehat{\delta}^{\varepsilon}_{\mathfrak{I}}\langle\widehat{\mathcal{J}}\rangle}$ is a unitary map.

**Proof** We define:

$$\begin{array}{rcl} \widehat{\delta}_1 : \widehat{\mathcal{J}} & \to & \widehat{\mathcal{J}} \otimes \mathfrak{T} \\ \psi & \mapsto & \psi \otimes \delta_{\epsilon} \end{array}, \qquad \begin{array}{rcl} \widehat{\delta}_2 : \widehat{\mathcal{J}} \otimes \mathfrak{T} & \to & \widehat{\mathcal{J}} \otimes \mathfrak{T} \\ \psi & \mapsto & \exp\left(-i\widehat{H^{\varepsilon}_{\mathcal{J}}} \otimes \widehat{T}\right) \psi \end{array} \quad \text{with } H^{\varepsilon} := \Pi_{\mathcal{J}}\, H\, \Pi_{\mathcal{J}} ,$$

$$\begin{array}{rcl} \widehat{\delta}_3 : \widehat{\mathcal{J}} \otimes \mathfrak{T} & \to & \widehat{\mathfrak{I} \cap \mathcal{J}^{\perp}} \otimes \widehat{\mathcal{J}} \otimes \mathfrak{T} \\ \psi & \mapsto & \zeta_{\mathfrak{I}\to\mathcal{J}} \otimes \psi \end{array}, \qquad \begin{array}{rcl} \widehat{\delta}_4 : \widehat{\mathfrak{I} \cap \mathcal{J}^{\perp}} \otimes \widehat{\mathcal{J}} \otimes \mathfrak{T} & \to & \widehat{\mathfrak{I}} \otimes \mathfrak{T} \\ \psi & \mapsto & \left(\widehat{\varphi}^{-1}_{\mathfrak{I}\to\mathcal{J}} \otimes \text{id}_{\mathfrak{T}}\right) \psi \end{array}.$$

We have $\widehat{\delta}_1 \left\langle \widehat{\mathcal{J}} \right\rangle = \widehat{\mathcal{J}} \otimes \text{Vect}\left\{\delta_\epsilon\right\} = \widehat{\mathcal{J}} \otimes \left\{\psi \in \mathcal{T} \,\middle|\, \widehat{C}_1\,\psi = \psi\right\}$, where $\widehat{C}_1 := \frac{1}{\epsilon^2}\,\widehat{E}^2 + \epsilon^2\,\widehat{T}^2$, and $\widehat{\delta}_1\Big|_{\widehat{\mathcal{J}}\to\widehat{\delta}_1\left\langle\widehat{\mathcal{J}}\right\rangle}$ is a unitary map.

$\widehat{\delta}_2$ is a unitary map, because $\widehat{H}^\varepsilon_{\mathcal{J}}$ and $\widehat{T}$ are essentially self-adjoint ($\forall N \in \mathbb{N}$, $\widehat{H}^\varepsilon_{\mathcal{J}}$ stabilize $\mathcal{H}^{\otimes N,\text{sym}}$ and the restriction of $\widehat{H}^\varepsilon_{\mathcal{J}}$ to $\mathcal{H}^{\otimes N,\text{sym}}$ is a bounded self-adjoint operator, for so is $H^\varepsilon|_{\mathcal{J}\to\mathcal{J}}$, by definition of $\mathcal{L}_H$). And we have:

$$\widehat{\delta}_2 \circ \widehat{\delta}_1 \left\langle \widehat{\mathcal{J}} \right\rangle = \left\{\psi \in \widehat{\mathcal{J}} \otimes \mathcal{T} \,\middle|\, \widehat{C}_2\,\psi = \psi\right\},$$

with:

$$\begin{aligned}
\widehat{C}_2 &:= \widehat{\delta}_2 \circ \left(\text{id}_{\widehat{\mathcal{J}}} \otimes \widehat{C}_1\right) \circ \widehat{\delta}_2^{-1} \\
&= \exp\left(-i\left[\widehat{H}^\varepsilon_{\mathcal{J}} \otimes \widehat{T},\,\cdot\,\right]\right)\,\left(\text{id}_{\widehat{\mathcal{J}}} \otimes \widehat{C}_1\right) \\
&= \frac{1}{\epsilon^2}\,\left(\text{id}_{\widehat{\mathcal{J}}} \otimes \widehat{E} - \widehat{H}^\varepsilon_{\mathcal{J}} \otimes \text{id}_{\mathcal{T}}\right)^2 + \epsilon^2\,\text{id}_{\widehat{\mathcal{J}}} \otimes \widehat{T}^2 .
\end{aligned}$$

Next, we compute:

$$\begin{aligned}
\widehat{\delta}_3 \circ \widehat{\delta}_2 \circ \widehat{\delta}_1 \left\langle \widehat{\mathcal{J}} \right\rangle &= \left\{\zeta_{\mathfrak{I}\to\mathcal{J}} \otimes \psi \in \widehat{\mathfrak{I}\cap\mathcal{J}^\perp} \otimes \widehat{\mathcal{J}} \otimes \mathcal{T} \,\middle|\, \widehat{C}_2\,\psi = \psi\right\} \\
&= \left\{\psi \in \widehat{\mathfrak{I}\cap\mathcal{J}^\perp} \otimes \widehat{\mathcal{J}} \otimes \mathcal{T} \,\middle|\, \text{id}_{\widehat{\mathfrak{I}\cap\mathcal{J}^\perp}} \otimes \widehat{C}_2\,\psi = \psi \quad \& \quad \left(\left|\zeta_{\mathfrak{I}\to\mathcal{J}}\right\rangle\!\left\langle \zeta_{\mathfrak{I}\to\mathcal{J}}\right| \otimes \text{id}_{\widehat{\mathcal{J}}\otimes\mathcal{T}}\right)\,\psi = \psi\right\},
\end{aligned}$$

and $\widehat{\delta}_3\Big|_{\widehat{\mathcal{J}}\otimes\mathcal{T}\to\widehat{\delta}_3\left\langle\widehat{\mathcal{J}}\otimes\mathcal{T}\right\rangle}$ is a unitary map.

Finally, $\widehat{\delta}_4$ is unitary (from prop. 3.14) and:

$$\widehat{\delta}^\varepsilon_{\mathfrak{I}} \left\langle \widehat{\mathcal{J}} \right\rangle = \widehat{\delta}_4 \circ \widehat{\delta}_3 \circ \widehat{\delta}_2 \circ \widehat{\delta}_1 \left\langle \widehat{\mathcal{J}} \right\rangle = \left\{\psi \in \widehat{\mathfrak{I}} \otimes \mathcal{T} \,\middle|\, \widehat{C}_4\,\psi = \psi \quad \& \quad \widehat{D}_4\,\psi = \psi\right\},$$

with:

$$\begin{aligned}
\widehat{C}_4 &:= \left(\widehat{\varphi}^{-1}_{\mathfrak{I}\to\mathcal{J}} \otimes \text{id}_{\mathcal{T}}\right)\,\left(\text{id}_{\widehat{\mathfrak{I}\cap\mathcal{J}^\perp}} \otimes \widehat{C}_2\right)\,\left(\widehat{\varphi}_{\mathfrak{I}\to\mathcal{J}} \otimes \text{id}_{\mathcal{T}}\right) \\
&= \frac{1}{\epsilon^2}\,\left(\text{id}_{\widehat{\mathfrak{I}}} \otimes \widehat{E} - \left[\widehat{\varphi}^{-1}_{\mathfrak{I}\to\mathcal{J}}\,\left(\text{id}_{\widehat{\mathfrak{I}\cap\mathcal{J}^\perp}} \otimes \widehat{H}^\varepsilon_{\mathcal{J}}\right)\,\widehat{\varphi}_{\mathfrak{I}\to\mathcal{J}}\right] \otimes \text{id}_{\mathcal{T}}\right)^2 + \epsilon^2\,\text{id}_{\widehat{\mathfrak{I}}} \otimes \widehat{T}^2 \\
&= \frac{1}{\epsilon^2}\,\left(\text{id}_{\widehat{\mathfrak{I}}} \otimes \widehat{E} - \widehat{H}^\varepsilon_{\mathfrak{I}} \otimes \text{id}_{\mathcal{T}}\right)^2 + \epsilon^2\,\text{id}_{\widehat{\mathfrak{I}}} \otimes \widehat{T}^2 \text{ (using eq. (3.16.1))},
\end{aligned}$$

and:

$$\begin{aligned}
\widehat{D}_4 &:= \left(\widehat{\varphi}^{-1}_{\mathfrak{I}\to\mathcal{J}} \otimes \text{id}_{\mathcal{T}}\right)\,\left(\left|\zeta_{\mathfrak{I}\to\mathcal{J}}\right\rangle\!\left\langle \zeta_{\mathfrak{I}\to\mathcal{J}}\right| \otimes \text{id}_{\widehat{\mathcal{J}}\otimes\mathcal{T}}\right)\,\left(\widehat{\varphi}_{\mathfrak{I}\to\mathcal{J}} \otimes \text{id}_{\mathcal{T}}\right) \\
&= \widehat{\sqcap}_{\mathfrak{I}|\mathcal{J}} \otimes \text{id}_{\mathcal{T}} \text{ (as was shown in the proof of prop. 3.17).}
\end{aligned}$$

□

**Proposition 3.19** We consider the same objects as in prop. 3.18. For $\varepsilon = (\mathcal{J}, \epsilon) \in \mathcal{E}$ and $\rho_{\mathcal{J}}$ a

(self-adjoint) positive semi-definite, traceclass operator on $\widehat{\mathcal{J}}$, we define:

$$\forall \mathcal{I} \in \mathcal{L}^\varepsilon,\ \widehat{\Delta}^\varepsilon_{\mathcal{I}}(\rho_{\mathcal{J}}) := \widehat{\delta}^\varepsilon_{\mathcal{I}}\, \rho_{\mathcal{J}} \left(\widehat{\delta}^\varepsilon_{\mathcal{I}}\right)^+ .$$

Then, $\left(\widehat{\Delta}^\varepsilon_{\mathcal{I}}(\rho_{\mathcal{J}})\right)_{\mathcal{I}\in\mathcal{L}^\varepsilon} \in \overline{\mathcal{S}}^{\otimes}_{(\mathcal{L}^\varepsilon, \widehat{\mathcal{M}}^{\text{KIN}}, \widehat{\varphi}^{\text{KIN}})}$.

Hence, for $\rho = (\rho_{\mathcal{J}})_{\mathcal{J}\in\mathcal{L}_H} \in \overline{\mathcal{S}}^{\otimes}_{(\mathcal{L}_H, \widehat{\mathcal{M}}^{\text{DYN}}, \widehat{\varphi}^{\text{DYN}})}$ (prop. 3.16), we can define:

$$\widehat{\Delta}^\varepsilon(\rho) = \widehat{\sigma}^{-1}\left(\left(\widehat{\Delta}^\varepsilon_{\mathcal{I}}(\rho_{\mathcal{J}})\right)_{\mathcal{I}\in\mathcal{L}^\varepsilon}\right),$$

where the map $\widehat{\sigma} : \overline{\mathcal{S}}^{\otimes}_{(\mathcal{L}, \widehat{\mathcal{M}}^{\text{KIN}}, \widehat{\varphi}^{\text{KIN}})} \to \overline{\mathcal{S}}^{\otimes}_{(\mathcal{L}^\varepsilon, \widehat{\mathcal{M}}^{\text{KIN}}, \widehat{\varphi}^{\text{KIN}})}$ is defined as in [11, prop. 2.6] (and is bijective, since $\mathcal{L}^\varepsilon$ is cofinal in $\mathcal{L}$).

**Proof** We need to prove that $\forall \mathcal{I}, \mathcal{I}' \in \mathcal{L}^\varepsilon$, with $\mathcal{I} \subset \mathcal{I}'$, $\text{Tr}_{\mathcal{I}'\to\mathcal{I}} \widehat{\Delta}^\varepsilon_{\mathcal{I}'}(\rho_{\mathcal{J}}) = \widehat{\Delta}^\varepsilon_{\mathcal{I}}(\rho_{\mathcal{J}})$. We have:

$$\forall \psi \in \widehat{\mathcal{J}},\ \widehat{\varphi}^{\text{KIN}}_{\mathcal{I}'\to\mathcal{I}} \circ \widehat{\delta}^\varepsilon_{\mathcal{I}'}(\psi) = \left(\left[\widehat{\varphi}_{\mathcal{I}'\to\mathcal{I}} \circ \widehat{\varphi}^{-1}_{\mathcal{I}'\to\mathcal{J}}\right] \otimes \text{id}_{\mathcal{T}}\right) \left[\zeta_{\mathcal{I}'\to\mathcal{J}} \otimes e^{-i\,\widehat{H}^\varepsilon_{\mathcal{J}}\otimes\widehat{T}}\,(\psi\otimes\delta_\epsilon)\right]$$

$$= \left(\left[\left(\text{id}_{\widehat{\mathcal{I}'\cap\mathcal{I}^\perp}} \otimes \widehat{\varphi}^{-1}_{\mathcal{I}\to\mathcal{J}}\right) \circ \left(\widehat{\varphi}_{\mathcal{I}'\to\mathcal{I}\to\mathcal{J}} \otimes \text{id}_{\widehat{\mathcal{J}}}\right)\right] \otimes \text{id}_{\mathcal{T}}\right) \left[\zeta_{\mathcal{I}'\to\mathcal{J}} \otimes e^{-i\,\widehat{H}^\varepsilon_{\mathcal{J}}\otimes\widehat{T}}\,(\psi\otimes\delta_\epsilon)\right]$$

$$= \left(\left(\text{id}_{\widehat{\mathcal{I}'\cap\mathcal{I}^\perp}} \otimes \widehat{\varphi}^{-1}_{\mathcal{I}\to\mathcal{J}}\right) \otimes \text{id}_{\mathcal{T}}\right) \left[\zeta_{\mathcal{I}'\to\mathcal{I}} \otimes \zeta_{\mathcal{I}\to\mathcal{J}} \otimes e^{-i\,\widehat{H}^\varepsilon_{\mathcal{J}}\otimes\widehat{T}}\,(\psi\otimes\delta_\epsilon)\right]$$

$$= \zeta_{\mathcal{I}'\to\mathcal{I}} \otimes \widehat{\delta}^\varepsilon_{\mathcal{I}}(\psi),$$

hence $\widehat{\varphi}^{\text{KIN}}_{\mathcal{I}'\to\mathcal{I}} \circ \widehat{\Delta}^\varepsilon_{\mathcal{I}'}(\rho_{\mathcal{J}}) \circ \widehat{\varphi}^{\text{KIN},-1}_{\mathcal{I}'\to\mathcal{I}} = |\zeta_{\mathcal{I}'\to\mathcal{I}} \rangle\langle \zeta_{\mathcal{I}'\to\mathcal{I}}| \otimes \widehat{\Delta}^\varepsilon_{\mathcal{I}}(\rho_{\mathcal{J}})$, therefore:

$$\text{Tr}_{\mathcal{I}'\to\mathcal{I}} \widehat{\Delta}^\varepsilon_{\mathcal{I}'}(\rho_{\mathcal{J}}) = \text{Tr}_{\widehat{\mathcal{I}'\cap\mathcal{I}^\perp}}\, |\zeta_{\mathcal{I}'\to\mathcal{I}} \rangle\langle \zeta_{\mathcal{I}'\to\mathcal{I}}| \otimes \widehat{\Delta}^\varepsilon_{\mathcal{I}}(\rho_{\mathcal{J}}) = \widehat{\Delta}^\varepsilon_{\mathcal{I}}(\rho_{\mathcal{J}}).$$

□

As a preparation for the study of convergence, we define a subset $\widehat{\mathcal{R}}$ of the space of states over the quantum projective structure. The motivation is to implement a quantum version of the regularity condition that was ensuring convergence on the classical side: at the classical level we have proved the convergence for normalized states, so in analogy we consider here states with a bounded expectation value for the total number of particles (which indeed corresponds to the quantization of the classical observable $\psi \mapsto \langle \psi, \psi \rangle$).

Note that, as we show in the following result, the regular states (the elements of $\widehat{\mathcal{R}}$) can be seen as states in the Fock space via the embedding of prop. 3.17. This is not really surprising, since we know that the Fock space quantization is appropriate for a basic non-interacting field theory like the Schrödinger equation.

**Proposition 3.20** We consider the same objects as in prop. 3.17 and we define:

$$\widehat{\mathcal{R}} := \left\{\rho \in \overline{\mathcal{S}}^{\otimes}_{(\mathcal{L}_H, \widehat{\mathcal{M}}^{\text{DYN}}, \widehat{\varphi}^{\text{DYN}})} \,\middle|\, \sup_{\mathcal{J}\in\mathcal{L}_H} \text{Tr}\left(\rho\,(\widehat{\sqcap_{\mathcal{J}}})_{\mathcal{L}_H}\right) < \infty\right\},$$

where $\mathrm{Tr}\left(\rho\,\widehat{(\sqcap_{\mathcal{J}})}_{\mathcal{L}_H}\right) := \sum_{n\in\mathbb{N}} n\,\mathrm{Tr}\left(\rho\;\; \mathbb{I}_{\{n\}}\left(\widehat{(\sqcap_{\mathcal{J}})}_{\mathcal{L}_H}\right)\right)$ and $\mathbb{I}_{\{n\}}\left(\widehat{(\sqcap_{\mathcal{J}})}_{\mathcal{L}_H}\right)$ denotes the spectral projector as in [11, prop. 2.5].

Then, $\widehat{\mathcal{R}} \subset \widehat{\sigma}_{\downarrow}\left\langle \overline{\mathcal{S}}_{\mathrm{Fock}}\right\rangle$.

**Proof** Let $\rho \in \widehat{\mathcal{R}}$ and $N = \sup_{\mathcal{J}\in\mathcal{L}_H} \mathrm{Tr}\left(\rho\,\widehat{(\sqcap_{\mathcal{J}})}_{\mathcal{L}_H}\right)$. If $\rho = 0$, then $\rho = \widehat{\sigma}_{\downarrow}(0)$. Otherwise, $\mathrm{Tr}\,\rho = r > 0$, hence $\rho = r\,\left(\frac{1}{r}\rho\right)$ with $\frac{1}{r}\rho \in \mathcal{S}^{\otimes}_{(\mathcal{L}_H,\widehat{\mathcal{M}}^{\text{DYN}},\widehat{\varphi}^{\text{DYN}})}$. Let $\mathcal{J},\,\mathcal{J}' \in \mathcal{L}_H$ with $\mathcal{J}\subset\mathcal{J}'$. Let $(e_i)_{i\in J}$ be an orthonormal basis of $\mathcal{J}$ and $(e_i)_{i\in J'\setminus J}$ ($J'\supset J$) be an orthonormal basis of $\mathcal{J}'\cap\mathcal{J}^{\perp}$. For $(n_i)_{i\in J'}\,,\,(m_i)_{i\in J'} \in \Lambda_{J'}$, we have:

$$\left\langle (n_i)_{i\in J'}\,,\,(e_i)_{i\in J'}\,\middle|\, \widehat{\left(\sqcap_{\mathcal{J}'\cap\mathcal{J}^{\perp}}|_{\mathcal{J}'\to\mathcal{J}'}\right)}\,\middle|\,(m_i)_{i\in J'}\,,\,(e_i)_{i\in J'}\right\rangle = \begin{cases}\sum_{i\in J'\setminus J} n_i & \text{if } \forall i\in J',\ n_i = m_i\\ 0 & \text{else}\end{cases},$$

and:

$$\left\langle (n_i)_{i\in J'}\,,\,(e_i)_{i\in J'}\,\middle|\, \left(\mathrm{id}_{\widehat{\mathcal{J}'}} - \widehat{\sqcap}_{\mathcal{J}'|\mathcal{J}}\right)\,\middle|\,(m_i)_{i\in J'}\,,\,(e_i)_{i\in J'}\right\rangle = \begin{cases}1 \text{ if } \forall i\in J',\ n_i = m_i \quad \& \quad \sum_{i\in J'\setminus J} n_i \geqslant 1\\ 0 \text{ else}\end{cases},$$

therefore:

$$\mathrm{Tr}_{\widehat{\mathcal{J}'}}\left(\rho_{\mathcal{J}'}\,\widehat{\sqcap}_{\mathcal{J}'|\mathcal{J}}\right) \geqslant r - \sum_{n\in\mathbb{N}} n\,\mathrm{Tr}_{\widehat{\mathcal{J}'}}\left[\rho_{\mathcal{J}'}\;\; \mathbb{I}_{\{n\}}\left(\widehat{\left(\sqcap_{\mathcal{J}'\cap\mathcal{J}^{\perp}}|_{\mathcal{J}'\to\mathcal{J}'}\right)}\right)\right] =: r - \mathrm{Tr}_{\widehat{\mathcal{J}'}}\left[\rho_{\mathcal{J}'}\,\widehat{\left(\sqcap_{\mathcal{J}'\cap\mathcal{J}^{\perp}}|_{\mathcal{J}'\to\mathcal{J}'}\right)}\right].$$

Now, $\mathrm{Tr}_{\widehat{\mathcal{J}'}}\left[\rho_{\mathcal{J}'}\,\widehat{\left(\sqcap_{\mathcal{J}'\cap\mathcal{J}^{\perp}}|_{\mathcal{J}'\to\mathcal{J}'}\right)}\right] = \sum_{n\in\mathbb{N}} n\,\mathrm{Tr}_{\widehat{\mathcal{J}'}}\left[\rho_{\mathcal{J}'}\;\; \mathbb{I}_{\{n\}}\left(\widehat{(\sqcap_{\mathcal{J}'})}_{\mathcal{J}'}\right)\right] - \sum_{n\in\mathbb{N}} n\,\mathrm{Tr}_{\widehat{\mathcal{J}'}}\left[\rho_{\mathcal{J}'}\;\; \mathbb{I}_{\{n\}}\left(\widehat{(\sqcap_{\mathcal{J}})}_{\mathcal{J}'}\right)\right] =$ $\mathrm{Tr}\left(\rho\,\widehat{(\sqcap_{\mathcal{J}'})}_{\mathcal{L}_H}\right) - \mathrm{Tr}\left(\rho\,\widehat{(\sqcap_{\mathcal{J}})}_{\mathcal{L}_H}\right)$, hence:

$$\inf_{\mathcal{J}'\supset\mathcal{J}} \mathrm{Tr}_{\widehat{\mathcal{J}'}}\left(\rho_{\mathcal{J}'}\,\widehat{\sqcap}_{\mathcal{J}'|\mathcal{J}}\right) \geqslant r - \sup_{\mathcal{J}'\supset\mathcal{J}} \mathrm{Tr}\left(\rho\,\widehat{(\sqcap_{\mathcal{J}'})}_{\mathcal{L}_H}\right) + \mathrm{Tr}\left(\rho\,\widehat{(\sqcap_{\mathcal{J}})}_{\mathcal{L}_H}\right).$$

Finally, $\left(\mathrm{Tr}\left(\rho\,\widehat{(\sqcap_{\mathcal{J}'})}_{\mathcal{L}_H}\right)\right)_{\mathcal{J}'\in\mathcal{L}_H}$ is increasing, so $\sup_{\mathcal{J}'\supset\mathcal{J}} \mathrm{Tr}\left(\rho\,\widehat{(\sqcap_{\mathcal{J}'})}_{\mathcal{L}_H}\right) = N$ and:

$$\sup_{\mathcal{J}\in\mathcal{L}_H}\inf_{\mathcal{J}'\supset\mathcal{J}} \mathrm{Tr}_{\widehat{\mathcal{J}'}}\left(\rho_{\mathcal{J}'}\,\widehat{\sqcap}_{\mathcal{J}'|\mathcal{J}}\right) \geqslant r - N + N = r\,.$$

On the other hand, $\forall \mathcal{J}\subset\mathcal{J}'$, $\mathrm{Tr}_{\widehat{\mathcal{J}'}}\left(\rho_{\mathcal{J}'}\,\widehat{\sqcap}_{\mathcal{J}'|\mathcal{J}}\right) \leqslant r$, thus, using prop. 3.17, $\frac{1}{r}\rho \in \widehat{\sigma}_{\downarrow}\left\langle \mathcal{S}_{\mathrm{Fock}}\right\rangle$, and therefore $\rho \in \widehat{\sigma}_{\downarrow}\left\langle \overline{\mathcal{S}}_{\mathrm{Fock}}\right\rangle$. □

Finally, we prove a convergence result at the quantum level. We define here two different notions of convergence, one stronger than the other, in both cases requiring convergence of the expectation values for a certain class of observables. To assess how exactly the convergence should be adjusted would require a closer study of which observables are really measured in practice, for these constitute the class of kinematical observables that we want to be able to transport on the dynamical side.

In addition, we need to introduce an $\epsilon$-dependent normalization parameter $\mathcal{N}$ that accounts for the fact that states solving the exact dynamics cannot be correctly normalized (they describe probability distributions invariant under a transformation running along the full time line from $t = -\infty$ to $t = +\infty$) so that it only makes sense to consider partial probability, measuring the probability of measuring the system in a certain state, knowing that the measurement takes place at a certain time. So, as we lift the $\epsilon$-regularization (that was making the gauge orbits compact and the solution of the quantum constraint normalizable), the probability of measuring the system in a certain time interval is dropping and needs to be accordingly compensated.

**Theorem 3.21** We consider the same objects as in props. 3.19 and 3.20. Let $\mathcal{I} \in \mathcal{L}$, $A$ be a bounded operator on $\widehat{\mathcal{I}}$ and $\varphi,\ \varphi' \in \mathcal{T}$. We additionally assume that $\varphi,\ \varphi'$ have compact support. On $\widehat{\mathcal{M}}^{\text{KIN}}_{\mathcal{I}}$, we define the operator:

$$R^{\mathcal{I}}_{A,\,\varphi,\,\varphi'} := A \otimes |\varphi \,\rangle\langle\, \varphi'|\,,$$

and, for $\varepsilon = (\mathcal{J},\ \epsilon) \in \mathcal{E}$ and $\rho \in \widehat{\sigma}_{\downarrow}\left\langle \overline{\mathcal{S}}_{\text{Fock}} \right\rangle$, we define:

$$R^{\mathcal{I},\varepsilon}_{A,\,\varphi,\,\varphi'}(\rho) := \frac{1}{\mathcal{N}(\epsilon,\ \varphi,\ \varphi')} \text{Tr}_{\widehat{\mathcal{M}}^{\text{KIN}}_{\mathcal{I}}} \left[\widehat{\Delta}^{\varepsilon}_{\mathcal{I}}(\rho)\, R^{\mathcal{I}}_{A,\,\varphi,\,\varphi'}\right],$$

where $\mathcal{N}(\epsilon,\ \varphi,\ \varphi') = \text{Tr}_{\mathcal{T}}\ |\varphi \,\rangle\langle\, \varphi'|\ |\delta_\epsilon \,\rangle\langle\, \delta_\epsilon| = \langle \varphi',\ \delta_\epsilon\rangle\ \langle \delta_\epsilon,\ \varphi\rangle$.

Then, the net $\left(R^{\mathcal{I},\varepsilon}_{A,\,\varphi,\,\varphi'}(\rho)\right)_{\varepsilon\in\mathcal{E}}$ converges.

For $e,\ f \in \mathcal{I}$, we also define on $\widehat{\mathcal{M}}^{\text{KIN}}_{\mathcal{I}}$ the operator:

$$R^{\mathcal{I}}_{e,\,f,\,\varphi,\,\varphi'} := \widehat{\text{a}}^{\mathcal{I},+}_{e}\ \widehat{\text{a}}^{\mathcal{I}}_{f} \otimes |\varphi \,\rangle\langle\, \varphi'|\,,$$

and, for $\varepsilon = (\mathcal{J},\ \epsilon) \in \mathcal{E}$ and $\rho \in \widehat{\mathcal{R}}$, we define:

$$R^{\mathcal{I},\varepsilon}_{e,\,f,\,\varphi,\,\varphi'}(\rho) := \frac{1}{\mathcal{N}(\epsilon,\ \varphi,\ \varphi')} \text{Tr}_{\widehat{\mathcal{M}}^{\text{KIN}}_{\mathcal{I}}} \left[\widehat{\Delta}^{\varepsilon}_{\mathcal{I}}(\rho)\, R^{\mathcal{I}}_{e,\,f,\,\varphi,\,\varphi'}\right].$$

Then, the net $\left(R^{\mathcal{I},\varepsilon}_{e,\,f,\,\varphi,\,\varphi'}(\rho)\right)_{\varepsilon\in\mathcal{E}}$ converges.

**Proof** *Bounded operator & Fock state.* Let $\rho \in \widehat{\sigma}_{\downarrow}\left\langle \overline{\mathcal{S}}_{\text{Fock}} \right\rangle$. For $\varepsilon = (\mathcal{J},\ \epsilon) \in \mathcal{E}$ and $\mathcal{J}' \in \mathcal{L}^{\varepsilon}$, we have:

$$\begin{aligned}
\widehat{\Delta}^{\varepsilon}_{\mathcal{J}'}(\rho) &= \widehat{\varphi}^{\text{KIN},-1}_{\mathcal{J}'\to\mathcal{J}} \left(|\zeta_{\mathcal{J}'\to\mathcal{J}} \,\rangle\langle\, \zeta_{\mathcal{J}'\to\mathcal{J}}| \otimes \left[e^{-i\,\widehat{H}^{\varepsilon}_{\mathcal{J}}\otimes\widehat{T}}\ (\rho_{\mathcal{J}} \otimes |\delta_\epsilon \,\rangle\langle\, \delta_\epsilon|)\ e^{i\,\widehat{H}^{\varepsilon}_{\mathcal{J}}\otimes\widehat{T}}\right]\right) \widehat{\varphi}^{\text{KIN}}_{\mathcal{J}'\to\mathcal{J}} \\
&= \widehat{\varphi}^{\text{KIN},-1}_{\mathcal{J}'\to\mathcal{J}}\ e^{-i\,\text{id}_{\widehat{\mathcal{J}'\cap\mathcal{J}^{\perp}}}\otimes\widehat{H}^{\varepsilon}_{\mathcal{J}}\otimes\widehat{T}}\ (|\zeta_{\mathcal{J}'\to\mathcal{J}} \,\rangle\langle\, \zeta_{\mathcal{J}'\to\mathcal{J}}| \otimes \rho_{\mathcal{J}} \otimes |\delta_\epsilon \,\rangle\langle\, \delta_\epsilon|)\ e^{i\,\text{id}_{\widehat{\mathcal{J}'\cap\mathcal{J}^{\perp}}}\otimes\widehat{H}^{\varepsilon}_{\mathcal{J}}\otimes\widehat{T}}\ \widehat{\varphi}^{\text{KIN}}_{\mathcal{J}'\to\mathcal{J}} \\
&= e^{-i\,\widehat{H}^{\varepsilon}_{\mathcal{J}'}\otimes\widehat{T}}\ \widehat{\varphi}^{\text{KIN},-1}_{\mathcal{J}'\to\mathcal{J}}\ (|\zeta_{\mathcal{J}'\to\mathcal{J}} \,\rangle\langle\, \zeta_{\mathcal{J}'\to\mathcal{J}}| \otimes \rho_{\mathcal{J}} \otimes |\delta_\epsilon \,\rangle\langle\, \delta_\epsilon|)\ \widehat{\varphi}^{\text{KIN}}_{\mathcal{J}'\to\mathcal{J}}\ e^{i\,\widehat{H}^{\varepsilon}_{\mathcal{J}'}\otimes\widehat{T}} \quad \text{(like in eq. (3.16.1))} \\
&= e^{-i\,\widehat{H}^{\varepsilon}_{\mathcal{J}'}\otimes\widehat{T}}\ \left(\left[\tau_{\mathcal{J}'\leftarrow\mathcal{J}}\ \rho_{\mathcal{J}}\ \tau^{+}_{\mathcal{J}'\leftarrow\mathcal{J}}\right] \otimes |\delta_\epsilon \,\rangle\langle\, \delta_\epsilon|\right)\ e^{i\,\widehat{H}^{\varepsilon}_{\mathcal{J}'}\otimes\widehat{T}}\,,
\end{aligned}$$

where $\tau_{\mathcal{J}'\leftarrow\mathcal{J}} = \widehat{\varphi}^{-1}_{\mathcal{J}'\to\mathcal{J}}\ (\zeta_{\mathcal{J}'\to\mathcal{J}} \otimes (\,\cdot\,))$. Hence, for $\mathcal{I} \subset \mathcal{J}'$:

$$R^{\mathcal{I},\varepsilon}_{A,\varphi,\varphi'}(\rho) = \frac{1}{\mathcal{N}(\epsilon,\,\varphi,\,\varphi')}\mathrm{Tr}_{\widehat{\mathcal{M}}^{\mathrm{KIN}}_{\mathcal{I}}}\left[\widehat{\Delta}^{\varepsilon}_{\mathcal{I}}(\rho)\,A \otimes |\varphi\,\rangle\langle\,\varphi'|\right] =$$

$$= \frac{\mathrm{Tr}_{\widehat{\mathcal{M}}^{\mathrm{KIN}}_{\mathcal{J}'}}\left[\widehat{\Delta}^{\varepsilon}_{\mathcal{J}'}(\rho)\;\left(\widehat{\varphi}^{-1}_{\mathcal{J}'\to\mathcal{I}}\;\left(\mathrm{id}_{\widehat{\mathcal{J}'\cap\mathcal{I}^{\perp}}}\otimes A\right)\;\widehat{\varphi}_{\mathcal{J}'\to\mathcal{I}}\right)\otimes|\varphi\,\rangle\langle\,\varphi'|\right]}{\mathcal{N}(\epsilon,\,\varphi,\,\varphi')}$$

$$= \frac{\mathrm{Tr}_{\widehat{\mathcal{M}}^{\mathrm{KIN}}_{\mathcal{J}'}}\left[e^{-i\,\widehat{H}^{\varepsilon}_{\mathcal{J}'}\otimes\widehat{T}}\;\left(\left[\tau_{\mathcal{J}'\leftarrow\mathcal{J}}\,\rho_{\mathcal{J}}\,\tau^{+}_{\mathcal{J}'\leftarrow\mathcal{J}}\right]\otimes|\delta_\epsilon\,\rangle\langle\,\delta_\epsilon|\right)\;e^{i\,\widehat{H}^{\varepsilon}_{\mathcal{J}'}\otimes\widehat{T}}\;\left(\widehat{\varphi}^{-1}_{\mathcal{J}'\to\mathcal{I}}\;\left(\mathrm{id}_{\widehat{\mathcal{J}'\cap\mathcal{I}^{\perp}}}\otimes A\right)\;\widehat{\varphi}_{\mathcal{J}'\to\mathcal{I}}\right)\otimes|\varphi\,\rangle\langle\,\varphi'|\right]}{\mathcal{N}(\epsilon,\,\varphi,\,\varphi')}$$

$$= \frac{\displaystyle\int_{-T}^{T} dt\,dt'\;\;\varphi(t)\varphi'^{*}(t')\,\delta_\epsilon(t')\delta^{*}_\epsilon(t)\;\;Z^{\varepsilon}_{\mathcal{J}'}(t,t')}{\displaystyle\int_{-T}^{T} dt\,dt'\;\;\varphi(t)\varphi'^{*}(t')\,\delta_\epsilon(t')\delta^{*}_\epsilon(t)},$$

where $T > 0$ is such that the support of $\varphi$ and $\varphi'$ is included in $[-T,\,T]$, and $Z^{\varepsilon}$ is defined as:

$$Z^{\varepsilon}(t,t') = \mathrm{Tr}_{\widehat{\mathcal{J}'}}\left[e^{-it'\,\widehat{H}^{\varepsilon}_{\mathcal{J}'}}\;\left(\tau_{\mathcal{J}'\leftarrow\mathcal{J}}\,\rho_{\mathcal{J}}\,\tau^{+}_{\mathcal{J}'\leftarrow\mathcal{J}}\right)\;e^{it\,\widehat{H}^{\varepsilon}_{\mathcal{J}'}}\;\left(\widehat{\varphi}^{-1}_{\mathcal{J}'\to\mathcal{I}}\;\left(\mathrm{id}_{\widehat{\mathcal{J}'\cap\mathcal{I}^{\perp}}}\otimes A\right)\;\widehat{\varphi}_{\mathcal{J}'\to\mathcal{I}}\right)\right]$$

$$= \mathrm{Tr}_{\widehat{\mathcal{J}}}\;\;\rho_{\mathcal{J}}\left[\tau^{+}_{\mathcal{J}'\leftarrow\mathcal{J}}\,e^{it\,\widehat{H}^{\varepsilon}_{\mathcal{J}'}}\;\left(\widehat{\varphi}^{-1}_{\mathcal{J}'\to\mathcal{I}}\;\left(\mathrm{id}_{\widehat{\mathcal{J}'\cap\mathcal{I}^{\perp}}}\otimes A\right)\;\widehat{\varphi}_{\mathcal{J}'\to\mathcal{I}}\right)\,e^{-it'\,\widehat{H}^{\varepsilon}_{\mathcal{J}'}}\;\tau_{\mathcal{J}'\leftarrow\mathcal{J}}\right]$$

$$= \mathrm{Tr}_{\widehat{\mathcal{J}}}\;\;\rho_{\mathcal{J}}\left[e^{it\,\widehat{H}^{\varepsilon}_{\mathcal{J}}}\;\tau^{+}_{\mathcal{J}'\leftarrow\mathcal{J}}\;\left(\widehat{\varphi}^{-1}_{\mathcal{J}'\to\mathcal{I}}\;\left(\mathrm{id}_{\widehat{\mathcal{J}'\cap\mathcal{I}^{\perp}}}\otimes A\right)\;\widehat{\varphi}_{\mathcal{J}'\to\mathcal{I}}\right)\;\tau_{\mathcal{J}'\leftarrow\mathcal{J}}\,e^{-it'\,\widehat{H}^{\varepsilon}_{\mathcal{J}}}\right],$$

for any $\mathcal{J}' \in \mathcal{L}$ such that $\mathcal{I},\,\mathcal{J} \subset \mathcal{J}'$.

Next, $\dfrac{\sqrt{\pi}}{\epsilon}\,\delta_\epsilon(t')\delta^{*}_\epsilon(t)$ converges uniformly to 1 for $t, t' \in [-T,\,T]$, when $\epsilon \to 0$. Therefore, we need to show that the net $\left(Z^{\varepsilon}(t,t')\right)_{\varepsilon\in\mathcal{E}}$ converges uniformly for $t, t' \in [-T,\,T]$.

Let $\rho_{\mathrm{Fock}} \in \overline{\mathcal{S}}_{\mathrm{Fock}}$ such that $\rho = \widehat{\sigma}_{\downarrow}\left(\rho_{\mathrm{Fock}}\right)$. Using the definition of $\widehat{\sigma}_{\downarrow}$, we can show:

$$Z^{\varepsilon}(t,t') = \mathrm{Tr}_{\widehat{\mathcal{H}}}\;\;\rho_{\mathrm{Fock}}\,\widehat{\varphi}^{-1}_{\mathcal{H}\to\mathcal{J}}\left[\mathbb{1}_{\widehat{\mathcal{J}^{\perp}}}\otimes e^{it\,\widehat{H}^{\varepsilon}_{\mathcal{J}}}\;\tau^{+}_{\mathcal{J}'\leftarrow\mathcal{J}}\;\left(\widehat{\varphi}^{-1}_{\mathcal{J}'\to\mathcal{I}}\;\left(\mathrm{id}_{\widehat{\mathcal{J}'\cap\mathcal{I}^{\perp}}}\otimes A\right)\;\widehat{\varphi}_{\mathcal{J}'\to\mathcal{I}}\right)\;\tau_{\mathcal{J}'\leftarrow\mathcal{J}}\,e^{-it'\,\widehat{H}^{\varepsilon}_{\mathcal{J}}}\right]\widehat{\varphi}_{\mathcal{H}\to\mathcal{J}}$$

$$= \mathrm{Tr}_{\widehat{\mathcal{H}}}\;\;\rho_{\mathrm{Fock}}\,e^{it\,\widehat{H}^{\varepsilon}}\;\widehat{\varphi}^{-1}_{\mathcal{H}\to\mathcal{J}}\left[\mathbb{1}_{\widehat{\mathcal{J}^{\perp}}}\otimes\tau^{+}_{\mathcal{J}'\leftarrow\mathcal{J}}\;\left(\widehat{\varphi}^{-1}_{\mathcal{J}'\to\mathcal{I}}\;\left(\mathrm{id}_{\widehat{\mathcal{J}'\cap\mathcal{I}^{\perp}}}\otimes A\right)\;\widehat{\varphi}_{\mathcal{J}'\to\mathcal{I}}\right)\;\tau_{\mathcal{J}'\leftarrow\mathcal{J}}\right]\widehat{\varphi}_{\mathcal{H}\to\mathcal{J}}\,e^{-it'\,\widehat{H}^{\varepsilon}}$$

(like in eq. (3.16.1))

$$= \mathrm{Tr}_{\widehat{\mathcal{J}}}\;\left[\mathrm{Tr}_{\mathcal{H}\to\mathcal{J}}\,e^{-it'\,\widehat{H}^{\varepsilon}}\;\rho_{\mathrm{Fock}}\,e^{it\,\widehat{H}^{\varepsilon}}\right]\left[\tau^{+}_{\mathcal{J}'\leftarrow\mathcal{J}}\;\left(\widehat{\varphi}^{-1}_{\mathcal{J}'\to\mathcal{I}}\;\left(\mathrm{id}_{\widehat{\mathcal{J}'\cap\mathcal{I}^{\perp}}}\otimes A\right)\;\widehat{\varphi}_{\mathcal{J}'\to\mathcal{I}}\right)\;\tau_{\mathcal{J}'\leftarrow\mathcal{J}}\right]$$

(by definition of $\mathrm{Tr}_{\mathcal{H}\to\mathcal{J}}$)

And using twice [11, eq. (2.1.1)] (for $\mathcal{J}, \mathcal{I} \subset \mathcal{J}' \subset \mathcal{H}$), we have, for any $\psi \in \mathcal{J}$:

$$\widehat{\varphi}_{\mathcal{H}\to\mathcal{I}}\;\circ\;\tau_{\mathrm{Fock}\leftarrow\mathcal{J}}(\psi) = \widehat{\varphi}_{\mathcal{H}\to\mathcal{I}}\;\circ\;\widehat{\varphi}^{-1}_{\mathcal{H}\to\mathcal{J}}\left(\zeta_{\mathrm{Fock}\to\mathcal{J}}\otimes\psi\right)$$

$$= \widehat{\varphi}_{\mathcal{H}\to\mathcal{I}}\;\circ\;\widehat{\varphi}^{-1}_{\mathcal{H}\to\mathcal{J}}\;\circ\;\left(\widehat{\varphi}^{-1}_{\mathcal{J}^{\perp}\to\mathcal{J}'\cap\mathcal{J}^{\perp}}\otimes\mathbb{1}_{\mathcal{J}}\right)\left(\zeta_{\mathrm{Fock}\to\mathcal{J}'}\otimes\zeta_{\mathcal{J}'\to\mathcal{J}}\otimes\psi\right)$$

$$= \widehat{\varphi}_{\mathcal{H}\to\mathcal{I}}\;\circ\;\widehat{\varphi}^{-1}_{\mathcal{H}\to\mathcal{J}'}\left(\zeta_{\mathrm{Fock}\to\mathcal{J}'}\otimes\tau_{\mathcal{J}'\leftarrow\mathcal{J}}(\psi)\right)$$

$$= \left(\widehat{\varphi}^{-1}_{\mathcal{I}^{\perp}\to\mathcal{J}'\cap\mathcal{I}^{\perp}}\otimes\mathrm{id}_{\mathcal{I}}\right)\left(\zeta_{\mathrm{Fock}\to\mathcal{J}'}\otimes\widehat{\varphi}_{\mathcal{J}'\to\mathcal{I}}\;\circ\;\tau_{\mathcal{J}'\leftarrow\mathcal{J}}(\psi)\right).$$

Hence, we get:

$$Z^{\varepsilon}(t,t') = \mathrm{Tr}_{\widehat{\mathcal{J}}}\ \left[\mathrm{Tr}_{\mathcal{H}\to\mathcal{J}}\, e^{-it'\widehat{H}^{\varepsilon}}\, \rho_{\mathrm{Fock}}\, e^{it\widehat{H}^{\varepsilon}}\right] \left[\tau^{+}_{\mathrm{Fock}\leftarrow\mathcal{J}}\ \left(\widehat{\varphi}^{-1}_{\mathcal{H}\to\mathcal{J}}\ \left(\mathrm{id}_{\widehat{\mathcal{J}^{\perp}}}\otimes A\right)\ \widehat{\varphi}_{\mathcal{H}\to\mathcal{J}}\right)\ \tau_{\mathrm{Fock}\leftarrow\mathcal{J}}\right]$$

$$= \mathrm{Tr}_{\widehat{\mathcal{H}}}\ \left[\tau_{\mathrm{Fock}\leftarrow\mathcal{J}}\ \left(\mathrm{Tr}_{\mathcal{H}\to\mathcal{J}}\, e^{-it'\widehat{H}^{\varepsilon}}\, \rho_{\mathrm{Fock}}\, e^{it\widehat{H}^{\varepsilon}}\right)\ \tau^{+}_{\mathrm{Fock}\leftarrow\mathcal{J}}\right] \left[\widehat{\varphi}^{-1}_{\mathcal{H}\to\mathcal{J}}\ \left(\mathrm{id}_{\widehat{\mathcal{J}^{\perp}}}\otimes A\right)\ \widehat{\varphi}_{\mathcal{H}\to\mathcal{J}}\right]$$

But we have:

$$\left\|\widehat{\varphi}^{-1}_{\mathcal{H}\to\mathcal{J}}\ \left(\mathrm{id}_{\widehat{\mathcal{J}^{\perp}}}\otimes A\right)\ \widehat{\varphi}_{\mathcal{H}\to\mathcal{J}}\right\| = \|A\| < \infty\,,$$

therefore, what remains to be shown is that the net:

$$\left(\tau_{\mathrm{Fock}\leftarrow\mathcal{J}}\ \left(\mathrm{Tr}_{\mathcal{H}\to\mathcal{J}}\, e^{-it'\widehat{H}^{\varepsilon}}\, \rho_{\mathrm{Fock}}\, e^{it\widehat{H}^{\varepsilon}}\right)\ \tau^{+}_{\mathrm{Fock}\leftarrow\mathcal{J}}\right)_{\varepsilon=(\mathcal{J},\epsilon)\in\mathcal{E}}$$

converges in trace norm (which was defined in [11, lemma 2.10]), uniformly for $t, t' \in [-T,\ T]$.

Let $\epsilon_o > 0$. We have:

$$\widehat{\mathcal{H}} = \overline{\bigoplus_{\mathcal{J}\in\mathcal{L}_H, N\geqslant 1} \widehat{\mathcal{J}}^{N}} \text{ where } \widehat{\mathcal{J}}^{N} = \bigoplus_{n\leqslant N} \widehat{\mathcal{J}}^{\otimes n,\,\mathrm{sym}}\,,$$

because $\mathcal{H} = \overline{\bigoplus_{\mathcal{J}\in\mathcal{L}_H}\mathcal{J}}$. Hence, we can prove, using the spectral decomposition of the self-adjoint traceclass operator $\rho_{\mathrm{Fock}}$ and the directed preorder on $\mathcal{L}_H$ and $\mathbb{N}$, that there exist $\mathcal{J}_o \in \mathcal{L}_H$ and $N_o \geqslant 1$ such that:

$$\|\rho_{\mathrm{Fock}} - \rho^{o}_{\mathrm{Fock}}\|_1 \leqslant \frac{\epsilon_o}{6}\,,$$

where $\rho^{o}_{\mathrm{Fock}} := \widehat{\Pi}_{\mathcal{J}_o,N_o}\,\rho_{\mathrm{Fock}}\,\widehat{\Pi}_{\mathcal{J}_o,N_o}$ (with $\widehat{\Pi}_{\mathcal{J}_o,N_o}$ the orthogonal projection on $\widehat{\mathcal{J}}_o^{N_o}$) and $\|\cdot\|_1$ denotes the trace norm.

Since $\widehat{\mathcal{J}}_o^{N_o}$ is finite dimensional, there exist vectors $\psi_\alpha \in \widehat{\mathcal{J}}_o^{N_o}$, $\alpha \in \{1,\dots,K\}$ (with $K\in\mathbb{N}$) such that:

$$\rho^{o}_{\mathrm{Fock}} = \sum_{\alpha=1}^{K} |\psi_\alpha \,\rangle\langle\, \psi_\alpha|\,.$$

We define:

$$\epsilon_1 := \frac{1}{N_o}\,\frac{1}{1+|T|}\,\log\left(1 + \frac{\epsilon_o}{12K\,\left(1+\max_\alpha \|\psi_\alpha\|^2\right)}\right) > 0\,,$$

and $H^{\epsilon_1} := \epsilon_1 \left\lfloor \frac{1}{\epsilon_1} H\right\rfloor$ (as in the proof of theorem 3.10). Then, since $H^{\epsilon_1}$ has discrete spectrum and $K,\ N_o < \infty$, we can construct $\mathcal{J}_1 \in \mathcal{L}_H$, such that $\mathcal{J}_1$ is stabilized by $H^{\epsilon_1}$ and:

$$\forall \alpha \leqslant K,\ \left\|\psi_\alpha - \widehat{\Pi}_{\mathcal{J}_1,N_o}\,\psi_\alpha\right\| \leqslant \frac{\epsilon_o}{12K\,(1+\max_\alpha\|\psi_\alpha\|)}\,.$$

Thus, we get:

$$\left\|\rho_{\mathrm{Fock}} - \widehat{\Pi}_{\mathcal{J}_1,N_o}\,\rho^{o}_{\mathrm{Fock}}\,\widehat{\Pi}_{\mathcal{J}_1,N_o}\right\|_1 \leqslant \frac{\epsilon_o}{6} + \sum_{\alpha=1}^{K}\left\||\psi_\alpha\,\rangle\langle\,\psi_\alpha| - \left|\widehat{\Pi}_{\mathcal{J}_1,N_o}\,\psi_\alpha\ \right\rangle\!\left\langle\ \widehat{\Pi}_{\mathcal{J}_1,N_o}\,\psi_\alpha\right|\right\|_1$$

$$\leqslant \frac{\epsilon_o}{6} + \sum_{\alpha=1}^{K} \left\| \psi_\alpha - \widehat{\Pi}_{\mathcal{J}_1, N_o}\, \psi_\alpha \right\| \, \|\psi_\alpha\| + \left\| \widehat{\Pi}_{\mathcal{J}_1, N_o}\, \psi_\alpha \right\| \, \left\| \psi_\alpha - \widehat{\Pi}_{\mathcal{J}_1, N_o}\, \psi_\alpha \right\| \leqslant \frac{\epsilon_o}{3}$$

Now, we consider $\mathcal{J}_2 \in \mathcal{L}_H$ such that $\mathcal{J}_o + \mathcal{J}_1 \subset \mathcal{J}_2$. For all $t, t' \in [-T,\, T]$, we have:

$$\left\| e^{-it' \left(\widehat{\Pi_{\mathcal{J}_2} H \Pi_{\mathcal{J}_2}}\right)}\, \rho_{\text{Fock}}\, e^{it \left(\widehat{\Pi_{\mathcal{J}_2} H \Pi_{\mathcal{J}_2}}\right)} - e^{-it' \widehat{H}^{\epsilon_1}}\, \widehat{\Pi}_{\mathcal{J}_1, N_o}\, \rho^o_{\text{Fock}}\, \widehat{\Pi}_{\mathcal{J}_1, N_o}\, e^{it \widehat{H}^{\epsilon_1}} \right\|_1 \leqslant$$

$$\leqslant \frac{\epsilon_o}{3} + \left\| e^{-it' \left(\widehat{\Pi_{\mathcal{J}_2} H \Pi_{\mathcal{J}_2}}\right)} - e^{-it' \left(\widehat{\Pi_{\mathcal{J}_2} H^{\epsilon_1} \Pi_{\mathcal{J}_2}}\right)} \right\| \, \|\rho^o_{\text{Fock}}\|_1 + \|\rho^o_{\text{Fock}}\|_1 \, \left\| e^{it \left(\widehat{\Pi_{\mathcal{J}_2} H \Pi_{\mathcal{J}_2}}\right)} - e^{it \left(\widehat{\Pi_{\mathcal{J}_2} H^{\epsilon_1} \Pi_{\mathcal{J}_2}}\right)} \right\|$$

(since $H^{\epsilon_1}$ stabilizes $\mathcal{J}_1 \subset \mathcal{J}_2$)

$$\leqslant \frac{\epsilon_o}{3} + 2 \left| e^{T\, N_o\, \epsilon_1} - 1 \right| \, \|\rho^o_{\text{Fock}}\|_1 \leqslant \frac{\epsilon_o}{2}\,,$$

and similarly:

$$\left\| e^{-it' \widehat{H}}\, \rho_{\text{Fock}}\, e^{it \widehat{H}} - e^{-it' \widehat{H}^{\epsilon_1}}\, \widehat{\Pi}_{\mathcal{J}_1, N_o}\, \rho^o_{\text{Fock}}\, \widehat{\Pi}_{\mathcal{J}_1, N_o}\, e^{it \widehat{H}^{\epsilon_1}} \right\|_1 \leqslant \frac{\epsilon_o}{2}\,.$$

Next, using again that $H^{\epsilon_1}$ stabilizes $\mathcal{J}_1 \subset \mathcal{J}_2$, we also have:

$$\tau_{\text{Fock}\leftarrow\mathcal{J}_2} \left[ \text{Tr}_{\mathcal{H}\to\mathcal{J}_2}\;\; e^{-it' \widehat{H}^{\epsilon_1}}\, \widehat{\Pi}_{\mathcal{J}_1, N_o}\, \rho^o_{\text{Fock}}\, \widehat{\Pi}_{\mathcal{J}_1, N_o}\, e^{it \widehat{H}^{\epsilon_1}} \right] \tau^+_{\text{Fock}\leftarrow\mathcal{J}_2} = e^{-it' \widehat{H}^{\epsilon_1}}\, \widehat{\Pi}_{\mathcal{J}_1, N_o}\, \rho^o_{\text{Fock}}\, \widehat{\Pi}_{\mathcal{J}_1, N_o}\, e^{it \widehat{H}^{\epsilon_1}}\,.$$

Hence, for any $\varepsilon = (\mathcal{J}_2,\, \epsilon_2) \in \mathcal{E}$ such that $(\mathcal{J}_o + \mathcal{J}_1,\, 1) \preccurlyeq \varepsilon$, and any $t, t' \in [-T,\, T]$, we have:

$$\left\| \tau_{\text{Fock}\leftarrow\mathcal{J}_2} \left[ \text{Tr}_{\mathcal{H}\to\mathcal{J}_2}\, e^{-it' \widehat{H}^{\varepsilon}}\, \rho_{\text{Fock}}\, e^{it \widehat{H}^{\varepsilon}} \right] \tau^+_{\text{Fock}\leftarrow\mathcal{J}_2} - e^{-it' \widehat{H}}\, \rho_{\text{Fock}}\, e^{it \widehat{H}} \right\|_1 \leqslant \epsilon_o\,,$$

which provides the desired convergence.

*Transition operator & regular state.* Let $\rho \in \widehat{\mathcal{R}}$, $\mathcal{I} \in \mathcal{L}$ and $e, f \in \mathcal{I}$. Since $\widehat{\mathcal{R}} \subset \widehat{\sigma}_\downarrow \left\langle \overline{\mathcal{S}}_{\text{Fock}} \right\rangle$ (prop. 3.20), there exists $\rho_{\text{Fock}} \in \overline{\mathcal{S}}_{\text{Fock}}$ such that $\rho = \widehat{\sigma}_\downarrow(\rho_{\text{Fock}})$. Like above, a sufficient condition for the convergence of the net $\left( R^{\mathcal{I},\varepsilon}_{e,\, f,\, \varphi,\, \varphi'}(\rho) \right)_{\varepsilon\in\mathcal{E}}$ is uniform convergence for $t, t' \in [-T,\, T]$ of the net:

$$\left( \text{Tr}_{\widehat{\mathcal{H}}}\;\; \left[ \tau_{\text{Fock}\leftarrow\mathcal{J}} \left( \text{Tr}_{\mathcal{H}\to\mathcal{J}}\, e^{-it' \widehat{H}^{\varepsilon}}\, \rho_{\text{Fock}}\, e^{it \widehat{H}^{\varepsilon}} \right) \tau^+_{\text{Fock}\leftarrow\mathcal{J}} \right] T_{e,f} \right)_{\varepsilon=(\mathcal{J},\epsilon)\in\mathcal{E}}\,,$$

where we define:

$$T_{e,f} := \widehat{\varphi}^{-1}_{\mathcal{H}\to\mathcal{I}} \left( \text{id}_{\widehat{\mathcal{I}^\perp}} \otimes \widehat{a}^{\mathcal{I},+}_e\, \widehat{a}^{\mathcal{I}}_f \right) \widehat{\varphi}_{\mathcal{H}\to\mathcal{I}}\,.$$

Choosing an orthonormal basis $(e_i)_{i\in I}$ of $\mathcal{I}$ and completing it into an orthonormal basis $(e_i)_{i\in\mathbb{N}}$ of $\mathcal{I}'$ ($I \subset \mathbb{N}$), we get:

$$T_{e,f} = \sum_{i,j\in I} \langle e_i,\, e\rangle \, \langle f,\, e_j\rangle \, \widehat{\varphi}^{-1}_{\mathcal{H}\to\mathcal{I}} \left( \text{id}_{\widehat{\mathcal{I}^\perp}} \otimes \widehat{a}^{\mathcal{I},+}_{e_i}\, \widehat{a}^{\mathcal{I}}_{e_j} \right) \widehat{\varphi}_{\mathcal{H}\to\mathcal{I}}$$

$$= \sum_{i,j\in I} \langle e_i,\, e\rangle \, \langle f,\, e_j\rangle \left( \widehat{a}^{\text{Fock},+}_{e_i}\, \widehat{a}^{\text{Fock}}_{e_j} \right) \text{ (like in the proof of prop. 3.16)}$$

$$= \widehat{a}^{\text{Fock},+}_e\, \widehat{a}^{\text{Fock}}_f\,.$$

Now, from the definition of the creation and annihilation operators, we have:

$$T_{e,f} = \sum_{n=0}^{\infty} \widehat{\Pi}^{(n)}\, \widehat{a}_e^{\text{Fock},+}\, \widehat{a}_f^{\text{Fock}}\, \widehat{\Pi}^{(n)}\,,$$

where, for all $n \in \mathbb{N}$, $\widehat{\Pi}^{(n)}$ is the orthogonal projection on the subspace $\mathcal{H}^{\otimes n,\,\text{sym}}$ of $\widehat{\mathcal{H}}$, and:

$$\left\| \widehat{\Pi}^{(n)}\, \widehat{a}_e^{\text{Fock},+}\, \widehat{a}_f^{\text{Fock}}\, \widehat{\Pi}^{(n)} \right\| \leqslant n\ \|e\|\ \|f\|\,.$$

On the other hand, we have, by definition of $\widehat{\mathcal{R}}$, $\sup_{\mathcal{J}\in\mathcal{L}_H} \operatorname{Tr}\left(\rho\,\widehat{(\Pi_{\mathcal{J}})}_{\mathcal{L}_H}\right) =: N_{\text{tot}} < \infty$, so:

$$\begin{aligned}
\sum_{n=0}^{\infty} n \operatorname{Tr}_{\widehat{\mathcal{H}}} \rho_{\text{Fock}}\, \widehat{\Pi}^{(n)} &= \sum_{n\in\mathbb{N}} n \operatorname{Tr}_{\widehat{\mathcal{H}}} \rho_{\text{Fock}}\, \mathbb{I}_{\{n\}} \left(\widehat{\langle \text{id}_{\mathcal{H}} \rangle}_{\text{Fock}}\right) \\
&= \sum_{n\in\mathbb{N}} n \sup_{\mathcal{J}\in\mathcal{L}_H} \operatorname{Tr}_{\widehat{\mathcal{H}}} \rho_{\text{Fock}}\, \mathbb{I}_{\{n\}} \left(\widehat{\langle \Pi_{\mathcal{J}} \rangle}_{\text{Fock}}\right) \text{ (using [11, lemma 2.10])} \\
&= \sum_{n\in\mathbb{N}} n \sup_{\mathcal{J}\in\mathcal{L}_H} \operatorname{Tr}_{\widehat{\mathcal{J}}} \rho_{\mathcal{J}}\, \mathbb{I}_{\{n\}} \left(\widehat{(\Pi_{\mathcal{J}})}_{\mathcal{J}}\right) = N_{\text{tot}} \text{ (from eq. (3.17.1) ).}
\end{aligned}$$

Let $\epsilon_o > 0$. Then, there exists $N_o \geqslant 1$ such that:

$$\sum_{n>N_o} n \operatorname{Tr}_{\widehat{\mathcal{H}}} \rho_{\text{Fock}}\, \widehat{\Pi}^{(n)} \leqslant \frac{\epsilon_o}{3}\,,$$

and therefore, for all $\varepsilon = (\mathcal{J},\, \epsilon) \in \mathcal{E}$:

$$\sum_{n>N_o} n \left\| \left[ \tau_{\text{Fock}\leftarrow\mathcal{J}} \left( \operatorname{Tr}_{\mathcal{H}\to\mathcal{J}} e^{-it'\widehat{H}^{\varepsilon}}\, \rho_{\text{Fock}}\, e^{it\widehat{H}^{\varepsilon}} \right) \tau^{+}_{\text{Fock}\leftarrow\mathcal{J}} \right] \widehat{\Pi}^{(n)} \right\|_1 =$$

$$= \sum_{n>N_o} n \left\| \tau_{\text{Fock}\leftarrow\mathcal{J}} \left( \operatorname{Tr}_{\mathcal{H}\to\mathcal{J}} e^{-it'\widehat{H}^{\varepsilon}}\, \rho_{\text{Fock}}\, e^{it\widehat{H}^{\varepsilon}} \right) \widehat{\Pi}^{(n)}_{\mathcal{J}}\, \tau^{+}_{\text{Fock}\leftarrow\mathcal{J}} \right\|_1$$

(where for all $n \in \mathbb{N}$, $\widehat{\Pi}^{(n)}_{\mathcal{J}}$ is the orthogonal projection on the subspace $\mathcal{J}^{\otimes n,\,\text{sym}}$ of $\widehat{\mathcal{J}}$)

$$\begin{aligned}
&= \sum_{n>N_o} n \left\| \left( \operatorname{Tr}_{\mathcal{H}\to\mathcal{J}} e^{-it'\widehat{H}^{\varepsilon}}\, \rho_{\text{Fock}}\, e^{it\widehat{H}^{\varepsilon}} \right) \widehat{\Pi}^{(n)}_{\mathcal{J}} \right\|_1 \\
&= \sum_{n>N_o} n \left\| \operatorname{Tr}_{\mathcal{H}\to\mathcal{J}}\ e^{-it'\widehat{H}^{\varepsilon}}\, \rho_{\text{Fock}}\, e^{it\widehat{H}^{\varepsilon}}\, \widehat{\varphi}^{-1}_{\mathcal{H}\to\mathcal{J}} \left( \text{id}_{\widehat{\mathcal{J}^{\perp}}} \otimes \widehat{\Pi}^{(n)}_{\mathcal{J}} \right) \widehat{\varphi}_{\mathcal{H}\to\mathcal{J}} \right\|_1 \\
&\leqslant \sum_{n>N_o} n \left\| e^{-it'\widehat{H}^{\varepsilon}}\, \rho_{\text{Fock}}\, e^{it\widehat{H}^{\varepsilon}}\, \widehat{\varphi}^{-1}_{\mathcal{H}\to\mathcal{J}} \left( \text{id}_{\widehat{\mathcal{J}^{\perp}}} \otimes \widehat{\Pi}^{(n)}_{\mathcal{J}} \right) \widehat{\varphi}_{\mathcal{H}\to\mathcal{J}} \right\|_1 \\
&\leqslant \sum_{n>N_o} \sum_{n'\geqslant 0} n \left\| e^{-it'\widehat{H}^{\varepsilon}}\, \rho_{\text{Fock}}\, e^{it\widehat{H}^{\varepsilon}}\, \widehat{\varphi}^{-1}_{\mathcal{H}\to\mathcal{J}} \left( \widehat{\Pi}^{(n')}_{\mathcal{J}^{\perp}} \otimes \widehat{\Pi}^{(n)}_{\mathcal{J}} \right) \widehat{\varphi}_{\mathcal{H}\to\mathcal{J}} \right\|_1 \\
&\leqslant \sum_{n>N_o} \sum_{n'\geqslant 0} (n+n') \left\| e^{-it'\widehat{H}^{\varepsilon}}\, \rho_{\text{Fock}}\, e^{it\widehat{H}^{\varepsilon}}\, \widehat{\varphi}^{-1}_{\mathcal{H}\to\mathcal{J}} \left( \widehat{\Pi}^{(n')}_{\mathcal{J}^{\perp}} \otimes \widehat{\Pi}^{(n)}_{\mathcal{J}} \right) \widehat{\varphi}_{\mathcal{H}\to\mathcal{J}} \right\|_1
\end{aligned}$$

$$= \sum_{n>N_o} \sum_{n' \geqslant 0} (n+n') \left\| e^{-it'\widehat{H}^{\varepsilon}} \rho_{\mathrm{Fock}} \,\widehat{\varphi}^{-1}_{\mathcal{H}\to\mathcal{J}} \left( \widehat{\Pi}^{(n')}_{\mathcal{J}^\perp} \otimes \widehat{\Pi}^{(n)}_{\mathcal{J}} \right) \widehat{\varphi}_{\mathcal{H}\to\mathcal{J}} \, e^{it\widehat{H}^{\varepsilon}} \right\|_1$$

(for $\widehat{H}^{\varepsilon}$ stabilizes the subspaces $\left[ \left(\mathcal{J}^\perp\right)^{\otimes n'} \otimes \mathcal{J}^{\otimes n} \right]^{\mathrm{sym}}$ for all $n, n'$)

$$\leqslant \sum_{n>N_o} \sum_{n' \geqslant 0} (n+n') \left\| \rho_{\mathrm{Fock}} \,\widehat{\varphi}^{-1}_{\mathcal{H}\to\mathcal{J}} \left( \widehat{\Pi}^{(n')}_{\mathcal{J}^\perp} \otimes \widehat{\Pi}^{(n)}_{\mathcal{J}} \right) \widehat{\varphi}_{\mathcal{H}\to\mathcal{J}} \right\|_1$$

$$= \sum_{n>N_o} \sum_{n' \geqslant 0} (n+n') \,\mathrm{Tr}_{\widehat{\mathcal{H}}} \,\rho_{\mathrm{Fock}} \,\widehat{\varphi}^{-1}_{\mathcal{H}\to\mathcal{J}} \left( \widehat{\Pi}^{(n')}_{\mathcal{J}^\perp} \otimes \widehat{\Pi}^{(n)}_{\mathcal{J}} \right) \widehat{\varphi}_{\mathcal{H}\to\mathcal{J}}$$

$$\leqslant \sum_{n''>N_o} n'' \,\mathrm{Tr}_{\widehat{\mathcal{H}}} \,\rho_{\mathrm{Fock}} \,\widehat{\Pi}^{(n'')} \leqslant \frac{\epsilon_o}{3} \,.$$

Next, from the previous point there exists $\varepsilon_o \in \mathcal{E}$ such that, for all $\varepsilon = (\mathcal{J}, \epsilon) \succcurlyeq \varepsilon_o$:

$$\left\| \tau_{\mathrm{Fock}\leftarrow\mathcal{J}} \left[ \mathrm{Tr}_{\mathcal{H}\to\mathcal{J}} \, e^{-it'\widehat{H}^{\varepsilon}} \rho_{\mathrm{Fock}} \, e^{it\widehat{H}^{\varepsilon}} \right] \tau^{+}_{\mathrm{Fock}\leftarrow\mathcal{J}} - e^{-it'\widehat{H}} \rho_{\mathrm{Fock}} \, e^{it\widehat{H}} \right\|_1 \leqslant \frac{\epsilon_o}{3\,N_o} \,,$$

thus:

$$\left| \mathrm{Tr}_{\widehat{\mathcal{H}}} \left[ \tau_{\mathrm{Fock}\leftarrow\mathcal{J}} \left( \mathrm{Tr}_{\mathcal{H}\to\mathcal{J}} \, e^{-it'\widehat{H}^{\varepsilon}} \rho_{\mathrm{Fock}} \, e^{it\widehat{H}^{\varepsilon}} \right) \tau^{+}_{\mathrm{Fock}\leftarrow\mathcal{J}} \right] T_{e,f} - \mathrm{Tr}_{\widehat{\mathcal{H}}} \; e^{-it'\widehat{H}} \rho_{\mathrm{Fock}} \, e^{it\widehat{H}} \, T_{e,f} \right|$$

$$\leqslant \left\| \left[ \tau_{\mathrm{Fock}\leftarrow\mathcal{J}} \left( \mathrm{Tr}_{\mathcal{H}\to\mathcal{J}} \, e^{-it'\widehat{H}^{\varepsilon}} \rho_{\mathrm{Fock}} \, e^{it\widehat{H}^{\varepsilon}} \right) \tau^{+}_{\mathrm{Fock}\leftarrow\mathcal{J}} \right] - e^{-it'\widehat{H}} \rho_{\mathrm{Fock}} \, e^{it\widehat{H}} \right\|_1 \times$$

$$\times \left\| \sum_{n\leqslant N_o} \widehat{\Pi}^{(n)} \,\widehat{a}^{\mathrm{Fock},+}_{e} \,\widehat{a}^{\mathrm{Fock}}_{f} \,\widehat{\Pi}^{(n)} \right\| + \sum_{n>N_o} \left\| \left[ \tau_{\mathrm{Fock}\leftarrow\mathcal{J}} \left( \mathrm{Tr}_{\mathcal{H}\to\mathcal{J}} \, e^{-it'\widehat{H}^{\varepsilon}} \rho_{\mathrm{Fock}} \, e^{it\widehat{H}^{\varepsilon}} \right) \tau^{+}_{\mathrm{Fock}\leftarrow\mathcal{J}} \right] \widehat{\Pi}^{(n)} \right\|_1 \times$$

$$\times \left\| \widehat{\Pi}^{(n)} \,\widehat{a}^{\mathrm{Fock},+}_{e} \,\widehat{a}^{\mathrm{Fock}}_{f} \,\widehat{\Pi}^{(n)} \right\| + \sum_{n>N_o} \left\| e^{-it'\widehat{H}} \rho_{\mathrm{Fock}} \, e^{it\widehat{H}} \,\widehat{\Pi}^{(n)} \right\|_1 \left\| \widehat{\Pi}^{(n)} \,\widehat{a}^{\mathrm{Fock},+}_{e} \,\widehat{a}^{\mathrm{Fock}}_{f} \,\widehat{\Pi}^{(n)} \right\|$$

$$\leqslant \frac{\epsilon_o}{3\,N_o} N_o \,\|e\|\,\|f\| + \frac{\epsilon_o}{3} \,\|e\|\,\|f\| + \sum_{n>N_o} \left\| e^{-it'\widehat{H}} \rho_{\mathrm{Fock}} \,\widehat{\Pi}^{(n)} \, e^{it\widehat{H}} \right\|_1 n \,\|e\|\,\|f\|$$

(for $e^{it\widehat{H}}$ stabilizes the subspaces $\mathcal{H}^{\otimes n,\,\mathrm{sym}}$ for all $n$)

$$\leqslant \epsilon_o \,\|e\|\,\|f\| \,,$$

which concludes the proof. □

# 4 Outlook

While it will be essential to play with more toy models (and especially with more sophisticated ones), in order to sharpen our still rather crude proposal for dealing with constraints, we have at least ascertained that this program can be applied to the most simple quantum field theory, where it satisfactorily reproduces established results. Indeed, we found that we can define a sensible convergence at the quantum level, on a subspace of states that can either be identified with the Fock space or with a subset of it. This is reassuring, for we know that the Fock space is the right arena to describe interaction-free theory (since such a theory preserves the subspaces of fixed particles number). It would be interesting to study whether more general quantum field theories can be translated in this language too.

On the classical side, we would like to develop systematic recipes to generate the input needed for the regularization. On the quantum side, we still have to provide a rigorous procedure, including rules for defining an effective and physically meaningful notion of convergence. As a general guiding principle, we should strive to reflect the concrete experimental implementation of the observables. In particular, when considering a theory of gravity, it might prove legitimate to define the convergence in a way that completely ignores the gravitational degrees of freedom: indeed, geometry is only probed by matter, and never measured directly.

Additionally, we might be able to gain a deeper understanding of the formalism considered here by studying its relations to approaches that incorporate similar ingredients, like lattice quantum field theory or other discretization techniques. This could help shed light on issues that are shared with these approaches, notably the problem of 'universality': in other words, the concern about how to ensure that the results we are getting are robust, and do not depend critically on some arbitrary choices entering the definition of the regularization scheme. We have displayed in section 2 a trick to circumvent this pitfall: by assembling all reasonable approximations into a huge label set $\mathcal{E}$, and ordering them by their respective quality, we can view a specific regularization prescription as simply selecting a cofinal subset in $\mathcal{E}$. However, it is not clear whether this could still be done for less trivial systems, because it could become difficult to arrange for $\mathcal{E}$ to be directed. Hence, we will probably need to invent more subtle ways of ensuring universal properties.

## Acknowledgements

This work has been financially supported by the Université François Rabelais, Tours, France.

This research project has been supported by funds to Emerging Field Project "Quantum Geometry" from the FAU Erlangen-Nuernberg within its Emerging Fields Initiative.

# A References